\documentclass[aps,twocolumn,prx,floatfix,preprintnumbers, 10pt]{revtex4-2}
\usepackage{xpatch}
\makeatletter
\patchcmd{\@ssect@ltx}
    {\addcontentsline{toc}{#1}{\protect\numberline{}#8}}
    {}
    {}
    {}
\makeatother
\pdfoutput=1
\usepackage{color}
\usepackage{enumitem}
\makeatletter

\renewcommand{\thesection}{\arabic{section}}
\renewcommand{\thesubsection}{\thesection.\arabic{subsection}}
\renewcommand{\thesubsubsection}{\thesubsection.\arabic{subsubsection}}

\renewcommand{\p@section}{}
\renewcommand{\p@subsection}{}
\renewcommand{\p@subsubsection}{}

\let\oldappendix\appendix
\renewcommand{\appendix}{%
  \oldappendix
  \renewcommand{\thesection}{\Alph{section}}%
  \renewcommand{\thesubsection}{\thesection.\arabic{subsection}}%
  \renewcommand{\thesubsubsection}{\thesubsection.\arabic{subsubsection}}%
  \renewcommand{\p@section}{}%
  \renewcommand{\p@subsection}{}%
  \renewcommand{\p@subsubsection}{}%
  \renewcommand{\theHsection}{appendix.\Alph{section}}%
  \renewcommand{\theHsubsection}{\theHsection.\arabic{subsection}}%
  \renewcommand{\theHsubsubsection}{\theHsubsection.\arabic{subsubsection}}%
}

\makeatother

\usepackage{booktabs}
\usepackage{makecell}
\usepackage{cancel} 

\newcommand{\paragraf}[1]{\par\medskip\noindent{\bf #1}}

\usepackage{mathtools}
\usepackage[dvipsnames]{xcolor}
\usepackage{float}

\usepackage{hyperref}
\definecolor{amaranth}{rgb}{0.9, 0.17, 0.31}
\definecolor{forestForestGreen(web)}{rgb}{0.13, 0.55, 0.13}
\definecolor{blue(munsell)}{HTML}{005567}
\definecolor{oxfordblue}{rgb}{0.0, 0.2, 0.4}
\definecolor{britishracinggreen}{rgb}{0.0, 0.26, 0.15}

\definecolor{bblue}{rgb}{0.0, 0.58, 0.71}
\hypersetup{
pdfstartview={FitH}, % fits the width of the page to the window
pdftitle={}, % title
colorlinks=true, % false: boxed links; true: colored links
linkcolor=oxfordblue, % color of internal links (change box color with linkbordercolor)
citecolor=oxfordblue, % color of links to bibliography
filecolor=oxfordblue, % color of file links
urlcolor=oxfordblue% color of external links
}

\usepackage{pgfplots}
\pgfplotsset{compat=1.18}
\usepackage{amssymb}
\usepackage{amsfonts}
\usepackage{graphicx}
\usepackage{epstopdf}
\usepackage{dcolumn}
\usepackage{amsmath}
\usepackage{latexsym,bm}
\usepackage{amsthm}
\usepackage{slashed}
\usepackage{color}
\usepackage{url}
\usepackage{rotating}
\usepackage[normalem]{ulem}
\usepackage{faktor}
\usepackage{tikz-cd}
\usepackage{tensor}
\usepackage{array}
\usepackage{multirow}
\usepackage{comment}

\usepackage{tabularx}
\usepackage{makecell}
\usepackage[normalem]{ulem}
\usepackage{changepage}
 \numberwithin{equation}{section}

\definecolor{ssnblue}{HTML}{006B8F}

\definecolor{darkgreen}{HTML}{006400}

\newcommand{\irrep}{\chi}

\DeclareMathOperator{\Core}{Core}

\usepackage{makecell}
\newtheorem{theorem}{Theorem}[section]
\newtheorem{proposition}[theorem]{Proposition}
\newtheorem{lemma}[theorem]{Lemma}
\newtheorem{corollary}[theorem]{Corollary}
\theoremstyle{definition}
\newtheorem{definition}[theorem]{Definition}
\theoremstyle{remark}

\usepackage{tikz}
\usepackage{diagbox}
\usetikzlibrary{positioning}
\usetikzlibrary{calc}
\usetikzlibrary{decorations.pathreplacing,calligraphy}

\usepackage{xstring}
\usetikzlibrary{decorations.pathmorphing} 
\usetikzlibrary{decorations.markings} 
\usetikzlibrary{arrows} 
\usetikzlibrary{decorations.pathreplacing,decorations.markings}
\usetikzlibrary{arrows.meta,decorations.markings}
\usetikzlibrary{shapes} 
\usetikzlibrary{matrix} 
\usetikzlibrary{positioning} 
\usepackage[english]{babel} 
\usepackage[autostyle]{csquotes}
\usepackage{pifont}
\usetikzlibrary{shapes.multipart}

\tikzset{->-/.style={decoration={
  markings,
  mark=at position .6 with {\arrow{Latex[length=1.5mm,width=1.5mm]}}},postaction={decorate}}}

\newcommand{\bea}{\begin{eqnarray}}
\newcommand{\eea}{\end{eqnarray}}
\newcommand{\be}{\begin{equation}}
\newcommand{\ee}{\end{equation}}
\newcommand{\ba}{\begin{aligned}}
\newcommand{\ea}{\end{aligned}}
\newcommand{\bit}{\begin{itemize}}
\newcommand{\eit}{\end{itemize}}
\newcommand{\ben}{\begin{enumerate}}
\newcommand{\een}{\end{enumerate}}

\newcommand{\id}{\text{id}}

\newcommand{\im}{\operatorname{Im}}
\newcommand{\Aut}{\operatorname{Aut}}

\newcommand{\Neu}{\text{Neu}}
\newcommand{\Dir}{\text{Dir}}

\newcommand{\lb}{\left(}
\newcommand{\rb}{\right)}

\newcommand{\Z}{{\mathbb Z}}
\newcommand{\R}{{\mathbb R}}
\newcommand{\bC}{{\mathbb C}}

\newcommand{\Fun}{\text{Fun}}

\newcommand{\cA}{\mathcal{A}}
\newcommand{\cB}{\mathcal{B}}

\newcommand{\cC}{\mathcal{C}}
\newcommand{\cD}{\mathcal{D}}
\newcommand{\cE}{\mathcal{E}}
\newcommand{\cF}{\mathcal{F}}
\newcommand{\cG}{\mathcal{G}}

\newcommand{\cL}{\mathcal{L}}
\newcommand{\cM}{\mathcal{M}}
\newcommand{\cN}{\mathcal{N}}

\newcommand{\cR}{\mathcal{R}}

\newcommand{\cT}{\mathcal{T}}

\newcommand{\cV}{\mathcal{V}}

\newcommand{\cZ}{\mathcal{Z}}

\newcommand{\fB}{\mathfrak{B}}
\newcommand{\fZ}{\mathfrak{Z}}

\newcommand{\Hom}{\text{Hom}}

\newcommand{\sVec}{\mathsf{sVec}}
\newcommand{\Rep}{\mathsf{Rep}}
\newcommand{\Mod}{\mathsf{Mod}}

\renewcommand{\dim}{\operatorname{dim}}

\newcommand{\Vect}{\mathsf{Vec}}
\newcommand{\TwoVec}{2\mathsf{Vec}}

\newcommand{\TY}{\mathsf{TY}}

\newcommand{\sym}{\text{sym}}
\newcommand{\phys}{\text{phys}}

\makeatother

\makeatletter
\renewcommand\subsection{\@startsection{subsection}{3}{\z@}%
                                     {-3.25ex\@plus -1ex \@minus -.2ex}%
                                     {2.5ex \@plus .2ex}%
                                     {\centering\itshape\bfseries\small}}
\makeatother

\usepackage{physics}

\renewcommand{\arraystretch}{1.7}

\def\unit{{1\kern-.65ex {\rm l}}}
\def\1{{1\kern-.65ex {\rm l}}}

\def\CA{{\cal A}}

\usepackage{makecell}

\makeatletter
\newcommand\xlabel[2][]{\phantomsection\def\@currentlabelname{#1}\label{#2}}
\makeatother

\usepackage[status=draft,inline,nomargin]{fixme}
\fxusetheme{color}
\FXRegisterAuthor{ki}{aki}{\color{orange}KI}
\FXRegisterAuthor{ofw}{aofw}{\color{green!40!black}OFW}
\usepackage{adjustbox}

\usepackage{faktor}

\newcommand\restr[2]{{% we make the whole thing an ordinary symbol
  \left.\kern-\nulldelimiterspace % automatically resize the bar with \right
  #1 % the function
  \right|_{#2} % this is the delimiter
}}

\makeatletter
\def\l@subsubsection#1#2{}
\makeatother
\begin{document}

\title{Proliferation Transitions for Quantum Doubles of Groups}

\author{Yuhan Gai and 
Sakura Sch\"afer-Nameki} 

\affiliation{Mathematical Institute, University of Oxford, Woodstock Road, Oxford, OX2 6GG, United Kingdom}

\begin{abstract} 
\noindent 
Proliferation transitions are phase transitions between a 2+1d topological order $\mathcal{T}$ and another $\mathcal{T}/\mathcal{A}$, obtained by condensing anyons in an algebra $\mathcal{A}\subset \mathcal{T}$. We consider such transitions in  quantum doubles of groups, $\mathcal{T}=D(G, \omega)$, for finite groups $G$ possibly with three-cocycle $\omega$, for which condensable algebras are fully classified. 
Given a condensable algebra $\mathcal{A}$, we represent $\mathcal{T}$ in terms of a Symmetry Topological Field Theory (SymTFT), that is tailored to the $\mathcal{A}$ proliferation transition. 
The SymTFT bulk is constructed from the M\"uger center of the
subcategory generated by the algebra, i.e. the category of
transparent lines, which can be bosons or fermions.
Properties of the transition can then be inferred directly from  the algebra classification data. 
For gauged Landau-Ginzburg (LG*) type transitions, we show that they do not only depend on the anyons in $\mathcal{A}$ through their field content, but the cubic interactions are fixed in terms of  the multiplication structure of $\mathcal{A}$.
Applied to twin algebras, which have the same anyon content but different multiplication structures, we show that twins result in distinct proliferation transitions.

\end{abstract}

%%%%%%%%%%%%%%%%%%%%%%%%%%%
\maketitle
%%%%%%%%%%%%%%%%%%%%%%%%%%%%%%%%%%%%%

\tableofcontents

\section{Introduction}
\label{sec:intro}

Consider a 2+1d topological order (TO) $\cT$, described by a modular tensor category (MTC), and a condensable algebra $\cA$ in $\cT$, i.e. a collection of mutually local bosonic anyons together with a multiplication that allows them to condense. 
Inserting a fine mesh of anyons in $\cA$ results in a new TO:
\be
\label{TtoTA}
\cT \to \cT/\cA \,.
\ee
This is the condensation of $\cA$, or the gauging of an anomaly-free, in general non-invertible, 1-form symmetry \cite{Moore:1989yh, Bais:2008ni, Kong:2014condensation, Lan:2017grs, Hsin:2018vcg, Burnell:2017otf} (for reviews of non-invertible symmetries see \cite{Schafer-Nameki:2023jdn, Shao:2023gho}). A natural question is whether this topological manipulation can be realized as a phase transition, in which scalar fields coupled to the anyons in $\cA$ condense. 
For abelian anyons such proliferation transitions were constructed in \cite{Cheng:2026qax, Schafer-Nameki:2026hgx, Cheng:2026dxn}, and for non-abelian anyons in a number of examples \cite{Zhang:2024bye, Ji:2026yfj, Cordova:2025eim, Cordova:2025zkz, Lu:2026fid, Garcia-Sepulveda:2026fud}. Related work on anyon condensation, gapped boundaries and lattice models includes \cite{Hung:2015hfa, Zhao:2024ilc, Zhao:2025zsb, Schafer-Nameki:2026hgx, Cheng:2026hgu, Hung:2025gcp} and, for Kitaev's quantum double models, \cite{Bombin:2007qv, Kang:2020avc, Flores-Calderon:2026bue}.

\paragraf{Proliferation Transitions from the SymTFT.}
A systematic construction was proposed in \cite{Schafer-Nameki:2026hgx}, using the Symmetry Topological Field Theory (SymTFT) for 2+1d theories \cite{Bhardwaj:2023ayw, Bhardwaj:2024qiv, Wen:2024qsg, Bhardwaj:2025piv, Bhardwaj:2025jtf}. 
The symmetry underlying the SymTFT is extracted from the algebra 
 as follows: let $\langle\cA\rangle$ be the minimal sub-category that is generated by $\cA$. The transparent lines in this generated subcategory, its so-called  M\"uger center $\cE_{\cA}$, specifies the bulk of the SymTFT.
When all transparent lines are bosons, $\cE_{\cA}\cong\Rep(G_{\cA})$ for a finite group $G_{\cA}$ \cite{Deligne:2002cat}, and $\cT$ is the interval compactification of the 3+1d $G_{\cA}$ gauge theory, with the MTCs $\cM_{\cA}=\langle\cA\rangle/\Rep(G_{\cA})$ and $\cN_{\cA}=\langle\cA\rangle'/\Rep(G_{\cA})$, where $\langle\cA\rangle'$ is the centralizer in $\cT$, stacked on the symmetry and physical boundary, respectively. 
This gives a SymTFT presentation of $\cT$ that depends on the algebra $\cA$ that will be condensed as follows: 
\be \label{SymTFTT}
\cT = \biggl( \fB^{\cM_{\cA}}_{\Neu(G_{\cA})} \bigg| \fZ(G_{\cA}) \bigg| \fB^{\cN_{\cA}}_{\Neu(G_{\cA})}\biggr)\,.
\ee
It was argued in \cite{Schafer-Nameki:2026hgx} that the proliferation transition for $\cA$ is driven by scalars on the symmetry boundary that couple to the anyons in $\cA$. Depending on the properties of the generated subcategory, we distinguish three types of proliferation transitions:
For type (i) algebras,  $\langle\cA\rangle=\cE_{\cA}$, it is a Landau-Ginzburg (LG) transition with gauged symmetry (LG*). For type (iii) algebras, $\langle\cA\rangle$ is  modular, and this case needs generically a case-by-case analysis. Finally, type (ii) is a hybrid, which allows for a partial LG* description. 

\paragraf{Scope of the Paper.}
In the present paper we apply the construction of \cite{Schafer-Nameki:2026hgx} to any  condensable algebras of the (twisted) group quantum double $D(G,\omega)$, i.e. the TO of the 2+1d Dijkgraaf-Witten theory \cite{Dijkgraaf:1989pz} of a finite group $G$ with twist $\omega\in H^3(G,U(1))$. 
This is a particularly nice setup, as all condensable algebras are classified: any condensable algebra is characterized in terms of four pieces of data,
$\cA(H,N,\gamma,\epsilon)$, given by two subgroups $N\trianglelefteq H\subseteq G$, with $N$ normal in $H$, cocycles $\gamma\in C^2(N,U(1))$ and $\epsilon:H\times N\to U(1)$, subject to compatibility conditions \cite{davydov2009modular, davydov2017lagrangian, Gai:2026hjk}. 
We will determine the complete map from 
\be
(H, N, \gamma, \epsilon) 
\ee
to the data of the generated subcategory, the type of the algebra and the SymTFT setup, i.e. a description in terms of (\ref{SymTFTT}).
The proliferation transition is then driven by dynamics on the symmetry boundary. For type (i) algebras, we construct LG*-descriptions (and this also covers part of type (ii)). For 
Tannakian $\cE_{\cA}$ this gives a presentation of $\cT=D(G,\omega)$ as in \eqref{SymTFTT}, with the inherited symmetry action and defect data understood. We discuss the super-Tannakian $\cE_{\cA}$ case, when there are transparent fermions, in section~\ref{sec:MANA_super_Tannaka}, we replace the two modular factors $\cN$ and $\cM$ by slightly degenerate categories (i.e. their M\"uger centers are $\sVec$) embedded in a common modular extension.

More concretely, in section~\ref{sec:DGProlif} we take as the sole input $\cA(H, N, \gamma, \epsilon)$ and $D(G, \omega)$, and then determine the required categorical input to construct (\ref{SymTFTT}): 
\be 
\langle\cA\rangle\,, \ \cE_{\cA}\,,\  G_{\cA}\,,\  \cM_{\cA}\,,\ \cN_{\cA} \,,
\ee
and hence the type of $\cA$ and its SymTFT, in terms of these data, see Table~\ref{tab:dictionary} for a summary of what these correspond to. 
To keep the paper readable, we have  relegated the technical proofs to the appendix \ref{sec:math}). 
We determine for each algebra what type it is, in particular for $D(G, \omega)$ all types can arise and have a precise characterization  in terms of $H, N, \gamma, \epsilon$. 
We also prove that $\cA$ is of type (i) if and only if it is purely electric, $\cA\cong\Fun(G'/H')$, 
in an equivalent presentation $D(G',\omega')\cong D(G,\omega)$, and subsequently, the proliferation transition is of LG* type.

\paragraf{Transparent Fermions.}
 We also study type (ii) algebras, where the generated subcategory contains transparent fermions, i.e. the M\"uger center is super-Tannakian, given by 
 \be 
 \cE_{\cA}\cong\Rep(G_{\cA},z_{\cA})
 \ee 
with nontrivial fermion parity $z_{\cA}$. To exemplify this, we study the smallest, non-trivial example, $D(\Z_2\times A_4)$ with $\omega=1$. 
Bosonic reduction replaces $\cM_{\cA}$, $\cN_{\cA}$ by two slightly degenerate categories $\cM_{\cA}^f$, $\cN_{\cA}^f$ and a common modular extension $\cT^f$. 
Here slightly degenerate means that their M\"uger center is $\sVec$. 

To make these general statement more tangible, we exemplify it with concrete groups section~\ref{sec:examples} where we discuss $D(S_3)$, $D(D_4)$, $D(D_4,\omega)$ and $D(\Z_2\times A_4)$. 

\paragraf{LG*-theory and Twin Algebras.}
With the SymTFT in place, section~\ref{sec:LG-algebra} constructs scalar EFTs, whose field content is specified by the anyons in the condensable algebra. 
We show that the multiplication structure of the algebra, plays an important physical role in that it fixes the cubic interactions, including relative coefficients. 
This is particularly interesting for so-called twin algebras \cite{Gai:2026hjk,Warman:2026gfz} (see also \cite{Pollmann:2012, Davydov:2013xov, Kobayashi:2025ykb, Kobayashi:2025pxs} for earlier constructions with Bogomolov multipliers): these are algebras that have the same anyon decomposition, but different algebra multiplication, and therefore their LG-theories will be in-equivalent. 
As an illustration, we determine the  LG-theories for twin algebras in the double of $G=G_{32}= \Z_8^*\ltimes\Z_8$, and show how their relative cubic signs produce opposite quartic threshold contributions, when massive multiplets are integrated out. 

Identifying a scalar field with every anyon endpoint degree of freedom in the algebra can lead to a redundant description. We discuss how to determime a minimal set, and show this in the context of the dihedral groups. 

{\bf Note added:} We thank the Xie Chen and Luisa Eck for letting us know of their upcoming paper on related questions and for coordinating submission. 

\section{Proliferation Transitions for $D(G)$}
\label{sec:DGProlif}

In a 2+1d topological order (TO) $\cT$, condensing anyons results in a reduced TO \cite{Bais:2008ni, Kong:2014condensation, Eliens:2013epa, Burnell:2017otf}. This is a topological operation, which relates to 2+1d TOs. The topological defects of $\cT$ form a modular tensor category (MTC). 
For a collection of anyons to consistently condense, they need to form an \'etale or condensable algebra $\cA$ in the MTC underlying $\cT$. 
Denote the reduced TO after condensation by $\cT/\cA$.

The question that was asked and in part answered in \cite{Cheng:2026qax, Schafer-Nameki:2026hgx} is, whether there is an actual phase transition -- the proliferation transition -- between these topological orders  
\begin{equation}
    \cT \rightarrow \cT/\cA \,.
\end{equation}
In \cite{Schafer-Nameki:2026hgx} it was proposed that based on the properties of the condensable algebra $\cA$ one can use the SymTFT approach to determine the possible phase transitions. 
The goal of this section is to first review this SymTFT proposal. Then we focus on the topological orders 
\be
\cT = D(G, \omega)
\ee
for any finite group $G$, with anomaly $\omega \in H^3 (G, U(1))$. In this case there is a  full classification of condensable algebras known 
\be
\cA (H, N, \gamma, \epsilon)\,,
\ee
which we will review shortly. We will determine the type of proliferation transition from the data that specifies the algebra as well as the possible transition theory, e.g. whether it is realized in terms of a (gauged) LG model or requires a specific other transition model.

\subsection{Proliferation Transitions via SymTFT}

Let us recall the salient features of the proposal in \cite{Schafer-Nameki:2026hgx}, the detailed definitions and more background are provided in that paper. 
The crucial point to note is that not every algebra in fact defines a subcategory of $\cT$.
However the concept of a {\bf generated subcategory} $\langle \cA \rangle$ is well-defined: it  is the fusion subcategory generated by anyons in $\cA$ consisting of all anyons closed under fusion of $\cA$ with itself. 
We denote the M\"uger center, i.e. the transparent lines, by 
\be
\label{EADef}
\cE_{\cA}=\cZ_2(\langle\cA\rangle)\,.
\ee
There are three distinguished types:
\begin{enumerate}
\item[(i)] $\langle\cA\rangle=\cE_{\cA}$ is Tannakian, i.e. $\Rep (G_\cA)$ for some finite group $G_\cA$, including the trivial algebra.
\item[(ii)] $\Vect\subsetneq\cE_{\cA}\subsetneq\langle\cA\rangle$.
\item[(iii)] $\cE_{\cA}=\Vect$ and $\langle\cA\rangle\neq\Vect$, i.e. it is modular. 
\end{enumerate}
Type (i) algebras were  shown to be completely characterized in terms of a LG*-type transition. Type (iii) requires a case by case description -- some cases can be treated using CS matter theories, but no general theory is known. 
 For type (ii), the intersection of $\cA$ with the bosonic transparent subcategory determines the possible LG$^*$ step. It can be trivial. Any remaining condensation needs a separate dynamical description, {and reduces to type (iii).}

To arrive at these statements the proposed general approach in 
\cite{Schafer-Nameki:2026hgx} is to realize the theory $\cT$ in terms of a SymTFT in one higher dimension. 
The M\"uger center determines the transparent line category (\ref{EADef}).
When these lines are bosons, $\cE_{\cA}\cong\Rep(G_{\cA})$. The modular categories
\be
\label{eq:anomaly-modular-factors}
\cM_{\cA}=\langle\cA\rangle/\Rep(G_{\cA})\,,\qquad
\cN_{\cA}=\langle\cA\rangle'/\Rep(G_{\cA})
\ee
carry categorical $G_{\cA}$-actions. Let $\nu_{\cM}$ and $\nu_{\cN}$ denote their obstruction classes in $H^4(G_{\cA},U(1))$ to braided crossed extensions \cite{Etingof2009FusionCA}. Equivalently, $\nu_{\cM}$ vanishes precisely when $\langle\cA\rangle$ admits a minimal nondegenerate extension, and similarly for its centralizer \cite{galindo2025categorical}.

De-equivariantizing the TO $\cT$ gives a crossed extension with uncharged component $\cM_{\cA}\boxtimes\cN_{\cA}$. Therefore
\be
\label{eq:anomaly-factor-cancellation}
\nu_{\cM}+\nu_{\cN}=0\,,
\ee
but neither class vanishes separately necessarily. 
For the following, assume $\nu_{\cM}=\nu_{\cN}=0$ and choose compatible crossed-extension data that gives rise to $\cT$. 
Then the boundary conditions of the SymTFT for $\cT$ are 
\be
\label{eq:anomaly-conditional-boundaries}
\ba
\fB^{\sym}&=(\fB_{\Dir}\boxtimes\cM_{\cA})/G_{\cA}^{(0)}\,,\\
\fB^{\phys}&=(\fB_{\Dir}\boxtimes\cN_{\cA})/G_{\cA}^{(0)}\,,
\ea
\ee
and
\be
\label{eqn:SymTFT_T}
\cT=\biggl(\fB^{\cM_{\cA}}_{\Neu(G_{\cA})}
\bigg|\fZ(G_{\cA})\bigg|
\fB^{\cN_{\cA}}_{\Neu(G_{\cA})}\biggr)\,.
\ee
If the individual obstructions are  nonzero, the boundaries must instead be treated as relative theories with anomaly. 
A case that occurs frequently is $\cN_{\cA}=\Vect$: then $\langle\cA\rangle'_{\cT}=\cE_{\cA}$, so $\cT$ itself is a minimal extension and $\nu_{\cM}=0$. Type (i) algebras also have vanishing individual obstructions, where $\cM_{\cA}=\Vect$. 

 The proposed transition is implemented on the symmetry boundary. For type (i), section~\ref{sec:LG-algebra} constructs a scalar potential realizing the required Higgs endpoint. For types (ii) and (iii), the remaining anyonic sector requires additional dynamics. The categorical construction alone does not establish a direct transition, its order, or a universal critical theory. 
We will discuss concretely the transitions, in particular those that are LG*-models, in section \ref{sec:LG-algebra}.

\subsection{Algebras and Subcategories of Quantum Doubles}

We will now realize the construction in \cite{Schafer-Nameki:2026hgx} for all condensable algebras of the twisted quantum double $D(G, \omega)$. First we recall that anyons in this TQFT are labeld by 
\be 
(x, \irrep) \in   \cT= D(G, \omega)\,,
\ee 
where $x$ stands for a conjugacy class $[x]$ of $x\in G$ and $\chi$ the character of a (projective) irreducible representation $\rho_{\chi}$ of the centralizer of $x$ in $G$, $C_G(x)$.  
This setup is particularly suitable to study proliferation transitions, as a complete classification of condensable algebras is known in terms of very simple group-theoretical data. Let us recall this briefly here, as in the following these data will also characterize the proliferation transitions.

\paragraf{Classification of Condensable Algebras.}
Condensable algebras in $D(G, \omega)$ have been classified in \cite{davydov2009modular, davydov2017lagrangian, Gai:2026hjk}.  
They are in 1-1 correspondence with the following data -- and we will denote them accordingly by 
\be\label{AHNge}
\cA(H, N, \gamma, \epsilon)\,.
\ee
A subgroup 
$H<G$, a normal subgroup $N\trianglelefteq H$ of $H$, and cocyle data $\gamma \in C^2(N, U(1))$ and $\epsilon:H\times N \rightarrow U(1)$ with various conditions \eqref{eqn:epsilon_compatibility} that need to be satisfied. The reduced TO is 
\be 
\cT/(\cA(H, N, \gamma, \epsilon)) = D(H/N, \alpha)
\ee
with $\alpha$ an 3-cocycle determined by $\gamma$ and $\epsilon$ (see \cite{Gai:2026hjk}).
An intuitive way to think about these algebras is that 
the charges in $\cA$ form $\Fun(G/H) = \operatorname{Ind}^G_H 1$, the induced representation, as a $G$-representation,  
the fluxes are determined by $N\trianglelefteq H$, $\gamma \in C^2(N, U(1))$ fixes the multiplication of the condensed flux operators, and $\epsilon:H\times N \rightarrow U(1)$ fixes the charge they carry. 
As a $G$-graded vector space, $\cA$ is spanned by ${}^gn$-graded component $v_{g,n}$, $g\in G$, $n\in N$, where grading corresponding to flux, with $G$-action 
\begin{equation}
    x\cdot v_{g,n}=\tau_{n}(\omega)(x,g)v_{xg, n}.
\end{equation}
See Appendix~\ref{sec:GeneratedSubcat} for more details on the algebra, including the multiplication.

\paragraf{Generated Subcategory.}
An algebra does not itself correspond to a symmetry, and so it is not clear how to a priori construct a SymTFT associated to this. We can however
consider the smallest sub-symmetry (i.e. sub-category) that contains it.  Given an algebra (\ref{AHNge}) $\cA = \cA (H, N, \gamma, \epsilon)$ we will define the generated subcategory of $ D(G, \omega)$
\be
\langle \cA \rangle = \text{fusion subcategory generated by $\cA$}
\ee
to be the smallest subcategory of $\cT$ that contains $\cA$.
We will determine $\langle\cA\rangle$ as  function of the characteric data $H, N, \gamma, \epsilon$. 

We will now determine the generated subcategory explicitly using some well-known results in group quantum doubles -- the mathematical background is relegated to  appendix \ref{sec:math}. 
The key result is that fusion-subcategories of $ D(G, \omega)$ are fully classified by Naidu-Nikshych-Witherspoon \cite{NNW:2009} in terms of a triple and we can express $\langle \cA \rangle$ explicitly in that context: let  
\be 
K, L \trianglelefteq G: \quad kl= lk \ \text{for all }k\in K,\,,  l\in L\,,
\ee
two normal subgroups $K, L \trianglelefteq G$ centralizing each other  and a $G$-invariant bicharacter 
\be 
B: K\times L \to U(1)
\ee 
multiplicative in both slots (for $\omega\neq1$ multiplicative only up to the factors $\beta_x$ in \eqref{eqn:defn_bicharacter}, i.e. $B$ is a $G$-invariant $\omega$-bicharacter, see section~\ref{sec:NNW}) and $G$-invariant, $B\big(gkg^{-1},glg^{-1}\big)=B(k,l)$.  
Then 
\be
\label{NNWcat}
S(K,L,B) = \Big\langle ([k],R) : 
k \in K\,,\, 
 R\big|_L = \dim (R)\cdot B(k,\,\cdot\,)\Big\rangle \,.
\ee
The category $S(K,L,B)$ contains the fluxes in $K$, dressed by charges whose $L$-holonomies are determined by $B$. In terms of anyons,  it is generated by 
\be 
([k], R) :\quad 
\left\{
\ba 
k &\in K\,, \cr 
R&\in \text{Irrep } C_G(k) \cr 
 \chi_R(h)&=B(k, h) \deg(\chi_R), \text { for all } h \in H\,,
\ea
\right.
\ee
where $\chi$ is the character of the (projective) irrep $R$.

The generated subcategory associated to an algebra $\langle \cA \rangle$ can be described in terms such such a triplet of data, which in turn will only depend on $H, N, \gamma, \epsilon$ of the algebra. 
We prove in Theorem~\ref{thm:fusion_subcat_from_alg} that the fusion subcategory $\langle \cA \rangle$ of $D(G, \omega)$ can be written as follows: for $\cA= \cA(H, N, \gamma, \epsilon)$ we have  
\begin{equation}
\label{eqn:A_gen_subcat}
    \langle \cA \rangle = S(K_{\cA}, L_{\cA}, B_{\cA})\,,
\end{equation}
where 
\begin{equation}
    K_{\cA} :=\langle {}^gn \mid n\in N, g\in G\rangle,
\end{equation}
and to determine $L_{\cA}$ and $B_{\cA}$, consider a $l\in G$ action on a $k$-graded component $v_{g_1, n_1} \otimes \cdots \otimes v_{g_r,n_r}\in \cA^{\otimes r}$, if $k$ commutes with each grading, it acts as scalar
\be\label{eq:audit-main-Lambda}
\ba
\Lambda_l(w):={}&\prod_{j=1}^{r-1}\beta_l(p_j,{}^{g_{j+1}}n_{j+1})\\
&\times\prod_{j=1}^{r}\frac{\tau_{n_j}(\omega)(l,g_j)}
{\tau_{n_j}(\omega)(g_j,l^{g_j})}\epsilon(l^{g_j},n_j)\,,
\ea
\ee
where $w=(g_1, n_1)\cdots (g_r, n_r)$ is a word in $G\times N$. Denote by $k(w) := {}^{g_r}n_r$, define
\begin{equation}
    L_{\cA}:= \{ l\in \Core_G(C_H(N)) \mid \Lambda_l(w) = 1 \text{ for }k(w)=1\}.
\end{equation}
The core is the largest normal subgroup of $G$ contained in $C_H(N)$, i.e., 
\be 
\Core_G(C_H(N))=\bigcap_{g\in G}{}^gC_H(N) \,.
\ee
For $k\in K_{\cA}$ and $l\in L_{\cA}$, let $w$ be any word in $G\times N$ with $k(w) = k$, 
\begin{equation}
    B_{\cA}(k,l) :=\Lambda_l (w)
\end{equation}
is well-defined (see Appendix~\ref{sec:GeneratedSubcat} for a proof).

\paragraf{Example: $G=S_3$.}
Let us consider this concretely for the well-known quantum double $D(S_3)$.
We generate the permutation group by 
$S_3=\langle a,b \mid a^3=b^2=1, \, abab=1 \rangle$.
The condensable algebras in $D(S_3)$ are summarized in Table~\ref{tab:S3example}. Here, we focus on the algebras $\cA_1=\cA(\Z_2^{\langle b \rangle},1,1,1)$ and $\cA_2=\cA(\Z_2^{\langle b \rangle}, \Z_2^{\langle b \rangle}, 1, 1)$ (the second one is Lagrangian) to illustrate the above notations. In both algebras, $\gamma=1$ and $\epsilon=1$, hence the action scalar $\Lambda_l=1$ is always well-defined.

For $\cA_1=\cA(\Z_2^{\langle b \rangle},1,1,1)$, $K_{\cA_1} = 1$, $L_{\cA_1} = 1$, so the generated fusion subcategory is
\begin{equation}
    \langle \cA_1 \rangle = S(1,1,1) = \Rep(S_3).
\end{equation}
The SymTFT bulk has $G_{\cA}=S_3$. The MTC stacked on both boundaries are trivial $\cM_{\cA}\cong \cN_{\cA}\cong \Vect$, hence both the symmetry and physical boundaries are the Neumann boundary.

For $\cA_2=\cA(\Z_2^{\langle b \rangle},\Z_2^{\langle b \rangle},1,1)$, $K_{\cA_2} = S_{3}$, $L_{\cA} =1$, so the generated fusion subcategory is
\begin{equation}
    \langle \cA_2 \rangle = S(S_3,1,1)
\end{equation}
The SymTFT bulk is trivial, $G_{\cA}=1$. The symmetry boundary has MTC $\cM_{\cA} \cong D(S_3)$ stacked, and $\cN_{\cA}\cong \Vect$, so the physical boundary is again the Neumann boundary.

We will consider examples with non-trivial $B_{\cA}$ later on.

\begin{table*}
\renewcommand{\arraystretch}{1.4}
\begin{tabular}{|c|c|c|}
\hline
Datum & Definition ($\omega=1$) & Physical Interpretation \\
\hline
$H\subseteq G$ & charge $\Fun(G/H)$ & Unbroken gauge group $H$ \\
$N\trianglelefteq H$ & fluxes ${}^gn$, $g\in G$, $n\in N$, in $\cA$ & Condensed fluxes \\
$\gamma$ & $V_nV_{n'}=\gamma(n,n')V_{nn'}$ & Multiplication of the condensed fluxes \\
$\epsilon$ & $U_hV_nU_h^{-1}=\epsilon(h,n)V_{{}^hn}$ & Charges carried by the condensed fluxes \\
\hline
$K_{\cA}$ & $\langle{}^gN\mid g\in G\rangle$ & Fluxes in $\langle\cA\rangle$ \\
$L_{\cA}$ & $\subseteq\Core_G(H)\cap C_G(K_{\cA})$, see \eqref{eqn:defn_LA} & Group elements acting on flux by scalar \\
$B_{\cA}$ & $B_{\cA}({}^gn,l)=\epsilon(l^g,n)$ & Phase of $l\in L_{\cA}$ on the flux sector ${}^gn$ \\
$R_{\cA}$ & $\{r\in K_{\cA}\cap L_{\cA}\mid B_{\cA}(r,s)B_{\cA}(s,r)=1\ \forall s\in K_{\cA}\cap L_{\cA}\}$ & Transparent fluxes \\
$P_{\cA}$ & $K_{\cA}L_{\cA}$ & Transparent charges $\Rep(G/P_{\cA})$ \\
$G_{\cA}$ & $\widehat R_{\cA}\rtimes_{\alpha_{\cA}}(G/P_{\cA})$ & Symmetry underlying the SymTFT $\fZ(G_{\cA})$ \\
$z_{\cA}$ & $B_{\cA}(r,r)$ & Fermion parity of the transparent lines \\
$\cM_{\cA}$ & $\langle\cA\rangle/\cE_{\cA}$ & MTC stacked on the symmetry boundary \\
$\cN_{\cA}$ & $\langle\cA\rangle'/\cE_{\cA}$ & MTC stacked on the physical boundary \\
$\cT/\cA$ & $D(H/N,\alpha)$ & Reduced TO \\
\hline
\end{tabular}
\caption{Dictionary between the classification data $(H,N,\gamma,\epsilon)$ of a condensable algebra $\cA$ of $\cT= D(G)$, the data of its proliferation transition, and their physical meaning. $V_n$ creates a condensed flux $n$ and $U_h$ is a gauge transformation. $K_{\cA}$, $L_{\cA}$, $B_{\cA}$ determine the generated subcategory $\langle\cA\rangle=S(K_{\cA},L_{\cA},B_{\cA})$, and $R_{\cA}$, $P_{\cA}$ and $\psi_{\cA}$ the M\"uger center $\cE_{\cA}=S(R_{\cA},P_{\cA},\psi_{\cA})\cong\Rep(G_{\cA},z_{\cA})$, see section~\ref{sec:math}. The extension class $\alpha_{\cA}$ of $G_{\cA}$ is given in \eqref{eqn:alpha_formula}. {The rows of $\cM_{\cA}$ and $\cN_{\cA}$} assume $z_{\cA}=1$. {The $z_{\cA}\neq1$ case  uses $\cM^f_{\cA}$, $\cN^f_{\cA}$} and their common modular extension from section~\ref{sec:MANA_super_Tannaka}.}
\label{tab:dictionary}
\end{table*}

\subsection{SymTFT Realization of $D(G, \omega)$}

We can now realize the twisted quantum double $D(G, \omega)$ in terms of a SymTFT, that takes into account the fact that we want to proliferate anyons in $\cA$. We summarize the map between these data in Table \ref{tab:dictionary}.

Each ingredient in the SymTFT \eqref{eqn:SymTFT_T} has an explicit expression in terms of algebra classification data $(H, N, \gamma, \epsilon)$ (a proof for each can be found in Appendix~\ref{sec:math}):

\paragraf{M\"uger Center and SymTFT Bulk.} 
The transparent lines, i.e. the M\"uger center, $\cE_{\cA}=\cZ_2(\langle \cA \rangle)$ of the generated subcategory is {(a proof can be found in Prop.~\ref{prop:Muger-center})}
    \begin{equation}\label{Mugg}
        \cE_{\cA} := S(R_{\cA}, \, K_{\cA}L_{\cA},\,  \psi_{\cA}),
    \end{equation}
    where $K, L$ are defined around (\ref{eqn:A_gen_subcat}) and 
\be
\ba
        R_{\cA} &:=\{a \in K_{\cA} \cap L_{\cA} \mid B_{\cA}(a,\cdot)B_{\cA}(\cdot, a) = 1 \}\cr 
        \psi_{\cA}(a, lk) &
        :=\beta_{a}(l,k)^{-1} B_{\cA}(a,l) B_{\cA}(k,a)^{-1}\,.
\ea\ee
As a super-Tannakian category, if there are transparent fermions, we have 
    \begin{equation}
        \cE_{\cA} \cong \Rep(G_{\cA}, z_{\cA})\,,
    \end{equation}
    where
    \begin{equation}
        G_{\cA} := \widehat R_{\cA} \rtimes_{\alpha_{\cA}} (G/(K_{\cA}L_{\cA}))
    \end{equation}
    and the central parity element
    \begin{equation}
        z_{\cA}:=(B_{\cA}\circ \Delta, 1)\,,
    \end{equation}
where $B_{\cA}\circ \Delta:R_{\cA} \rightarrow U(1)$ maps $r\in R_{\cA}$ to $(B_{\cA}\circ \Delta)(r) = B_{\cA}(r,r)$. 
    
The M\"uger center is bosonic if and only if $B_{\cA}(r,r)=1$ for all $r\in R_{\cA}$. In this case, 
    \begin{equation}
        \cE_{\cA}\cong \Rep(G_{\cA}).
    \end{equation}
The SymTFT bulk is a 3+1d $G_{\cA}$-Dijgraaf-Witten theory in the Tannakian case. In the super-Tannakian case, it is still a bosonic bulk, but with emergent fermionic charges. We explain this in section~\ref{sec:MANA_super_Tannaka}.

\paragraf{Symmetry Boundary.} When $\cE_{\cA}\cong \Rep(G_{\cA})$ is bosonic, the MTC $\cM_\cA = S(K_{\cA}, L_{\cA}, B_{\cA})/\cE_{\cA}$ can be expressed as
    \begin{equation}
        \cM_{\cA} \cong\lb  \left. \Vect_{K_{\cA}/R_{\cA}}^{\overline{\omega}_{\cM}} \rb \right/({K_{\cA}/I_{\cA}}) \,,
    \end{equation}
    where $\overline{\omega}_{\cM}$ can be found in Proposition~\ref{prop:MANA}.
    
\paragraf{Physical Boundary.} When $\cE_{\cA}\cong \Rep(G_{\cA})$ is bosonic, the MTC $\cN = S(L_{\cA}, K_{\cA}, (B_{\cA}^{\operatorname{op}})^{-1})/\cE_{\cA}$ can be expressed as
    \begin{equation}
        \cN_{\cA} \cong \lb \left.\Vect_{L_{\cA}/R_{\cA}}^{\overline{\omega}_{\cN}} \rb \right/ ({L_{\cA}/I_{\cA}})\,, 
    \end{equation}
    where $\overline{\omega}_{\cN}$ can be found in Proposition~\ref{prop:MANA}.

\paragraf{Example: $G=S_3$.}
Again, we focus on the algebras $\cA_1=\cA(\Z_2^{\langle b \rangle},1,1,1)$ and $\cA_2=\cA(\Z_2^{\langle b \rangle}, \Z_2^{\langle b \rangle}, 1, 1)$ in $D(S_3)$ to illustrate the above notations. 

For $\cA_1=\cA(\Z_2^{\langle b \rangle},1,1,1)$, $G_{\cA_1}=S_3$, the SymTFT bulk is a $S_3$-gauge theory. The MTC stacked on both boundaries are trivial $\cM_{\cA}\cong \cN_{\cA}\cong \Vect$, hence both the symmetry and physical boundaries are the Neumann boundary as discussed before.

For $\cA_2=\cA(\Z_2^{\langle b \rangle},\Z_2^{\langle b \rangle},1,1)$, $G_{\cA_2}=1$, the SymTFT bulk is trivial. The symmetry boundary has MTC $\cM_{\cA} \cong D(S_3)$ stacked, and $\cN_{\cA}\cong \Vect$, so the physical boundary is again the Neumann boundary as before.

% \paragraf{Phenomena Only in the Non-Abelian Case.}
% There are two phenomena that only occur when $G$ is non-abelian:
% \begin{enumerate}
%     \item \textbf{Twins.} For non-abelian $G$, there can be different condensable algebras, consisting of the same anyons but differing in the algebra multiplications.

% \end{enumerate}

\subsection{Types of Condensable Algebras}
\label{sec:algebra_types}
Given a condensable algebra $\cA$ in $D(G, \omega)$ we would now like to know the type of algebra as this determines what the proliferation transition is realized by. 
The three types can be characterized in terms of the categorical data as follows \cite{Schafer-Nameki:2026hgx}:
\begin{enumerate}
    \item[(i)] (Tannakian) All lines in $\langle \cA \rangle$ are transparent %and $\langle\cA\rangle$ is (super-)Tannakian, 
    i.e., 
    $\langle \cA \rangle = \cE_{\cA} \cong \Rep(G_{\cA})$ for some finite group $G_{\cA}$. 
    \item[(ii)] (Mixed type) The transparent lines of $\langle \cA \rangle$ form a proper non-trivial fusion subcategory of $\langle \cA \rangle$, i.e., $\cE_{\cA} \subset \langle \cA \rangle$, $\cE_{\cA}\not \cong \Vect$ and $\cE_{\cA} \not \cong \langle \cA\rangle$.
    \item[(iii)] (Modular type) $\langle \cA\rangle$ is non-degenerate, i.e. and MTC, and thus $\cE_{\cA}=\Vect$ (and $\langle \cA \rangle \not\cong \Vect$).
\end{enumerate}

One can immediately determine the algebra type of condensable algebras in $D(G, \omega)$ by comparing the data $(K_{\cA}, L_{\cA}, B_{\cA})$ that specifies the generated subcategory around (\ref{eqn:A_gen_subcat}) with $(R_{\cA}, P_{\cA}, \psi_{\cA})$ that determine the M\"uger center (\ref{Mugg}), and examining the resulting (super-)Tannakian data $(G_{\cA}, z_{\cA})$:
\begin{itemize}
    \item[(i)] $\cA$ is type (i) if and only if $(R_{\cA},\,  K_{\cA}L_{\cA}) = (K_{\cA},\,  L_{\cA})$;
    \item[(iii)] $\cA$ is type (iii) if and only if $R_{\cA}=1$ and $P_{\cA}=G$ and $\cA$ is not the trivial algebra;
    \item[(ii)] $\cA$ is type (ii) otherwise.
\end{itemize}
Note the SymTFT construction in \cite{Schafer-Nameki:2026hgx} assumes $\cE_{\cA}$ to be Tannakian. The super-Tannakian possibility is realized for algebras in $D(G)$ with $G=A_4\times \Z_2$, which we study in section~\ref{sec:supertannakian-example}. 
A super-Tannakian $\cE_{\cA}$ only appears when $\cA$ is of type (ii) (see argument below Proposition~\ref{prop:Tannakian_criterion}). Indeed, the  super-Tannakian examples in section~\ref{sec:supertannakian-example} are all from algebras of type (ii).
It is useful to highlight three special instances of algebras:
\begin{itemize}
    \item Condensable algebras that are pure charges are of the form 
    \be 
    \label{eqn:H_typei}
    \cA = \cA(H,1,1,1)
    \ee 
    are of type (i) since
    \be \ba 
    \label{eqn:G/CoreH}
    \langle \cA\rangle =& S(1, \Core_G(H), 1) \cr 
    =& \cZ_2(\langle \cA \rangle)\cong \Rep(G/\Core_G(H))\,.
    \ea \ee
    \item Condensable algebra of the form 
    \be 
    \cA = \cA(H, N,1,1)\,,
    \ee 
     in the untwisted theory $D(G)$, with $H$ abelian and normal in $G$, is of type (i). Indeed, $K_{\cA}\subseteq H$ and $L_{\cA}=H$, so
\be\label{eq:audit_abelian_normal}
\langle\cA\rangle=S(K_{\cA},H,1)=\cE_{\cA}\,.
\ee
For $\omega\neq1$, the cocycle factors must instead be retained in determining $L_{\cA}$ and $B_{\cA}$.

    \item For group $G$ of the form $G=K\times L$, if $H=A\times L$ for subgroup $A\subsetneq K$ and $N=N\times 1$, where $1\neq N\trianglelefteq A$ such that $\Core_K(C_A(N))=1$. If $K_{\cA}=K\times1$, then algebra $\cA(H,N,1,1)$ in $D(G)$ is of type (iii).
\end{itemize}

\paragraf{Braiding.}
Physically, all anyons in a type (i) algebra braid trivially with each other, and the transition Higgses $\Rep(G_{\cA})\to \Rep(H_{\cA})$ with $\cA\cong\Fun(G_{\cA}/H_{\cA})$ inside $\langle\cA\rangle=\cE_{\cA}\cong\Rep(G_{\cA})$, where $H_{\cA}$ is the stabilizer of a minimal idempotent of $\cA$, see section~\ref{sec:LG-type-i}. $G_{\cA}/H_{\cA}$ is in general not a group.
For type (i) algebra $\cA$, $\langle\cA\rangle\subseteq\langle\cA\rangle'$ is symmetric, and conversely. For type (iii) no line generated by the condensate is transparent, and $\cT\cong\langle\cA\rangle\boxtimes\langle\cA\rangle'$. 

For $\omega=\epsilon=1$, $B_{\cA}=1$ and type (iii) forces $G$ to decompose $G=K_{\cA}\times L_{\cA}$. This is never the case if $G$ is nilpotent:
$K_{\cA}$ is then a nilpotent direct factor of $G$, and since the normal closure of $N$ is $K_{\cA}$, centralized by $L_{\cA}$, and every maximal subgroup of a nilpotent group is normal, $N=K_{\cA}\subseteq H$. Then $1\neq Z(K_{\cA})\subseteq\Core_G(H)\cap C_G(K_{\cA})=L_{\cA}$, contradict with with $K_{\cA}\cap L_{\cA}=1$. For example, among all the algebras in $D(D_4)$ listed in Table~\ref{tab:D4examplenonLag} and Table~\ref{tab:D4exampleLag}, none of them is type (iii).

\subsection{Type (i) Algebras as LG Transitions}
\label{sec:type-i-Potts}
The proliferation transition of type (i) algebras are of LG*-type {(here, we allow $\omega$ to be non-trivial)}. More precisely, we prove Theorem~\ref{thm:type-i_LG}, which shows that:
$\cA$ in $ D(G, \omega)$ is of type (i) if and only if 
\begin{equation}
    \cA \cong \Fun(G'/H')
\end{equation}
in $\Rep(G') \subset  D(G', \omega')\cong  D(G, \omega)$, for some finite groups $H'\subseteq G'$ and $\omega'\in H^3(G', U(1))$. This implies that the proliferation transition induced by a type (i) algebra is always a Higgsing and has a Landau-Ginzburg description, see section \ref{sec:LG-algebra}.

\subsection{Super-Tannakian M\"uger Centers}
\label{sec:MANA_super_Tannaka}

We will now discuss the case where there are transparent fermions, i.e. 
the M\"uger center is  $\cE_{\cA}=\cZ_2(\langle\cA\rangle)\cong\Rep(G_{\cA},z_{\cA})$ with $z_{\cA}\neq1$. The element $z_{\cA}$ is a central involution of the reconstructed group, not an object of the category. {An irreducible representation on which it acts by $-1$ has spin $-1$, but need not have quantum dimension 1.} For example, in $\Rep(Q_8,-1)$ all four one-dimensional representations are even and the two-dimensional irreducible representation is odd. Thus a super-Tannakian category need not contain a subcategory $\sVec$.

Set $G_b=G_{\cA}/\langle z_{\cA}\rangle$ and take the maximal Tannakian subcategory $\cE^b_{\cA}=\Rep(G_b)$. With $A_b=\Fun(G_b)$, define
\be
\label{eq:super-neutral-definitions}
\ba
\cM^f_{\cA}&=(\langle\cA\rangle)_{A_b} {=\langle\cA\rangle/\cE_{\cA}^b}\cr 
\cN^f_{\cA} &=(\langle\cA\rangle')_{A_b}
{=\langle\cA\rangle'/\cE_{\cA}^b}\cr
\cT^f&=\cT_{A_b}^{\operatorname{loc}}
       \cong(\cE^b_{\cA})'/\cE^b_{\cA}\,.
\ea
\ee
The first two module categories are braided because $A_b$ is transparent in their respective parent categories. They embed in the modular category $\cT^f$. Their common M\"uger center is
\be
\label{eq:super-center-after-even-condensation}
\cZ_2(\cM^f_{\cA})=\cZ_2(\cN^f_{\cA})
\cong\cE_{\cA}/\cE^b_{\cA}\cong\sVec\,.
\ee
Write $\psi$ for its unique nontrivial simple object. This fermion is transparent in $\cM^f_{\cA}$ and $\cN^f_{\cA}$, but not in $\cT^f$. The full category $\cT_{A_b}$ is $G_b$-crossed braided and has neutral component $\cT^f$. Condensing $A_b$ therefore gives $\cT^f$, whereas reconstructing $\cT$ by gauging requires the full {crossed category structure}, including its {non-trivially graded components} \cite{DGNO}.

The central extension 
\be 
1\to\langle z_{\cA}\rangle\to G_{\cA}\to G_b\to1
\ee 
{is classified by an extension} class $\alpha\in H^2(G_b,\Z_2)$. The induced categorical $G_b$-actions retain this class in their coherence on $\psi$. They are fermionic $(G_{\cA},z_{\cA})$-actions, not merely permutations of anyons. Equivariantization recovers $\langle\cA\rangle$ and $\langle\cA\rangle'$ \cite{galindo2025categorical}. We discuss in appendix~\ref{append:fermion_de-equiv} the relevant coherence equation.

From this we can now realize $\cT^f$  as follows:
\begin{corollary}
The subcategories $\cM^f_{\cA}$ and $\cN^f_{\cA}$ are mutual centralizers in $\cT^f$. Moreover,
\be
\label{eqn:MAfboxNAf}
\ba
\cM^f_{\cA}\boxtimes_{\sVec}\cN^f_{\cA}
&:=\Mod_{1\oplus(\psi\boxtimes\psi)}
       (\cM^f_{\cA}\boxtimes\cN^f_{\cA})\\
&\cong\cM^f_{\cA}\vee\cN^f_{\cA}
 =\psi'\subset\cT^f\,.
\ea
\ee
This subcategory has half the dimension of $\cT^f$, so $\cT^f$ is its minimal modular extension.
\end{corollary}
\begin{proof}
Put $d=|G_b|$. The usual module and centralizer dimension formulas give
\be
\label{eq:super-dimension-check}
\ba
\dim\cM^f_{\cA}\,\dim\cN^f_{\cA}
&=\frac{\dim\langle\cA\rangle\,\dim\langle\cA\rangle'}{d^2}\\
&=\frac{\dim\cT}{d^2}=\dim\cT^f\,.
\ea
\ee
The two subcategories centralize one another, so this equality makes them mutual centralizers \cite{Muger:2003structure}. Their intersection is therefore $\sVec$. The tensor product functor to their join identifies the two copies of $\psi$ and descends to the relative product in \eqref{eqn:MAfboxNAf}. Its domain and image both have dimension $\dim\cT^f/2$, giving the equivalence. Alternatively, the double-centralizer theorem directly gives
\be
\label{eq:super-join-centralizer}
(\cM^f_{\cA}\vee\cN^f_{\cA})'
=(\cM^f_{\cA})'\cap(\cN^f_{\cA})'
=\sVec\,.
\ee
Taking centralizers again identifies the join with $\psi'$. Its own M\"uger center is $\sVec$, and its centralizer inside $\cT^f$ is $\sVec$. This is the defining minimality condition \cite{Lan:2016rcq}.
\end{proof}
Unlike the Tannakian case, the two reduced boundary categories do not fill the neutral modular category by an ordinary tensor product. They share one fermion, and the remaining sectors of $\cT^f$ braid non-trivially with that fermion.

\paragraf{SymTFT Bulk for Super-Tannakians.}
The symmetric category $\cE_{\cA}$ defines a $2$-category $\Sigma\cE_{\cA}$ and canonical bulk defect category $ D(\Sigma\cE_{\cA})$, realized by the Walker-Wang model with input $\cE_{\cA}$ \cite{Walker:2012mcd}. This specifies a candidate bulk for the boundary construction, a schematic SymTFT sandwich takes the form
\begin{equation}
    \cT = \Bigl( \fB^{\sym}_{\cA} \Big| D(\Sigma\cE_{\cA}) \Big| \fB^{\phys}_{\cA} \Bigr).
\end{equation}
Its bulk lines are $\cE_{\cA}$. The intrinsic Walker-Wang bulk associated with the full generated category instead has defect category $ D(\Sigma\langle\cA\rangle)$. Equality of bulk lines does not identify the two full defect categories \cite{JohnsonFreydReutter:2024}.

For a super-Tannakian M\"uger center, the precise extension test is given by \cite{galindo2025categorical}: choose a minimal modular extension of the slightly degenerate (i.e. its M\"uger center is $\sVec$) reduction {$\cM^f_{\cA} \boxtimes_{\sVec} \cN^f_{\cA}$, extend its inherited fermionic symmetry action}, and check that the resulting $G_b$-action has zero $H^4$ obstruction. {This is the case for the $\Z_2\times A_4$ examples that we discuss below.}

As a sanity check, if $\cE_{\cA} \cong \Rep(G_{\cA})$ is Tannakian, $\Sigma \cE_{\cA}$ is Morita equivalent to $\TwoVec_{G_{\cA}}$ and
\begin{equation}
     D(\Sigma\Rep(G_{\cA})) =  D(\TwoVec_{G_{\cA}}),
\end{equation}
recovering the $3+1$d $G_{\cA}$ DW theory $\fZ(G_{\cA})$.

When $z_{\cA}\neq 1$, $\cE_{\cA} \cong \Rep(G_{\cA}, z_{\cA})$. For a simple $V\in \Rep(G_{\cA}, z_{\cA})$, write 
\begin{equation}
    z_{\cA}|_V = (-1)^{p(V)}\id_V.
\end{equation}
Then $V$ and $W$ braid as
\begin{equation}
    c_{V,W}(v\otimes w) = (-1)^{p(V)p(W)}w\otimes v,
\end{equation}
and $\theta_V = (-1)^{p(V)}$. This encodes the fermionic statistics. The $V$'s with $\theta_V=-1$ are not local fermions, but rather are gauge charges, and we do not require a spacetime spin structure.

\vspace{2mm}
\noindent\textbf{Example: $\cE_{\cA} = \sVec$.}
For $G_{\cA}=\Z_2^f$, the bulk $ D(\Sigma\sVec)$ is the $(3+1)$-dimensional fermionic-charge toric code. This is a bosonic topological order with an emergent fermionic gauge charge, realizable by the $\sVec$ Walker-Wang model, and requires no microscopic spin structure \cite{Walker:2012mcd,Fidkowski:2022fermioniccode}.

\vspace{2mm}
\noindent\textbf{Split Extensions and Super-Tannakian Examples.}
If $G_{\cA} = G_b \times \langle z_{\cA} \rangle$, $G_b := G_{\cA} /\langle z_{\cA} \rangle$, i.e., $G_{\cA}$ is a split extension of $G_b$, then
\begin{equation}
    \Rep(G_{\cA}, z_{\cA}) \cong \Rep(G_b) \boxtimes \sVec.
\end{equation}
The SymTFT bulk then have topological defects
\begin{equation}
     D(\TwoVec_{G_b}) \boxtimes  D(\Sigma \sVec).
\end{equation}
An example will appear in section~\ref{sec:supertannakian-example} where we study examples with super-Tannakian $\cE_{\cA}$ 
\begin{equation}
    \cE_{\cA} \cong \Rep(\Z_6, z) \cong \Rep(\Z_3) \boxtimes \sVec.
\end{equation}
The bulk will be a 3+1d $\Z_3$ gauge theory stacked with the $\Z_2$ toric code.  Here, the toric code $ D(\Sigma\sVec)$ with fermionic electric charge $f$ is again obtained by gauging fermion parity of the trivial 3+1d fermionic phase.

In the split case, when $\cM^f_{\cA} = \sVec \boxtimes \cL_{\cA}$ and $\cN^f_{\cA} = \sVec \boxtimes\cR_{\cA}$, where $\cL_{\cA}$ and $\cR_{\cA}$ are MTCs, we write down the explicit SymTFT boundaries
\begin{equation}
    \begin{split}
        \fB^{\sym}_{\cA} & = (\fB_{\Dir(G_b)}\boxtimes \fB_f \boxtimes \cL_{\cA})/G_{b}^{(0)},\\
        \fB^{\phys}_{\cA} & = (\fB_{\Dir(G_b)}\boxtimes \fB_f \boxtimes \cR_{\cA})/G_{b}^{(0)},
    \end{split}
\end{equation}
where $\fB_f$ is the bosonic Neumann boundary of $D(\Sigma \sVec)$. The resulting SymTFT sandwich is
\begin{equation}
    \cT = \Bigl( \fB^{\sym}_{\cA} \Big| D(\Sigma\Rep(G_{\cA}, z_{\cA})) \Big| \fB^{\phys}_{\cA} \Bigr).
\end{equation}
This will indeed be the case when we study examples in $D(\Z_2\times A_4)$ in section~\ref{sec:supertannakian-example}.

\section{Examples: Tannakian and Super-Tannakian Algebras}
\label{sec:examples}

The framework we developed in section \ref{sec:DGProlif} is applicable to any finite group $G$, with 't Hooft anomaly $\omega$, i.e. any $\cT= D(G, \omega)$. As the formulation maybe rather abstract we will not discuss examples without and with anomaly. This provides the formulation of $D(G, \omega)$ that is most suitable to study the proliferation transition, which we discuss in the next section. 

\begin{table*}
    \centering
    \begin{tabular}{|c|c|c|c|c|c|c|c|}
    \bottomrule
        $(H,N)$ & object & $(K_{\cA}, L_{\cA})$ &  $(R_{\cA}, P_{\cA})$ & $G_{\cA}$ & type & $\cM_{\cA}$ & $\cN_{\cA}$ \\
        \hline\hline
        $(1,1)$ & $1\oplus 1^- \oplus 2E$ & $(1,1)$ & $(1,1)$ & $S_3$ & (i) & $\Vect$ & $\Vect$ \\
        $(\Z_2^{\langle b\rangle},1)$ & $1\oplus E$ & $(1,1)$ & $(1,1)$ & $S_3$ & (i) & $\Vect$ & $\Vect$ \\
        $(\Z_2^{\langle b \rangle},\Z_2^{\langle b \rangle})$ & $1\oplus b \oplus E$ & $(S_3,1)$ & $(1,S_3)$ & $1$ & (iii) & $D(S_3)$ & $\Vect$ \\
        $(\Z_3^{\langle a \rangle},1)$ & $1\oplus 1^{-}$ & $(1, \Z_3^{\langle a \rangle})$ & $(1,\Z_3^{\langle a \rangle})$ & $\Z_2$ & (i) & $\Vect$ & $D(\Z_3)$ \\
        $(\Z_3^{\langle a \rangle},\Z_3^{\langle a \rangle})$ & $1\oplus 1^{-} \oplus 2a$ & $(\Z_3^{\langle a \rangle}, \Z_3^{\langle a \rangle})$ & $(\Z_3^{\langle a \rangle},\Z_3^{\langle a \rangle})$ & $S_3$ & (i) & $\Vect$ & $\Vect$ \\
        $(S_3,1)$ & $1$ & $(1,S_3)$ & $(1,S_3)$ & $1$ & (i) & $\Vect$ & $D(S_3)$ \\
        $(S_3,\Z_3^{\langle a \rangle})$ & $1\oplus a$ & $(\Z_3^{\langle a \rangle},\Z_3^{\langle a \rangle})$ & $(\Z_3^{\langle a \rangle}, \Z_3^{\langle a \rangle})$ & $S_3$ & (i) & $\Vect$ & $\Vect$ \\
        $(S_3,S_3)$ & $1 \oplus a \oplus b$ & $(S_3,1)$ & $(1,S_3)$ & $1$ & (iii) & $D(S_3)$ & $\Vect$ \\
        \toprule
    \end{tabular}
    \caption{Condensable algebras in $D(S_3)$. Column 1 consists of $(H,N)$ in the classification of condensable algebras $\cA=\cA(H,N,1,1)$ ($\gamma$ and $\epsilon$ both trivial for condensable algebras in $D(S_3)$). The second column is the corresponding decomposition into anyons. The third column indicates the data that classifies the generated subcategoryconsists of $(K_{\cA}, L_{\cA})$ in the triple $(K_{\cA}, L_{\cA}, B_{\cA}=1)=\langle \cA \rangle$. The fourth column consists of $(R_{\cA}, P_{\cA})$ in the triple for the M\"uger center $(R_{\cA}, P_{\cA}, \psi_{\cA}=1)=\cZ_2(\langle \cA \rangle)$. The fifth column has $G_{\cA}$ such that $\cZ_2(\langle \cA \rangle)\cong \Rep(G_{\cA})$ (in this case, all M\"uger centers are Tannakian). The sixth column consists of the type of each algebra. The final two columns consists of $\cM_{\cA}$ and $\cN_{\cA}$, all of the form $D(\Gamma)$ for some finite group $\Gamma$.}
    \label{tab:S3example}
\end{table*}

\subsection{$D(S_3)$}
\label{sec:alg_S3}

To start with lets consider the case $G= S_3$ with $\omega=1$. This is well-studied \cite{Schafer-Nameki:2026hgx}, so this is largely to familiarize the reader with the notations of the general framework. 
Let 
$S_3=\langle a,b \mid a^3=b^2=1, \, abab=1 \rangle$. 
The condensable algebras $\cA(H,N, \gamma, \epsilon)$ in $D(S_3)$ have trivial $\gamma$ and $\epsilon$. Hence the $\lambda$, $\Lambda$, $B_{\cA}$ and $\psi_{\cA}$ previously defined are all trivial. 

Let $1^-$ denote the 1d sign irrep  and $E$ denote the 2d irrep of $S_3$.
For notational simplicity, we denote the simple object in $D(S_3)$ purely by a group element or a irrep if either the irrep or group element labeling data is trivial, respectively.

In this example, for any condensable algebra, the $S_3$-invariant bicharacters $B$ and $\psi$ in its generated fusion subcategory of $D(S_3)$ and its M\"uger center are trivial. Hence the M\"uger center is Tannakian. We list the condensable algebras, their generated fusion subcategories and their M\"uger centers in Table~\ref{tab:S3example}.

In terms of transitions theories, the type (i) algebras of $D(S_3)$ give LG$^*$ transitions \cite{Schafer-Nameki:2026hgx}: $1\oplus1^-$ fully breaks $G_{\cA}=\Z_2$ (Ising$^*$), $1\oplus E$ and $1\oplus a$ break $G_{\cA}=S_3$ to $\Z_2$ (three-state Potts$^*$, with the electric and the magnetic $\Rep(S_3)$, exchanged by the electric-magnetic duality of $D(S_3)$), and $1\oplus1^-\oplus 2E$, $1\oplus1^-\oplus2a$ break $S_3$ completely.  The two Lagrangian algebras $1\oplus b\oplus E$ and $1\oplus a\oplus b$ are of type (iii): $\langle\cA\rangle=D(S_3)$ is modular. 

\subsection{$D(D_4)$}
\label{sec:example_D_4}
Consider the dihedral group of order 8 generated by 
\begin{equation}
    D_4=\langle r,s \mid r^4=s^2=1, \, srs = r^{-1} \rangle\,.
\end{equation}
In $D(D_4)$, there exist condensable algebras such that the $D_4$-invariant bicharacters $B$ of the generated fusion subcategories (and those of the M\"uger center $\psi$) are non-trivial. However, the M\"uger centers turn out to be Tannakian for any condensable algebras. See Table~\ref{tab:D4examplenonLag} see the list of non-maximal condensable algebras, and Table~\ref{tab:D4exampleLag} for Lagrangian algebras, in $D(D_4)$.

\begin{table*}
    \setlength{\tabcolsep}{1.3pt}
    \begin{tabular}{|c|c|c|c|c|c|c|c|}
    \bottomrule
        $(H,N, \gamma, \epsilon)$ & Anyon Decomposition & $(K_{\cA}, L_{\cA}, B_{\cA})$ &  $(R_{\cA}, P_{\cA}, \psi_{\cA})$ & $G_{\cA}$ & type & $\cM_{\cA}$ & $\cN_{\cA}$ \\
        \hline\hline
        $(D_4,1,1,1)$ & $1$ & $(1, D_4, 1)$ & $(1, D_4, 1)$ & $1$ & (i) & $\Vect$ & $D(D_4)$ \\ %0 GAP:1
        \hline
        $(\langle r \rangle,1,1,1)$ & $1 \oplus (1, 1_r)$ & $(1, \langle r \rangle, 1)$ & $(1, \langle r \rangle, 1)$ & $\Z_2$ & (i) & $\Vect$ & $D(\Z_4)$ \\ %1 
        $(D_4,\langle r^2 \rangle,1,\epsilon_{s})$ & $1 \oplus (r^2, 1_{s})$ & $(\langle r^2 \rangle, D_4, B_{s})$ & $(\langle r^2 \rangle, D_4, \psi_{s})$ & $\Z_2$ & (i) & $\Vect$ & $D((\Z_2)^2, \omega)$ \\ %2 GAP:3/4/5
        $(D_4,\langle r^2 \rangle,1,\epsilon_{sr})$ & $1 \oplus (r^2, 1_{sr})$ & $(\langle r^2 \rangle, D_4, B_{sr})$ & $(\langle r^2 \rangle, D_4, \psi_{sr})$ & $\Z_2$ & (i) & $\Vect$ & $D((\Z_2)^2, \omega)$ \\ %3 GAP:3/4/5
        $(\langle s, r^2 \rangle,1,1,1)$ & $1 \oplus (1, 1_{s})$ & $(1,\langle s, r^2 \rangle, 1)$ & $(1, \langle s, r^2 \rangle, 1)$ & $\Z_2$ & (i) & $\Vect$ & $D((\Z_2)^2)$ \\ %4
        $(\langle sr, r^2 \rangle,1,1,1)$ & $1 \oplus (1, 1_{sr})$ & $(1, \langle sr, r^2 \rangle, 1)$ & $(1, \langle sr, r^2 \rangle, 1)$ & $\Z_2$ & (i) & $\Vect$ & $D((\Z_2)^2)$ \\ %5
        $(D_4,\langle r^2 \rangle,1,\epsilon_{r})$ & $1 \oplus (r^2, 1_{r})$ & $(\langle r^2 \rangle, D_4, B_{r})$ & $(\langle r^2 \rangle, D_4, \psi_{r})$ & $\Z_2$ & (i) & $\Vect$ & $D((\Z_2)^2)$ \\ %6 GAP:3/4/5
        $(D_4,\langle r^2 \rangle,1,1)$ & $1 \oplus (r^2, 1)$ & $(\langle r^2 \rangle, D_4, 1)$ & $(\langle r^2 \rangle, D_4, 1)$ & $\Z_2$ & (i) & $\Vect$ & $D((\Z_2)^2)$ \\ %7 GAP:2
        % ==================================
        \hline
        $(\langle r \rangle,\langle r^2 \rangle,1,\epsilon_{r^2})$ & $1 \oplus (1, 1_{r}) \oplus (r^2, 1_{s}) \oplus (r^2, 1_{sr})$ & $(\langle r^2 \rangle, \langle r \rangle, B_{r^2})$ & $(\langle r^2 \rangle, \langle r \rangle, \psi_{r^2})$ & $(\Z_2)^2$ & (i) & $\Vect$ & $D(\Z_2, \omega)$ \\ %8 GAP:24
        $(\langle s, r^2 \rangle,\langle s \rangle,1,1)$ & $1 \oplus (1, 1_{s}) \oplus (s, 1_{++})$ & $(\langle s, r^2 \rangle, \langle s, r^2 \rangle, 1)$ & $(\langle s, r^2 \rangle, \langle s, r^2 \rangle, 1)$ & $D_4$ & (i) & $\Vect$ & $\Vect$ \\ %9 GAP:17
        $(\langle s, r^2 \rangle,\langle s \rangle,1,\epsilon_{s})$ & $1 \oplus (1, 1_{s}) \oplus (s, 1_{+-})$ & $(\langle s, r^2 \rangle, \langle s, r^2 \rangle, B_{s})$ & $(\langle s, r^2 \rangle, \langle s, r^2 \rangle, \psi_{s})$ & $D_4$ & (i) & $\Vect$ & $\Vect$ \\ %10 GAP:18
        $(\langle s \rangle,1,1,1)$ & $1 \oplus (1, 1_{s}) \oplus (1, E)$ & $(1,1,1)$ & $(1,1,1)$ & $D_4$ & (i) & $\Vect$ & $\Vect$ \\ %11 GAP:25 (or 27)
        $(\langle sr, r^2 \rangle,\langle sr \rangle,1,1)$ & $1 \oplus (1, 1_{sr}) \oplus (sr, 1_{++})$ & $(\langle sr, r^2 \rangle, \langle sr, r^2 \rangle, 1)$ & $(\langle sr, r^2 \rangle, \langle sr, r^2 \rangle, 1)$ & $D_4$ & (i) & $\Vect$ & $\Vect$ \\ %12 GAP:21
        $(\langle sr, r^2 \rangle,\langle sr \rangle,1,\epsilon_{sr})$ & $1 \oplus (1, 1_{sr}) \oplus (sr, 1_{+-})$ & $(\langle sr, r^2 \rangle, \langle sr, r^2 \rangle, B_{sr})$ & $(\langle sr, r^2 \rangle, \langle sr, r^2 \rangle, \psi_{sr})$ & $D_4$ & (i) & $\Vect$ & $\Vect$ \\ %13 GAP:22
        $(\langle sr\rangle,1,1,1)$ & $1 \oplus (1, 1_{sr}) \oplus (1,E)$ & $(1,1,1)$ & $(1,1,1)$ & $D_4$ & (i) & $\Vect$ & $\Vect$ \\ %14 GAP:27
        $(D_4,\Z_4^{\langle r\rangle},1,\epsilon_{15})$ & $1 \oplus (r,1) \oplus (r^2, 1_{r})$ & $(\Z_4^{\langle r\rangle}, \Z_4^{\langle r\rangle}, B_{15}=1)$ & $(\Z_4^{\langle r\rangle}, \Z_4^{\langle r\rangle}, \psi_{15}=1)$ & $D_4$ & (i) & $\Vect$ & $\Vect$ \\ %15
        $(D_4,\langle sr, r^2 \rangle,\gamma_{16},\epsilon_{16})$ & $1 \oplus (r^2,1_{r}) \oplus (sr, 1_{+-})$ & $(\langle sr, r^2 \rangle,\langle sr, r^2 \rangle, B_{16})$ & $(\langle sr, r^2 \rangle,\langle sr, r^2 \rangle, \psi_{16})$ & $D_4$ & (i) & $\Vect$ & $\Vect$ \\ %16
        $(D_4,\langle s, r^2 \rangle,\gamma_{17},\epsilon_{17})$ & $1 \oplus (r^2,1_{r}) \oplus (s, 1_{+-})$ & $(\langle s, r^2 \rangle,\langle s, r^2 \rangle, B_{17})$ & $(\langle s, r^2 \rangle,\langle s, r^2 \rangle, \psi_{17})$ & $D_4$ & (i) & $\Vect$ & $\Vect$ \\ %17
        $(D_4,\Z_4^{\langle r\rangle},1,1)$ & $1 \oplus (r,1) \oplus (r^2, 1)$ & $(\Z_4^{\langle r\rangle}, \Z_4^{\langle r\rangle}, 1)$ & $(\Z_4^{\langle r\rangle}, \Z_4^{\langle r\rangle}, 1)$ & $D_4$ & (i) & $\Vect$ & $\Vect$ \\ %18
        $(D_4,\langle s, r^2 \rangle,1,1)$ & $1 \oplus (r^2,1) \oplus (s, 1_{++})$ & $(\langle s, r^2 \rangle,\langle s, r^2 \rangle,1)$ & $(\langle s, r^2 \rangle, \langle s, r^2 \rangle, 1)$ & $D_4$ & (i) & $\Vect$ & $\Vect$\\ %19
        $(D_4,\langle sr, r^2 \rangle,1,1)$ & $1 \oplus (r^2,1) \oplus (sr, 1_{++})$ & $(\langle sr, r^2 \rangle, \langle sr, r^2 \rangle, 1)$ & $(\langle sr, r^2 \rangle, \langle sr, r^2 \rangle, 1)$ & $D_4$ & (i) & $\Vect$ & $\Vect$ \\ %20
        $(\langle r^2 \rangle,1,1,1)$ & $1 \oplus (1,1_{r}) \oplus (1,1_{s}) \oplus (1, 1_{sr})$ & $(1, \langle r^2 \rangle, 1)$ & $(1, \langle r^2 \rangle, 1)$ & $(\Z_2)^2$ & (i) & $\Vect$ & $D(\Z_2)$ \\ %21 GAP:26
        $(\langle sr, r^2 \rangle,\langle r^2 \rangle,1,\epsilon_{r^2})$ & $1 \oplus (1,1_{sr}) \oplus (r^2,1_{s}) \oplus (r^2, 1_{r})$ & $(\langle r^2 \rangle, \langle sr, r^2 \rangle, B_{r^2})$ & $(\langle r^2 \rangle, \langle sr, r^2 \rangle, \psi_{r^2})$ & $(\Z_2)^2$ & (i) & $\Vect$ & $D(\Z_2)$ \\ %22 GAP:20
        $(\langle s, r^2 \rangle,\langle r^2 \rangle,1,\epsilon_{r^2})$ & $1 \oplus (1,1_{s}) \oplus (r^2,1_{sr}) \oplus (r^2, 1_{r})$ & $(\langle r^2 \rangle, \langle s, r^2 \rangle, B_{r^2})$ & $(\langle r^2 \rangle, \langle s, r^2 \rangle, \psi_{r^2})$ & $(\Z_2)^2$ & (i) & $\Vect$ & $D(\Z_2)$ \\ %23 GAP:16
        $(\langle s, r^2 \rangle,\langle r^2 \rangle,1,1)$ & $1 \oplus (1,1_{s}) \oplus (r^2,1) \oplus (r^2, 1_{s})$ & $(\langle r^2 \rangle, \langle s, r^2 \rangle, 1)$ & $(\langle r^2 \rangle, \langle s, r^2 \rangle, 1)$ & $(\Z_2)^2$ & (i) & $\Vect$ & $D(\Z_2)$ \\ %24 GAP:15 (or 19?)
        $(\langle sr, r^2 \rangle,\langle r^2 \rangle,1,1)$ & $1 \oplus (1,1_{sr}) \oplus (r^2,1) \oplus (r^2, 1_{sr})$ & $(\langle r^2 \rangle, \langle sr, r^2 \rangle, 1)$ & $(\langle r^2 \rangle, \langle sr, r^2 \rangle, 1)$ & $(\Z_2)^2$ & (i) & $\Vect$ & $D(\Z_2)$ \\ %25 GAP:19
        $(\Z_4^{\langle r\rangle}, \langle r^2 \rangle,1,1)$ & $1 \oplus (1,1_{r}) \oplus (r^2,1) \oplus (r^2, 1_{r})$ & $(\langle r^2 \rangle, \Z_4^{\langle r\rangle}, 1)$ & $(\langle r^2 \rangle, \Z_4^{\langle r\rangle}, 1)$ & $(\Z_2)^2$ & (i)  & $\Vect$ & $D(\Z_2)$\\ %26 GAP:23
        \toprule
    \end{tabular}   
\caption{The non-maximal condensable algebras in $D(D_4)$ and the associated data for the proliferation transitions: The first column lists the data classifying the algebra $\cA=\cA(H,N,\gamma, \epsilon)$. 
The generated subcategory is $S(K_{\cA}, L_{\cA}, B_{\cA})=\langle \cA \rangle$. The fourth column consists the triple for the M\"uger center $(R_{\cA}, P_{\cA}, \psi_{\cA})=\cZ_2(\langle \cA \rangle)$. 
 $G_{\cA}$ determined the M\"uger center $\cZ_2(\langle \cA \rangle)\cong \Rep(G_{\cA})$. We also list the type of algebra and the inputs into the SymTFT $\cM_{\cA}$ and $\cN_{\cA}$. Labels: $1_x$, $x\in\{r,s,sr\}$, is the one-dimensional irreducible representation of $D_4$ whose kernel contains $x$, and $E$ the two-dimensional irrep. $(s,1_{ab})$ is the flux $s$ with the character of $C_{D_4}(s)=\langle s,r^2\rangle$ that takes the value $a$ on $s$ and $b$ on $r^2$, and similarly for $sr$ with $C_{D_4}(sr)=\langle sr,r^2\rangle$. In $\cA(H,\Z_2^{\langle g\rangle},\gamma,\epsilon_x)$, $\epsilon_x(\cdot,g)$ is the non-trivial character of $H$ whose kernel contains $x$. The data $\epsilon_{15}$ (with $\epsilon_{15}(s,r)=i$), $\gamma_{16},\epsilon_{16},\gamma_{17},\epsilon_{17}$ are those of \cite{Gai:2026hjk}. For $\cA(\langle r\rangle,\langle r^2\rangle,1,\epsilon_{r^2})$, $\cN_{\cA}$ is the double semion $D(\Z_2,\overline\omega)$.}
    \label{tab:D4examplenonLag}
\end{table*}

\begin{table*}
\renewcommand{\arraystretch}{2}
    \begin{tabular}{|c|c|c|c|c|c|c|c|}
    \bottomrule
        $(H,\gamma)$ & Anyon Decomposition & $(K_{\cA}, L_{\cA}, B_{\cA})$ &  $(R_{\cA}, P_{\cA}, \psi_{\cA})$ & $G_{\cA}$ & type & $\cM_{\cA}$ & $\cN_{\cA}$ \\
        \hline
        $(1,1)$ & \makecell{$1 \oplus (1,1_r) \oplus (1, 1_{s})$\\$ \oplus (1, 1_{sr})\oplus 2E$} & $(1,1,1)$ & $(1,1,1)$ & $D_4$ & (i) & $\Vect$ & $\Vect$ \\ %27
        $(\langle r \rangle,1)$ & \makecell{$1 \oplus (1,1_r) \oplus (r^2, 1) $\\$\oplus (r^2, 1_{r})\oplus 2(r, 1)$} & $(\langle r \rangle, \langle r \rangle, 1)$ & $(\langle r \rangle, \langle r \rangle, 1)$ & $D_4$ & (i) & $\Vect$ & $\Vect$ \\ %28 GAP:34
        $(\langle s , r^2 \rangle,1)$ & \makecell{$1 \oplus (1,1_{s}) \oplus (r^2, 1) $\\$\oplus (r^2, 1_{s})\oplus 2(s, 1_{++})$} & $(\langle s , r^2 \rangle, \langle s , r^2 \rangle, 1)$ & $(\langle s , r^2 \rangle, \langle s , r^2 \rangle, 1)$ & $D_4$ & (i) & $\Vect$ & $\Vect$ \\ %29
        $(\langle s , r^2 \rangle,\gamma_{s,r^2})$ & \makecell{$1 \oplus (1,1_{s}) \oplus (r^2, 1_{sr}) $\\$\oplus (r^2, 1_{r})\oplus 2(s, 1_{+-})$} & $(\langle s , r^2 \rangle, \langle s , r^2 \rangle, B_{s,r^2})$ & $(\langle s , r^2 \rangle, \langle s , r^2 \rangle, \psi_{s,r^2})$ & $D_4$ & (i) & $\Vect$ & $\Vect$ \\ %30
        $(\langle sr , r^2 \rangle,1)$ & \makecell{$1 \oplus (1,1_{sr}) \oplus (r^2, 1) $\\$\oplus (r^2, 1_{sr})\oplus 2(sr, 1_{++})$} & $(\langle sr , r^2 \rangle, \langle sr , r^2 \rangle, 1)$ & $(\langle sr , r^2 \rangle,\langle sr , r^2 \rangle, 1)$ & $D_4$ & (i) & $\Vect$ & $\Vect$ \\ %31
        $(\langle sr , r^2 \rangle,\gamma_{sr,r^2})$ & \makecell{$1 \oplus (1,1_{sr}) \oplus (r^2, 1_{s}) $\\$\oplus (r^2, 1_{r})\oplus 2(sr, 1_{+-})$} & $(\langle sr , r^2 \rangle, \langle sr , r^2 \rangle, B_{sr,r^2})$ & $(\langle sr , r^2 \rangle, \langle sr , r^2 \rangle, \psi_{sr,r^2})$ & $D_4$ & (i) & $\Vect$ & $\Vect$ \\ %32
        $(D_4,1)$ & \makecell{$1 \oplus (s, 1_{++}) \oplus (r, 1) $\\$\oplus (r^2, 1)\oplus (sr, 1_{++})$} & $(D_4, \langle r^2 \rangle, 1)$ & $(\langle r^2 \rangle, D_4, 1)$ & $\Z_2$ & (ii) & $D((\Z_2)^2)$ & $\Vect$ \\ %33 GAP:28
        $(D_4,\gamma_{D_8})$ & \makecell{$1 \oplus (s, 1_{+-}) \oplus (r, 1) $\\$\oplus (r^2, 1_{r})\oplus (sr, 1_{+-})$} & $(D_4, \langle r^2 \rangle, B_{D_4})$ & $(\langle r^2 \rangle, D_4, \psi_{D_4})$ & $\Z_2$ & (ii) & $D((\Z_2)^2)$ & $\Vect$ \\ %34 GAP:29
        $(\langle s\rangle,1)$ & \makecell{$1 \oplus (1, 1_{s}) \oplus (1, E) $\\$\oplus (s, 1_{++})\oplus (s, 1_{+-})$} & $(\langle s, r^2 \rangle, 1, 1)$ & $(1, \langle s, r^2 \rangle, 1)$ & $\Z_2$ & (ii) & $D((\Z_2)^2)$ & $\Vect$ \\ %35 GAP:35 (or 37)
        $(\langle sr\rangle,1)$ & \makecell{$1 \oplus (1, 1_{sr}) \oplus (1, E) $\\$\oplus (sr, 1_{++})\oplus (sr, 1_{+-})$} & $(\langle sr , r^2 \rangle, 1, 1)$ & $(1, \langle sr , r^2 \rangle, 1)$ & $\Z_2$ & (ii) & $D((\Z_2)^2)$ & $\Vect$ \\ %36 GAP:37
        $(\langle r^2 \rangle,1)$ & \makecell{$1 \oplus (1, 1_{r}) \oplus (1, 1_{s}) $\\ $ \oplus(1, 1_{sr})\oplus (r^2, 1) \oplus (r^2, 1_{r})$\\ $\oplus (r^2, 1_{s}) \oplus (r^2, 1_{sr})$} & $(\langle r^2 \rangle, \langle r^2 \rangle, 1)$ & $(\langle r^2 \rangle, \langle r^2 \rangle, 1)$ & $(\Z_2)^3$ & (i) & $\Vect$ & $\Vect$ \\ %37 GAP:36
        \toprule
    \end{tabular}
    \caption{Lagrangian algebras in $D(D_4)$. $(H,\gamma)$ indicates  $\cA=\cA(H,H, \gamma, \epsilon_{\gamma})$, where $\epsilon$ here is determined uniquely by $\gamma$ as for all Lagrangian algebras. The remaining labels are as before.  }
    \label{tab:D4exampleLag}
\end{table*}

\begin{table*}
    \centering
    \begin{tabular}{|c|c|c|c|c|c|c|c|}
    \bottomrule
        $(H,N)$ & object & $(K_{\cA}, L_{\cA}, B_{\cA})$ &  $(R_{\cA}, P_{\cA})$ & $G_{\cA}$ & type & $\cM_{\cA}$ & $\cN_{\cA}$ \\
        \hline
        $(D_4,1)$ & $1$ & $(1,D_4, 1)$ & $(1,D_4)$ & $1$ & (i) & $\Vect$ & $ D(D_4, \omega)$ \\ %1
        $(\langle s, r^2\rangle,1)$ & $1\oplus (1, \chi_{-+})$ & $(1,\langle s, r^2\rangle, 1)$ & $(1,\langle s, r^2\rangle)$ & $\Z_2$ & (i) & $\Vect$ & $ D(\langle s, r^2\rangle)$ \\ %2
        $(\langle sr, r^2\rangle,1)$ & $1\oplus (1, \chi_{--})$ & $(1, \langle sr, r^2\rangle, 1)$ & $(1,\langle sr, r^2\rangle, 1)$ & $\Z_2$ & (i) & $\Vect$ & $ D(\langle sr, r^2\rangle, \omega)$ \\ %3
        $(\langle r \rangle,1)$ & $1\oplus (1, \chi_{+-})$ & $(1, \langle r \rangle, 1)$ & $(1,\langle r \rangle)$ & $\Z_2$ & (i) & $\Vect$ & $ D(\langle r \rangle, \omega)$ \\ %4
        $(\langle s, r^2\rangle, \langle r^2\rangle)$ & $1\oplus (1, \chi_{-+}) \oplus (r^2, \pi)$ & $(\langle r^2\rangle, \langle r^2\rangle, 1)$ & $(\langle r^2\rangle,\langle r^2\rangle)$ & $D_4$ & (i) & $\Vect$ & $\Vect$ \\ %5 
        $(\langle s, r^2\rangle, \langle s\rangle)$ & $1\oplus (1, \chi_{-+}) \oplus (s, \sigma_{+})$ & $(\langle s, r^2\rangle, \langle s, r^2\rangle, B_{+})$ & $(1, \langle s, r^2\rangle)$ & $\Z_2$ & (ii) & $ D(\Z_2)$ & $ D(\Z_2)$ \\ %6 GAP7
        $(\langle s, r^2\rangle, \langle s\rangle)_{\epsilon}$ & $1\oplus (1, \chi_{-+}) \oplus (s, \sigma_{-})$ & $(\langle s, r^2\rangle, \langle s, r^2\rangle, B_{-})$ & $(1, \langle s, r^2\rangle)$ & $\Z_2$ & (ii) & $ D(\Z_2)$ & $ D(\Z_2)$ \\ %7 GAP8
        $(\langle sr, r^2\rangle, \langle r^2\rangle)$ & $1\oplus (1, \chi_{--}) \oplus (r^2, \pi))$ & $(\langle r^2\rangle, \langle r^2\rangle, 1)$ & $(\langle r^2\rangle,\langle r^2\rangle)$ & $D_4$ & (i) & $\Vect$ & $\Vect$ \\ %8 GAP6
        $(\langle s\rangle,1)$ & $1 \oplus (1, \chi_{-+}) \oplus (1, E)$ & $(1,1)$ & $(1,1)$ & $D_4$ & (i) & $\Vect$ & $\Vect$ \\ %9
        $(\langle r^2\rangle,1)$ & $\bigoplus_{a,b=\pm}(1, \chi_{ab})$ & $(1,\langle r^2\rangle, 1)$ & $(1,\langle r^2\rangle)$ & $\Z_2^2$ & (i) & $\Vect$ & $ D(\langle r^2\rangle)$ \\ %10
        $(\langle sr\rangle,1)$ & $1 \oplus (1, \chi_{--}) \oplus (1, E)$ & $(1,1, 1)$ & $(1,1)$ & $D_4$ & (i) & $\Vect$ & $\Vect$ \\ %11
        \hline
        $(\langle s, r^2\rangle,\langle s, r^2\rangle)$ & $1 \oplus (1, \chi_{-+})\oplus (r^2, \pi) \oplus (s, \sigma_\pm) $ & $(\langle s, r^2\rangle,1, 1)$ & $(1,\langle s, r^2\rangle)$ & $\Z_2$ & (ii) & $ D(\langle s, r^2\rangle)$ & $\Vect$ \\ %12
        $(\langle s\rangle,\langle s\rangle)$ & $1 \oplus (1, \chi_{-+})\oplus (1, E) \oplus (s, \sigma_\pm) $ & $(\langle s, r^2\rangle,1, 1)$ & $(1,\langle s, r^2\rangle)$ & $\Z_2$ & (ii) & $ D(\langle s, r^2\rangle)$ & $\Vect$ \\ %13 GAP14
        $(\langle r^2\rangle,\langle r^2\rangle)$ & $\bigoplus_{a,b=\pm}(1, \chi_{ab}) \oplus 2 (r^2, \pi)$ & $(\langle r^2\rangle,\langle r^2\rangle, 1)$ & $(\langle r^2\rangle,\langle r^2\rangle)$ & $D_4$ & (i) & $\Vect$ & $\Vect$ \\ %14 GAP13
        $(1, 1)$ & $\bigoplus_{a,b=\pm}(1, \chi_{ab}) \oplus 2 (1, E)$ & $(1, 1, 1)$ & $(1, 1)$ & $D_4$ & (i) & $\Vect$ & $\Vect$ \\ %15
        \toprule
    \end{tabular}
    \caption{Condensable algebras proposed for the twisted $D(D_4,\omega)$ example.
   {The only pair of algebras having the same $(H,N)$ data have both trivial $\gamma$, one with trivial $\epsilon$, the other with non-trivial $\epsilon$: $\epsilon(r^2,s)=\epsilon(sr^2,s)=-1$ (all other entries evaluate to 1). $B_{\pm}$ are non-equivalent bicharacters whose value can be computed using GAP \cite{GAP4}.} The other notation is as in the earlier tables.}
    \label{tab:TY4example}
\end{table*}

All algebras of $D(D_4)$ are of type (i) except the four Lagrangian algebras $\cA(D_4,D_4,1)$, $\cA(D_4,D_4,\gamma_{D_8})$, $\cA(\langle s\rangle,\langle s\rangle,1)$ and $\cA(\langle sr\rangle,\langle sr\rangle,1)$, of type (ii). For the type (ii) algebras $G_{\cA}=\Z_2$, generated by the transparent central flux $r^2$ for $H=D_4$ and by the transparent charge $1_s$ (respectively $1_{sr}$) for $H=\langle s\rangle$ (respectively $\langle sr\rangle$), and $\cM_{\cA}\cong D(\Z_2^2)$ consists of two toric codes. 
A possible {2-step condensation} first condenses the transparent boson, giving the Ising$^*$ Higgs step when realized dynamically, and then condenses the residual Lagrangian algebra in $\cM_{\cA}$.

\subsection{$D(D_4, \omega)$}

When $\omega$ is non-trivial, $D(D_4, \omega)$ can realize the Drinfeld center of $\Z_2\times\Z_2$ TY-categories \cite{TambaraYamagami}. Here, we focus on the only $\TY(\Z_2\times \Z_2)$-category which does not admit a fiber functor.
{Using the GAP code \cite{AlgebrasTQD}, we produce}
Table~\ref{tab:TY4example}, which records the proposed algebra data for this example. 
The data for the 3-cocycle $\omega$ are generated using the GAP code \cite{AlgebrasTQD}. The $\gamma$ and $\epsilon$ data are produced by the same GAP code.
 For the anyon labels, $\chi_{ab}$, $a,b\in \{\pm 1\}$ denotes the 1-dim character with $\chi_{ab}(r)=a$, $\chi_{ab}(s)=b$; $\pi$ denotes the 2-dim projective character; $\sigma_{\pm}$ denotes a 1-dim character of $\langle s, r^2 \rangle$ with $\sigma_{\pm}(s)=1$ and $\sigma_{\pm}(r^2)=\pm1$.

\subsection{Super-Tannakian M\"uger Center}
\label{sec:supertannakian-example}

We used the GAP code \cite{AlgebrasTQD} scanning through the small groups, and found that the smallest group such that $D(G)$ admits condensable algebras whose generated fusion subcategories are super-Tannakian is 
\be 
G=\Z_2\times A_4\,.
\ee 
 Using the same GAP code, we found that there are in total five such algebras in $D(\Z_2\times A_4)$.
Their defining data and generated categories are given below and summarized in Table~\ref{tab:superexample}. All five are of type (ii).

In the current section, we fix a presentation of $\Z_2\times A_4$. In what follows, $[g,h]:=g^{-1}h^{-1}gh$ denotes the commutator. $\Z_2\times A_4$ has generators $f_1$, $f_2$, $f_3$ and $f_4$, with relations: $f_1^2=f_3^2 = f_4^2 = f_2^3=1$; 
%for $i=2,3,4$, $f_i$ commutes with $f_1$;
$[f_4, f_3]=[f_i, f_1]=1$ for $i=2,3,4$ and $[f_3,f_2]f_4f_3=[f_4, f_2] f_3=1$. 

\subsubsection{Algebras of Dimension 12}
% algebra 48 in the GAP output
Consider condensable algebra $\cA_1:=\cA(H_1, N_1, 1, \epsilon_1)$ in $D(\Z_2\times A_4)$, where
\begin{equation}
    H_1 = \Z_2^{\langle f_1 \rangle}\times \Z_2^{\langle f_3 \rangle} \times \Z_2^{\langle f_4 \rangle}, \quad N_{1} = \Z_2^{\langle f_4 \rangle} \times \Z_2^{\langle f_1f_3 \rangle}
\end{equation}
with trivial $\gamma$ %: $\gamma_2(f_4,f_4)=\gamma_2(f_4, f_1f_3f_4)=\gamma_2(f_1f_3, f_4) = \gamma_2(f_1 f_3, f_1 f_3 f_4)=-1$.
and non-trivial $\epsilon_1$, with non-trivial entries $\epsilon_{1}(g, h)=-1$ for $g\in \{f_1, f_3, f_1 f_4, f_3 f_4\}$ and $h\in \{f_1 f_3, f_1 f_3 f_4\}$. $\cA_1$ has quantum dimension $12$ and decomposes into simples:
\begin{equation}
    \begin{split}
        \cA_1 & \cong 1 \oplus (1, \chi_1) \oplus (1, \chi_2) \oplus (f_3, 1) \\
        & \quad \oplus (f_1 f_3, \chi^{f_1f_3}_1) \oplus (f_1 f_3, \chi^{f_1f_3}_2).
    \end{split}
\end{equation}
Here $\chi_1,\chi_2$ are the two non-trivial one-dimensional representations of $\Z_2\times A_4$ that are trivial on $f_1$ (the $\Z_3$ charges of $A_4$), and $\chi^{f_1f_3}_1$, $\chi^{f_1f_3}_2$ are the characters of $C_G(f_1f_3)=H_1$ with $\chi(f_1)=\chi(f_3)=-1$ and $\chi(f_4)=-1$, $+1$, respectively.
The fusion subcategory 
\begin{equation}
    \langle \cA_1\rangle = S(K_{\cA_1}, L_{\cA_1}, B_{\cA_1})
\end{equation}
has
\begin{equation}
    K_{\cA_1} = H_1, \quad L_{\cA_1} = \Z_2^{\langle f_1 \rangle},
\end{equation}
$B_{\cA_1}$ is non-trivial with non-trivial entries: $B_{\cA_1}(f_1, f_1) = B_{\cA_1}(f_1 f_3, f_1) =B_{\cA_1}(f_1 f_4, f_1) = B_{\cA_1}(f_1 f_3 f_4, f_1) = -1$ (all other entries are 1). The M\"uger center $\cZ_2(\langle \cA_1\rangle) =S(R_{\cA_1}, P_{\cA_1}, \psi_{\cA_1})$ has
\begin{equation}
    R_{\cA_1} = \Z_2^{\langle f_1 \rangle}, \quad P_{\cA_1} = H_1,
\end{equation}
and non-trivial $\psi_{\cA_1}$ with non-trivial entries: $\psi_{\cA_1}(f_1, g)=-1$ for every $g\in \{f_1, f_1 f_3, f_1 f_4, f_1 f_3 f_4\}$ (all other entries are 1). It is super-Tannakian, instead of Tannakian, 
\begin{equation}
    \cZ_2(\langle \cA_1\rangle ) \cong \Rep(G_{\cA_1}, z_{\cA_1})
\end{equation}
where 
\begin{equation}
    G_{\cA_1} = \Z_2 \rtimes \Z_3 = \Z_6,
\end{equation}
where the first $\Z_2 = \widehat R_{\cA_1}$
and $z_{\cA_1}$ is the non-trivial character on $R_{\cA_1}=\Z_2$.
Note that $\Core_G(C_{H_1}(N_1))=H_1$.

% algebra 54 from GAP output

The second such algebra $\cA_2 = \cA(H_2, N_2, \gamma_2, \epsilon_2)$ with $H_2=H_1$, $N_2=N_1$, a non-trivial $\gamma_2$ with non-trivial entries 
\begin{equation}
    \gamma_2(g, h)=-1
\end{equation}
if and only if $g\in \{f_4, f_1 f_3\}$ and $h\in \{ f_4, f_1 f_3 f_4 \}$, and it takes the value of 1 for any other pair of elements, and
\begin{equation}
    \epsilon(g, h) = -1
\end{equation}
if and only if
\begin{equation}
    \begin{split}
        (g,h)\in \{ & (f_1^{j}f_4^{1-j}, f_1 f_3 f_4^{i}), (f_3(f_1f_4)^j, (f_1 f_3)^{i} f_4^{1-i}), \\
        & (f_3 (f_1 f_4)^j, (f_1 f_3)^i f_4) \mid i, j=0,1 \}
    \end{split}
\end{equation}
and takes value 1 for any other pair.
In the decomposition below, $\chi^{f_3}_1$ is trivial on $f_1,f_3$ and takes value $-1$ on $f_4$. We obtain
\begin{equation}
    \begin{split}
        \cA_2 & \cong 1 \oplus (1, \chi_1) \oplus (1, \chi_2) \oplus (f_3, \chi^{f_3}_1) \\
        & \quad \oplus (f_1 f_3, \chi^{f_1f_3}_1) \oplus (f_1 f_3, \chi^{f_1f_3}_2).
    \end{split}
\end{equation}
$\cA_2$ generates the same fusion subcategory as $\cA_1$, 
\begin{equation}
    \langle \cA_2 \rangle = \langle \cA_1\rangle = S(H_1, \Z_2^{\langle f_1 \rangle}, B_{\cA_1})
\end{equation}
hence
\begin{equation}
    \cZ_2(\langle \cA_2\rangle) = \cZ_2(\langle \cA_1 \rangle) \cong \Rep(G_{\cA_1}, z_{\cA_1}).
\end{equation}

% algebra 61 from GAP
On the other hand, there exists condensable algebra $\cA_3=\cA(H_3, N_3, 1, \epsilon_3)$ where 
\begin{equation}
    H_3 = \Z_2^{\langle f_1 \rangle}\times \Z_2^{\langle f_4 \rangle}, \quad N_{3} = \Z_2^{\langle f_1 f_4 \rangle},
\end{equation}
and $\epsilon_3$ has non-trivial entries $\epsilon_3(f_1, f_1 f_4)= \epsilon_3(f_4, f_1 f_4) = -1$. $\cA_3$ decomposes into simples:
\begin{equation}
    \begin{split}
        \cA_3 & \cong 1 \oplus (1, \chi_1) \oplus (1, \chi_2) \oplus (1, \chi_3) \\
        & \quad \oplus (f_1 f_3, \chi^{f_1f_3}_1) \oplus (f_1 f_3, \chi^{f_1f_3}_2).
    \end{split}
\end{equation}
Here, $\chi_3$ is the character of a $(\Z_2\times A_4)$-representation of dimension $\chi_3(1)=3$. Again, the generated fusion subcategory is the same as the previous examples 
\begin{equation}
    \langle \cA_3\rangle = \langle \cA_1\rangle = S(H_1, \Z_2^{\langle f_1 \rangle}, B_{\cA_1})
\end{equation}
and 
\begin{equation}
    \cZ_2(\langle \cA_3 \rangle) = \cZ_2(\langle \cA_1 \rangle) \cong \Rep(G_{\cA_1}, z_{\cA_1}).
\end{equation}

\begin{table*}
\renewcommand{\arraystretch}{2}
    \centering
    \begin{tabular}{|c|c|c|c|c|}
    \bottomrule
        $i$ & $\cA_i=\cA(H,N, \gamma, \epsilon)$ & object & $(K_{\cA_i}, L_{\cA_i}, B_{\cA_i})$ & type  \\
        \hline
        $1$ & $(\Z_2^3=\langle f_1, f_3, f_4 \rangle, \Z_2^2 = \langle f_4, f_1f_3\rangle, 1, \epsilon_1)$ & \makecell[c]{$1 \oplus (1, \chi_1) \oplus (1, \chi_2) \oplus (f_3, 1)$ \\ $\oplus (f_1 f_3, \chi^{f_1f_3}_1) \oplus (f_1 f_3, \chi^{f_1f_3}_2)$} & $(\Z_2^3 = \langle f_1, f_3, f_4 \rangle, \Z_2=\langle f_1 \rangle, B_{\cA_1})$ & (ii) \\[7pt] %1
        $2$ & $(\Z_2^3 = \langle f_1, f_3, f_4 \rangle, \Z_2^2 = \langle f_4, f_1f_3\rangle, \gamma_2, \epsilon_2)$ & \makecell[c]{$1 \oplus (1, \chi_1) \oplus (1, \chi_2) \oplus (f_3, \chi^{f_3}_1)$ \\ $\oplus (f_1 f_3, \chi^{f_1f_3}_1) \oplus (f_1 f_3, \chi^{f_1f_3}_2))$} & $(\Z_2^3 = \langle f_1, f_3, f_4 \rangle, \Z_2 = \langle f_1 \rangle, B_{\cA_1})$ & (ii)  \\[7pt] %2
        $3$ & $(\Z_2^2 = \langle f_1, f_4 \rangle, \Z_2 = \langle f_1f_4\rangle, 1, \epsilon_3)$ & \makecell[c]{$1 \oplus (1, \chi_1) \oplus (1, \chi_2) \oplus (1, \chi_3)$ \\ $\oplus (f_1 f_3, \chi^{f_1f_3}_1) \oplus (f_1 f_3, \chi^{f_1f_3}_2)$} & $(\Z_2^3 = \langle f_1, f_3, f_4 \rangle, \Z_2 = \langle f_1 \rangle, B_{\cA_1})$ & (ii) \\[7pt] %3
        $4$ & $(\Z_2^3 = \langle f_1, f_3, f_4 \rangle, \Z_2 = \langle f_1f_4\rangle, 1, \epsilon_4)$ & $1 \oplus (1, \chi_1) \oplus (1, \chi_2) \oplus (f_1 f_3, \chi^{f_1f_3}_1)$ & $(\Z_2^3 = \langle f_1, f_3, f_4 \rangle, \langle f_1, f_3, f_4 \rangle, B_{\cA_4})$ & (ii) \\[7pt] %4
        $5$ & $(\Z_2^3 = \langle f_1, f_3, f_4 \rangle, \Z_2^{\langle f_1f_4\rangle}, 1, \epsilon_5)$ & $1 \oplus (1, \chi_1) \oplus (1, \chi_2) \oplus (f_1 f_3, \chi^{f_1f_3}_2)$ & $(\Z_2^3 = \langle f_1, f_3, f_4 \rangle, \langle f_1, f_3, f_4 \rangle, B_{\cA_5})$ & (ii) \\ %5 
        \toprule
        \bottomrule
        $i$ & $(R_{\cA_i}, P_{\cA_i}, \psi_{\cA_i})$ & $(G_{\cA_i}, z_{\cA_i})$ & $\cM_{\cA_i}^f$ & $\cN_{\cA_i}^f$ \\
        \hline
        $1$ & $(\Z_2 = \langle f_1 \rangle, \Z_2^3 = \langle f_1, f_3, f_4 \rangle, \psi_{\cA_1})$ & $(\Z_6^{\langle a\rangle}, a^3)$ & $\sVec \boxtimes D(\Z_2^2)$ & $\sVec$ \\[7pt] %1
        $2$ & $(\Z_2 = \langle f_1 \rangle, \Z_2^3 = \langle f_1, f_3, f_4 \rangle, \psi_{\cA_1})$ & $(\Z_6^{\langle a\rangle}, a^3)$ & $\sVec \boxtimes D(\Z_2^2)$ & $\sVec$ \\[7pt] %2
        $3$ & $(\Z_2 = \langle f_1 \rangle, \Z_2^3 = \langle f_1, f_3, f_4 \rangle, \psi_{\cA_1})$ & $(\Z_6^{\langle a\rangle}, a^3)$ & $\sVec \boxtimes D(\Z_2^2)$ & $\sVec$ \\[7pt] %3
        $4$ & $(\Z_2 = \langle f_1 \rangle, \Z_2^3 = \langle f_1, f_3, f_4 \rangle, \psi_{\cA_1})$ & $(\Z_6^{\langle a\rangle}, a^3)$ & $\sVec \boxtimes \cF_4$ & $\sVec \boxtimes \cF_5$ \\[7pt] %4
        $5$ & $(\Z_2 = \langle f_1 \rangle, \Z_2^3 = \langle f_1, f_3, f_4 \rangle, \psi_{\cA_1})$ & $(\Z_6^{\langle a\rangle}, a^3)$ & $\sVec \boxtimes \cF_5$ & $\sVec \boxtimes \cF_4$ \\ %5 
        \toprule
    \end{tabular}
    \caption{Condensable algebras with super-Tannakian M\"uger center in $D(\Z_2\times A_4)$. The top table consists of $(H,N, \gamma, \epsilon)$ in the classification of condensable algebras $\cA_i=\cA(H,N,\gamma, \epsilon)$, its anyon decomposition and generated subcategory $\langle \cA_i\rangle = S(K_{\cA_i}, L_{\cA_i}, B_{\cA_i})$. The bottom table consists of the M\"uger center $\cZ_2(\langle \cA_i \rangle) = S(R_{\cA_i}, P_{\cA_i}, \psi_{\cA_i}) \cong \Rep(G_{\cA_i}, z_{\cA_i})$ and the almost degenerate braided categories $\cM^f_{\cA_i}$ and $\cN^f_{\cA_i}$. Here, $\epsilon_i$, $B_{\cA_1}$, $\psi_{\cA_i}$ can be found in the main text. $\cF_4$ and $\cF_5$ denotes the 3-fermion MTCs consisting of anyons \eqref{eqn:F4F5-3fermMTC}. }
    \label{tab:superexample}
\end{table*}

\subsubsection{Algebras of Dimension 6}
% algebra 13 from GAP
Consider condensable algebra $\cA_4:=\cA(H_4, N_4, 1, \epsilon_4)$:
\begin{equation}
    H_{4}= H_1, \quad N_4 = N_3
\end{equation}
and $\epsilon_4$ has non-trivial entries $\epsilon_4(g, f_1 f_4)=-1$ for $g\in \{ f_1, f_4, f_1 f_3, f_3f_4 \}$ (note that $\epsilon_3 = \epsilon_4|_{H_3\times N_3}$). $\cA_4$ decomposes into simples
\begin{equation}
    \begin{split}
        \cA_4 & \cong 1 \oplus (1, \chi_1) \oplus (1, \chi_2) \oplus (f_1 f_3, \chi^{f_1f_3}_1).
    \end{split}
\end{equation}
The fusion subcategory $\langle \cA_4 \rangle = S(K_{\cA_4}, L_{\cA_4}, B_{\cA_4})$ has
\begin{equation}
    K_{\cA_4} = L_{\cA_4} = H_4 = \Z_2^{\langle f_1 \rangle}\times \Z_2^{\langle f_3 \rangle} \times \Z_2^{\langle f_4 \rangle},
\end{equation}
and $B_{\cA_4}$ such that $B_{\cA_4}(f_3, f_4) = B_{\cA_4}(g,g)=-1$ for $g\in \{f_1, f_3, f_4\}$. Its M\"uger center is the same as the M\"uger center of the previous examples $\cZ_2(\langle \cA_4\rangle ) = \cZ_2(\langle \cA_1 \rangle)$ and hence
\begin{equation}
    \cZ_2(\langle \cA_4\rangle) \cong \Rep(G_{\cA_1}, z_{\cA_1}).
\end{equation}

% algebra 15
Similarly, there is one more condensable algebra $\cA_5:=\cA(H_5, N_5, 1, \epsilon_5)$ with $H_5=H_4$ and $N_5=N_4$, while $\epsilon_5$ has non-trivial entries $\epsilon_5(g, f_1 f_4)=-1$ for $g\in \{f_1, f_3, f_4, f_1 f_3 f_4\}$. $\cA_5$ decomposes as
\begin{equation}
    \begin{split}
        \cA_5 & \cong 1 \oplus (1, \chi_1) \oplus (1, \chi_2) \oplus (f_1 f_3, \chi^{f_1f_3}_2).
    \end{split}
\end{equation}
The fusion subcategory $\langle \cA_5 \rangle = S(K_{\cA_5}, L_{\cA_5}, B_{\cA_5})$ has
\begin{equation}
    K_{\cA_5} = L_{\cA_5} = H_5
\end{equation}
and $B_{\cA_5}$ such that $B_{\cA_5}(f_4, f_3) = B_{\cA_5}(g,g)=-1$ for $g\in \{f_1, f_3, f_4\}$. Its M\"uger center is again the same as the previous examples,
\begin{equation}
    \cZ_2(\langle \cA_5 \rangle) = \cZ_2(\langle \cA_1 \rangle) \cong \Rep(G_{\cA_1}, z_{\cA_1}).
\end{equation}

\subsubsection{SymTFT Construction}
The five algebras have the same super-Tannakian M\"uger center.
Note that $D(\Z_2 \times A_4) = D(\Z_2) \boxtimes D(A_4)$, and denote by $\psi_1 = m_1 e_1$ the fermion in the first toric code factor, then
\be
\begin{split}
    \cE_{\cA_1} & =S(R_{\cA_1}, P_{\cA_1}, \psi_{\cA_1})\\
    & = \{ \psi_1^p \boxtimes (1, \rho_k) \mid p\in \Z_2, k\in \Z_3 \} \\
    & \cong \sVec \boxtimes \Rep(Z_3).
\end{split}
\ee
Here $\rho_k$ are the 1d irreps of $A_4$, trivial on $\Z_2^{\langle f_3\rangle} \times \Z_2^{\langle f_4\rangle}$. 
The spin of $\psi_1^p \boxtimes (1, \rho_k)$ is $\theta_{\psi_1^p \boxtimes (1, \rho_k)} = (-1)^p$. 

The odd lines $\psi_1\boxtimes(1,\rho_k)$ have twist $-1$ and are {not subobjects of $\cA_1$}. They occur in the {decomposition of $X^{\otimes 3}$,} $X=(f_1f_3,\chi_1^{f_1f_3}){\in \cA_1}$.  Here $\psi_j=m_je_j$ in the toric-code notation introduced below.

% Here $G_b=\Z_3$ and the extension by parityt splits. 
% The modular factors $\cL_i,\cR_i$ displayed below are preserved by the inherited $\Z_3$-action. The categories $D(\Z_2)_1\boxtimes\cL_i$ and $D(\Z_2)_1\boxtimes\cR_i$ are minimal modular extensions of $\cM_i^f=\sVec\boxtimes\cL_i$ and $\cN_i^f=\sVec\boxtimes\cR_i$, respectively. Restrict the inherited categorical action, including its coherence, to these invariant full subcategories. It is trivial on the first toric code and fixes $\psi_1$ with trivial parity-extension class. Since
% \be
% \label{eq:super-Z3-anomaly-vanishing}
% H^4(\Z_3,U(1))=0\,,
% \ee
% the remaining obstruction vanishes \cite[Theorem~4.9(2)]{galindo2025categorical}. With crossed-extension data chosen to match the untwisted parent, all five examples therefore admit the bulk
% \be
% \label{eq:super-verified-product-bulk}
% \fZ_{\cE_{\cA_1}}=\fZ(\Z_3)\boxtimes\fZ_f\,,
% \ee
% where $\fZ_f$ is the $(3+1)$-dimensional fermionic-charge toric code. The vanishing of $H^4$ establishes existence of the gauging, while its $H^3$ choice is fixed by matching the parent crossed extension.

Recall subgroup $H_1=\langle f_1, f_3, f_4\rangle$, the bosonic regular algebra
\begin{equation}
    \Fun(G/H_1) \cong \Fun(\Z_3) \cong 1 \oplus \rho_1 \oplus \rho_2
\end{equation}
is a subalgebra of every $\cA_i$ (for $i=3$ it is a proper subalgebra of the charge part $\Rep(G/H_3)$). Condensing this regular algebra gives
\begin{equation}
    \cT^f \cong D(H_1) = D(\Z_2)_1 \boxtimes D(\Z_2)_3 \boxtimes D(\Z_2)_4,
\end{equation}
where the subscript $j=1,3,4$ labels the generator $f_j$. An anyon of $\cT^f$ is of the form $\prod_{j=1,3,4}m_j^{x_j}e_j^{y_j}$, $x_j,y_j =0,1$. 

For $i=1,2,3$, gauging $\Rep(G/H_1)$ in the common generated fusion subcategory $\langle \cA_i\rangle$ and $\langle \cA_i\rangle'$ gives
\begin{equation}
    \begin{split}
        \cM^f_i & = \{1,\psi_1 \}'_{\cT^f} = \sVec \boxtimes D(\Z_2^2),\\
        \cN^f_i & = \sVec,
    \end{split}
\end{equation}
where $\{1,\psi_1 \}'_{\cT^f}$ denotes the centralizer of $\psi_1$ (and 1) in $\cT^f$. The SymTFTs for these three algebras have the same symmetry and physical boundaries and the same bulk.

For $\cA_i$, $i=4,5$, define two modular categories
\begin{equation}
\label{eqn:F4F5-3fermMTC}
    \cF_4:=\langle m_3 e_3 e_4, m_4 e_4 \rangle, \qquad \cF_5 := \langle m_3 e_3, m_4 e_3 e_4 \rangle.
\end{equation}
Each is the three-fermion MTC \cite{Teo:2015xla}. Every object of $\cF_4$ centralizes that of $\cF_5$ and we have the decomposition
\begin{equation}
    D(\Z_2)_3 \boxtimes D(\Z_2)_4 \cong \cF_4 \boxtimes \cF_5.
\end{equation}
The slightly degenerate categories associated with the symmetry and physical boundaries are
\begin{equation}
    \begin{split}
        \cM_4^f & = \sVec \boxtimes \cF_4, \quad \cN^f_{4}= \sVec \boxtimes \cF_5,\\
        \cM^f_5 & = \sVec \boxtimes \cF_5, \quad \cN_5 = \sVec \boxtimes \cF_4.
    \end{split}
\end{equation}

\paragraf{SymTFT Realizations.}
Let $\fB_f$ denote the bosonic Neumann boundary of $\fZ_f$, with boundary category $\sVec$. We stack the modular categories $\cL_i$ and $\cR_i$ defined below on this relative boundary and obtain
\be
\ba
\fB_i^{\sym} & = (\fB_{\Dir(G/H_1)} \boxtimes \fB_f \boxtimes \cL_i)/(G/H_1),\\
\fB_i^{\phys} & = (\fB_{\Dir(G/H_1)} \boxtimes \fB_f \boxtimes \cR_i)/(G/H_1),
\ea
\ee
where
\begin{align}
    \cL_i & = \begin{cases}
        D(\Z_2)_3 \boxtimes D(\Z_2)_4, & i = 1,2,3,\\
        \cF_i, & i=4,5,
    \end{cases}\\
    \cR_i & = \begin{cases}
        \Vect, & i =1,2,3,\\
        \cF_{i+(-1)^{i}}, & i=4,5.
    \end{cases}
\end{align}
We use the full $G/H_1$-crossed extension inherited from $D(\Z_2\times A_4)$. On the anyon labels of $\cT^f$, its generator $f_2^{-1}$ acts as
\begin{equation}
    \begin{split}
        m_3 \mapsto m_4, & \quad m_4 \mapsto m_3 m_4,\\
        e_3 \mapsto e_3 e_4, & \quad e_4 \mapsto e_3,
    \end{split}
\end{equation}
and acts trivially on the first toric code. The SymTFT sandwich is then
\begin{equation}
    \bigl( \fB^{\sym}_i \big| \fZ_{\cE_{\cA_1}} \big| \fB^{\phys}_i \bigr).
\end{equation}
In each sandwich, proliferate $\cA_i$ on the symmetry boundary gives
\begin{equation}
    \cT/\cA_i \cong \bigl( \fB^{\sym}_i/\cA_i \big| \fZ_{\cE_{\cA_1}} \big| \fB^{\phys}_i \bigr),
\end{equation}
where $\fB^{\sym}_i/\cA_i$ denotes the boundary after condensation, which can be computed in two steps: first condense the regular algebra $\Fun(G/H_1)$, then condense the algebra $\widehat \cA_i$ (which is $\cA_i$ regarded as a $\Fun(G/H_1)$-module) in $\cT^f$. We compute their anyon content and $\cT/\cA_i$ summarized in Table~\ref{tab:A_ihat}.

\begin{table}
    \centering
    \begin{tabular}{|c|c|c|}
\bottomrule
$i$ & generating anyons of $\widehat\cA_i$ & $\cT/\cA_i$\\
\hline
1 & $\langle m_4, \psi_1 \psi_3 \rangle$ & $D(\Z_2)$\\
2 & $\langle m_4 e_3, \psi_1 \psi_3 e_4 \rangle$ & $D(\Z_2)$\\
3 & $\langle e_3, \psi_1 \psi_4 \rangle$ & $D(\Z_2)$\\
4 & $\langle \psi_1 \psi_4 \rangle$ & $D(\Z_2^2)$\\
5 & $\langle \psi_1 \psi_4 e_3 \rangle$ & $D(\Z_2^2)$\\
\toprule
\end{tabular}
    \caption{Table of the anyons generating $\widehat \cA_i$, and $\cT/\cA_i$, for $i=1,\cdots, 5$.}
    \label{tab:A_ihat}
\end{table}

We identify the anyons in $D(\Z_2^3)$ with $V=\Z_2^6$ and write $q(x,y)=x\cdot y$ for their spins. 
The generators in Table~\ref{tab:A_ihat} span isotropic subgroups $L_i$ {(i.e., $q|_{L_i}=1$)} of dimension two for $i=1,2,3$ and dimension one for $i=4,5$. The condensed metric group is $L_i^\perp/L_i$ {\cite{WittGroup}}. Since $V$ is hyperbolic, this quotient is hyperbolic of dimension two and four, respectively. This proves the braided equivalences to $D(\Z_2)$ and $D(\Z_2^2)$. As independent checks, their dimensions are $4$ and $16$, with boson/fermion counts $(3,1)$ and $(10,6)$.

\paragraf{Implications for the Proliferation Transition.}
Every summand of the condensable algebra has trivial twist, and its local-module category $\cT^0_{\cA}$ is modular. Thus a super-Tannakian $\cE_{\cA}$ does not force a spin topological order {for the reduced TO}. The odd objects are transparent only within $\langle\cA\rangle$, not in the ambient modular category. 

There is a useful further distinction between simultaneous condensation and multi-step condensation, where we take a sequence of algebras and condense them one after another. First condense $\Fun(G/H)$, as in \eqref{eqn:two-step-condensation}, and let $\widetilde\cA=\cA(H,N,\gamma,\epsilon)$ denote the resulting algebra in $D(H,\omega|_H)$. Since $N\trianglelefteq H$, its generated flux subgroup is $K_{\widetilde\cA}=N$. For $r\in R_{\widetilde\cA}\subseteq N$, the defining action scalar and braided commutativity give
\be
\label{eq:second-stage-tannakian-proof}
B_{\widetilde\cA}(r,r)=\epsilon(r,r)
=\frac{\gamma(r,r)}{\gamma({}^{r}r,r)}=1\,.
\ee
Proposition~\ref{prop:Tannakian_criterion} therefore proves that the second-stage M\"uger center is Tannakian, even if the second-stage algebra is of type (ii).

The super-Tannakian center still changes the SymTFT and its boundary data. In the five examples above, $\psi_1$ is already nontransparent in the intermediate modular category $\cT^f=D(H_1)$ and remains a deconfined fermionic anyon of the final toric-code factors. This intermediate theory should not be confused with the fully condensed theory $\cT^0_{\cA}$.

\section{Landau-Ginzburg Theory and the Algebra Structure}
\label{sec:LG-algebra}

In the SymTFT formulation of the actual proliferation transition, we fix the 
 physical boundary, and drive the transition from the symmetry boundary \cite{Schafer-Nameki:2026hgx}. For type (i) algebras, we first retain scalar multiplets 
 for every nontrivial anyon in $\cA$ and use multiplicative structure on $\cA$ to specify their cubic interaction tensor. 
This is particularly insightful to consider, when {there are} twin condensations, which only differ in their multiplication rules. 
 We then examine reductions to fewer light fields, using the dihedral family as a test. The prescription fixes relative tensor components, while masses, overall strengths and other allowed interactions remain dynamical input.

\subsection{The LG-Theory on the Symmetry Boundary}
\label{sec:LG-type-i}

For type (i), let the generated subcategory be Tannakian for a finite group $G_\cA$ 
\be 
\langle\cA\rangle\cong\Rep(G_{\cA})  \,.
\ee 
A connected condensable algebra in 
this category is the algebra of complex functions on a finite transitive
$G_{\cA}$-set:
\be\label{eq:LG-coset-algebra}
\cA\cong\Fun(X),\qquad
X=G_{\cA}/H_{\cA},\qquad q=|X|=d_{\cA}\,.
\ee
Transitivity here means that the group action maps any point of 
$X$ to any other point. 
The stabilizer $H_{\cA}$ for each point is fixed up to conjugation.
The multiplication on $\cA$ is pointwise multiplication of functions. 
This description applies also to type (i) algebras containing magnetic or dyonic anyons: the equivalence identifies their generated symmetric category with charges of $G_{\cA}$, which need not be the original gauge group $G$ of $D(G, \omega)$. 

The scalar theory used in the construction is a  $2+1$d
LG theory with global $G_{\cA}$ symmetry. We denote the theory by
$\mathcal S_{\mathrm{LG}}(r)$ and its action by $S_{\mathrm{LG}}(r)$, where $r$ is a tunable parameter. It must have a symmetric, trivially gapped phase
and an ordered phase whose vacua form $X$. Thus the ordered phase has $q$
vacua before gauging, each invariant under $H_{\cA}$. 
We require that there are no SPTs as they will not be captured in this simple setup.
Below we give an explicit potential
realizing these vacua. 

This LG-theory is input to the SymTFT construction of the transition as follows:
For type (i), $\cM_{\cA}=\Vect$ on the symmetry boundary, whereas on the physical boundary we have stacked 
\be\label{eq:LG-spectator}
\cN_{\cA}=\langle\cA\rangle'/\Rep(G_{\cA})\,.
\ee
Here the quotient condenses the regular algebra of the transparent
$\Rep(G_{\cA})$ in the centralizer. We retain its symmetry action and
defect data inherited from $\cT$. Following
\cite{Schafer-Nameki:2026hgx}, put the scalar theory on the Dirichlet
symmetry boundary and gauge the diagonal $G_{\cA}$ 0-form symmetry:
\be\label{eq:LG-boundary-construction}
\ba
\fB^{\sym}(r)
&=\bigl(\fB_{\Dir}\boxtimes\mathcal S_{\mathrm{LG}}(r)\bigr)
  /G_{\cA}^{(0)},\\
\fB^{\phys}
&=\fB^{\cN_{\cA}}_{\Neu(G_{\cA})}
 =\bigl(\fB_{\Dir}\boxtimes\cN_{\cA}\bigr)/G_{\cA}^{(0)}\,.
\ea\ee
Only the symmetry boundary $\fB^{\sym}(r)$ in the first line depends on the LG theory, and on the parameter $r$, which we will discuss momentarily.
The endpoint boundary conditions follow directly. In the symmetric phase,
integrating out the scalar theory leaves
$\fB_{\Dir}/G_{\cA}^{(0)}=\fB_{\Neu(G_{\cA})}$. In the ordered phase,
write $\mathcal Q_X$ for the gapped theory with vacua labeled by $X$.
Gauging identifies the vacua related by $G_{\cA}$. The vacua after gauging form a $G_{\cA}$-set $G_{\cA}/H_{\cA}$. After choosing the
vacuum labeled by $H_{\cA}{\in G_{\cA}/H_{\cA}}$, the remaining gauge transformations on the vacuum are
exactly $H_{\cA}$. The resulting TO is $D(H_{\cA})$ only if $\cN_{\cA}=\Vect$. In general it is the ordered endpoint of the sandwich \eqref{eq:LG-sandwich} below, e.g. $D(\Z_3)$ for $\cA=1\oplus1^-$ in $D(S_3)$, for which $H_{\cA}=1$.
The two extremes are then 
\be\label{eq:LG-boundary-endpoints}
\fB^{\sym}_{\mathrm{symmetric}}=\fB_{\Neu(G_{\cA})}
\ee
which is the symmetry boundary for $\cT$ and 
\be\ba 
\fB^{\sym}_{\mathrm{ordered}}
&=\bigl(\fB_{\Dir}\boxtimes\mathcal Q_X\bigr)/G_{\cA}^{(0)}
 =\fB_{\Dir}/H_{\cA}^{(0)}\cr 
& =\fB_{\Neu(H_{\cA})}\,.
\ea\ee
This is the symmetry boundary for $\cT/\cA$.
The full interpolation is the SymTFT sandwich
\be\label{eq:LG-sandwich}
\mathcal P_{\cA}(r)=
\left(\fB^{\sym}(r)\,\middle|\,
\fZ(G_{\cA})\,\middle|\,\fB^{\phys}\right)\,.
\ee
With the two boundary conditions in \eqref{eq:LG-boundary-endpoints}, it
has starting point $\cT$ and endpoint $\cT/\cA$, respectively
\cite{Schafer-Nameki:2026hgx}. 
The number $q$ counts vacua of the scalar theory before gauging, which corresponds necessarily to the number of elementary scalar fields.

There is also a useful description entirely in $2+1$d. Shrinking
the interval between the two boundaries gives
\be\label{eq:LG-collapsed-theory}
\mathcal P_{\cA}(r)
=\bigl(\mathcal S_{\mathrm{LG}}(r)\boxtimes\cN_{\cA}\bigr)
 /G_{\cA}^{(0)}\,.
\ee
This is the interval-compactification identity used in
\cite{Schafer-Nameki:2026hgx}. 
The appearance of $\cN_{\cA}$ encodes the fixed physical boundary condition in the SymTFT. 
The two factors share a finite gauge field, as is explicit in the partition function on a closed Euclidean three-manifold $M$:
\be\label{eq:LG-finite-gauging}
Z_{\mathcal P_{\cA}}(M;r)
=\sum_{[a]}\frac{Z_{\cN_{\cA}}[M,a]}{|\operatorname{Aut}(a)|}
\int\!\mathcal D\varphi\,
e^{-S_{\mathrm{LG}}[\varphi;a,r]}\,.
\ee
The sum runs over isomorphism classes $[a]$ of flat principal
$G_{\cA}$-bundles. Here $\Aut(a)$ is the group of
bundle automorphisms preserving the flat connection.  For the trivial bundle, this factor is $1/|G_{\cA}|$.
The scalar fields are
sections of the associated representation bundle, and
$Z_{\cN_{\cA}}[M,a]$ includes the inherited symmetry defects and
topological weights. These data, rather than the modular category
$\cN_{\cA}$ alone, specify the gauging. The superscript star in LG$^*$
refers to this gauged theory. It changes the allowed local operators,
line operators and global sectors: a charged scalar is an ordinary local
order parameter in the input LG theory, but not a gauge-invariant local
operator of $\mathcal P_{\cA}$.

\paragraf{Purely Electric Algebra.}
For a purely electric algebra $\cA=\Fun(G/H)$ in $D(G,\omega)$, as in \eqref{eqn:G/CoreH},
\be 
G_{\cA}=G/\Core_G(H) \,, \qquad 
H_{\cA}=H/\Core_G(H) \,.
\ee
Equivalently, one can keep the possibly
non-faithful action of $G$ on $G/H$ and Higgs the original finite gauge theory. Choosing a coset vacuum restricts its flat connections to $H$ and restricts the topological action to $\omega|_H$. Its endpoints are $D(G,\omega)$ and $D(H,\omega|_H)$.

\subsection{From Algebras to LG-Theories}
\label{sec:LG-multiplication}

For type (i) algebras, the symmetric equivalence with $\Rep(G_{\cA})$ turns
each simple anyon $a$ in $\cA=\bigoplus_a n_a a$ into a
representation $V_a$. 
Introduce $n_a$ scalar multiplets in $V_a$,
with the reality condition inherited from $\Fun(X,\R)$, where
$X=G_{\cA}/H_{\cA}$. The full field space has $q=d_{\cA}$ real
components.  The constant field (identity component)
can be retained as an invariant scalar with a chosen positive mass.

We consider the real pairing $Q(f,g)=\sum_{a\in X}f(a)g(a)$
and  real basis vectors $e_I$, orthonormal with respect to $Q$, and normalization $e_0=\mathbf1/\sqrt q$. 
The multiplicative structure $m$ on $\cA$ gives an invariant,
completely symmetric cubic tensor 
\be\label{eqn:c_cubic_form}
\ba
C_{IJK}&=Q\bigl(m(e_I,e_J),e_K\bigr),\\
m(e_I,e_J)&=\sum_K C_{IJK}e_K,\qquad
C_{0IJ}=\frac{\delta_{IJ}}{\sqrt q}\,.
\ea\ee
Thus the tensor components are precisely the multiplication
coefficients in normalized coordinates. Expanding a field in this basis, 
\be 
\phi=\sum_{i=1}^{q-1}\phi_i e_i
\ee
we define 
\be\label{eq:LG-full-cubic}
c_{\cA}(\phi)=\sum_{i,j,k=1}^{q-1}C_{ijk}\phi_i\phi_j\phi_k\,.
\ee
The algebra then constrains the interaction terms of the LG-theory as follows 
\be\label{eq:LG-full-EFT}
\ba
\mathcal{L}= &\frac12(\partial_\mu\phi_i)^2
+\frac12\phi_i(M^2)_{ij}\phi_j\\
&\ -\frac w3c_{\cA}(\phi)+V_4(\phi)+\cdots\,.
\ea\ee
The algebra fixes the multiplets and the tensor $c_\cA$, whereas the dynamics fixes $w, M$ and $V_4$. 

This prescription has a simple realization of the   Higgs phase. In coset coordinates, $\sum_{a\in X}\phi_a=\sqrt{q}Q(e_0, \phi)=0$ and
$c_{\cA}=\sum_{a\in X}\phi_a^3$. For $q>2$, the choice
$M^2=t\mathbf1$, $V_4=uR^4/4$, $u,w>0$, with
$R^2=Q(\phi,\phi)$, selects the $q$ directions
$e_a-\mathbf1/q$.
Their stabilizers are conjugate to $H_{\cA}$. 

\subsection{Proliferation Transitions for Twin Algebras}

An interesting instance of proliferation transitions are those that are based on twin algebras \cite{Gai:2026hjk}, which have the same anyon content, but a different multiplication. In the above identification this means that the field content of type (i) twin algebras will be the same, but the cubic interaction $c_\cA$ is different. We will consider twins whose reduced TOs are the same -- this is not always the case -- but we will show that the resulting transitions are inequivalent.

\paragraf{Example: Proliferation Transition for {$G_{32}$} Twins.}
We now apply this to the twins of \cite{Gai:2026hjk} 
with 
\be \ba 
G &=G_{32}= \Z_8^*\ltimes\Z_8 \cr 
& = \{A, C, T| T^8=A^2=C^2=1\,, \ AC=CA\,, \cr 
& \qquad\quad  ATA=T^3 \,, \ CTC=T^{-1} \}\,.
\ea\ee  
We consider the algebras
\be 
\cA_i = \cA ( H_i, 1, 1, 1)  \,,
\ee 
with the subgroups 
\be \ba 
H_1&=\langle A,C\rangle\cr 
H_2&=\langle T^4A,C\rangle \,.
\ea\ee 
Both groups are isomorphic to $\Z_2\times\Z_2$, but crucially are not conjugate to each other. This means that the associated algebras are inequivalent. 
The anyon contents in the algebras $\cA_i$ are the same 
\be 
\cA_i|_{\text{anyons}}= 1\oplus R_1\oplus R_2\oplus R_4\,, 
\ee 
in terms of irreps $R_d$ of dim $d$, however the product structure is different, i.e. they are twins.  
For the transition, we couple the anyons to 7 real scalars: $\sigma$ in $R_1$, a complex
scalar $z_2$ in $R_2$, and two complex scalars $z_1,z_3$ together
in $R_4$. 
The group acts as follows on these fields: 
\be\label{eq:LG-G32-full-action}
\begin{array}{|c||c|c|c|}\hline 
 &\sigma&z_2&(z_1,z_3)\cr  \hline\hline 
T&-\sigma&-iz_2&(e^{-\pi i/4}z_1,e^{-3\pi i/4}z_3)\cr 
A&\sigma&\bar z_2&(z_3,z_1)\cr 
C&\sigma&\bar z_2&(\bar z_1,\bar z_3) \cr \hline 
\end{array}
\ee
There are three independent invariant masses,
\be\label{eq:LG-G32-full-masses}
V_2=\frac{r_1}2\sigma^2+\frac{r_2}2|z_2|^2
+\frac{r_4}2\bigl(|z_1|^2+|z_3|^2\bigr)\,,
\ee
and exactly four independent cubic operators:
\be\label{eq:LG-G32-full-cubic-basis}
\ba
K_0&=\sigma\operatorname{Re}(z_2^2),\\
K_1&=\operatorname{Re}\bigl[\bar z_2(z_1^2+\bar z_3^{\,2})\bigr],\\
K_2&=\operatorname{Re}(\bar z_2\bar z_1z_3),\\
K_3&=\sigma\operatorname{Re}(z_1z_3)\,.
\ea\ee
Each is invariant under \eqref{eq:LG-G32-full-action}. The invariant
symmetric cubic space has dimension four, so the general cubic
potential is $V_3=\sum_{\alpha=0}^3g_\alpha K_\alpha$.
In the orthonormal coset normalization, the multiplicative structure instead give
\be\label{eq:LG-G32-full-cubics}
c_\pm=\frac34\bigl(\sqrt2K_0+K_1
\pm2K_2\pm2\sqrt2K_3\bigr)\,.
\ee
The signs $+$ and $-$ correspond to $H_1$ and $H_2$. The
cubic couplings are then 
\be 
V_3^\pm=-{wc_\pm\over 3} \,,
\ee 
therefore fixes
\be\label{eq:LG-G32-full-couplings}
(g_0,g_1,g_2,g_3)_\pm
=-\frac w4\bigl(\sqrt2,1,\pm2,\pm2\sqrt2\bigr)\,.
\ee
In particular, $g_2/g_1=\pm2$ and $g_3/g_0=\pm2$.
These relative signs cannot be removed while preserving the group
action.  The cubic tensors are
inequivalent even though the matter multiplets agree. 

\paragraf{A Direct EFT Consequence.}
For the cubic couplings in \eqref{eq:LG-G32-full-couplings},
suppose $\sigma$ and $z_2$ have positive squared masses
$M_\sigma^2$ and $M_2^2$, while $z_1,z_3$ remain light.
Integrating out the massive fields at tree level gives a quartic potential -- that depends on the specific twin algebra:
\be\label{eq:LG-G32-threshold}
\ba
\Delta V_4^\pm={}&-\frac{w^2}{32M_2^2}
\left|z_1^2+\bar z_3^{\,2}\pm2\bar z_1z_3\right|^2\\
&-\frac{w^2}{4M_\sigma^2}
\bigl[\operatorname{Re}(z_1z_3)\bigr]^2\,.
\ea\ee
 The cross term between the two
$z_2$ exchange vertices contributes
\be\label{eq:LG-G32-interference}
\ba
J&=\operatorname{Re}(z_1^3\bar z_3+z_1\bar z_3^{\,3}),\\
\Delta\lambda_J^\pm&=\mp\frac{w^2}{8M_2^2}\,.
\ea\ee
For independent cubic coefficients $g_1,g_2$, the same contribution
is $-g_1g_2/M_2^2$. 
The multiplication structure in the algebra therefore has direct consequences 
and fixes the relative couplings and
hence this threshold correction. 

The field redefinition $z_3\mapsto-z_3$ exchanges $c_\pm$
and maps $J\mapsto-J$. 
It intertwines the group action
\eqref{eq:LG-G32-full-action} with an outer automorphism $\phi$ of $G_{32}$: 
\be 
\phi(T)=T\,,\ \phi(A)=T^4A\,,\ \phi(C)=C\,.
\ee
Since $\phi(H_1)=H_2$, the induced braided autoequivalence of $D(G_{32})$ exchanges the twin algebras while fixing each of their
simple summands. 

The two transitions are mapped into each other with the automorphism and field redefinitions. However, by considering them in a single parent theory, with a fixed reference $G_{32}$ action, these comprise distinct phase transitions.

In this example the reduced TO for the twins was the same. There are examples where this is not the case \cite{Gai:2026hjk}, and it is even more obvious that the twin condensations are inequivalent. 

\subsection{Proliferations for $D(D_{2N})$}
\label{sec:LG-coset-vacua}

In many instances there are field redundancies that can be reduced. To exemplify this, we consider the quantum doubles of  $D_{2N}=\langle r,s\mid r^{2N}=s^2=1,\ srs=r^{-1}\rangle$, i.e.  the dihedral group of  order $4N$, with $N\geq2$. Consider the subgroup $H=\langle s\rangle$.
The electric algebra is
\be\label{eq:LG-dihedral-algebras}
\cA=\Fun(D_{2N}/H)
\cong1\oplus(1,\chi_+)\oplus\bigoplus_{k=1}^{N-1}(1,E_k)\,.
\ee
Here $E_k$ is the rotation-frequency-$k$ doublet and
$\chi_+(r)=-1$, $\chi_+(s)=1$. The full nontrivial scalar space
has $2N-1$ real components. A reduced description can use just combination $z=x+iy$
\be\label{eq:LG-dihedral-action}
r:z\longmapsto e^{\pi i/N}z,\qquad s:z\longmapsto\bar z\,.
\ee
On its $2N$-point ordered orbit, the operators
$z,z^2,\ldots,z^{N-1},\operatorname{Re}z^N$, together with the
constant, span all the charge channels, as $\chi_+$ is in the fusion outcome. This establishes a possible
two-field realization. Obtaining it as the low-energy limit of a
particular full theory requires the omitted fields to remain massive
and their effects to be matched onto higher interactions. It does
not establish the critical spectrum of every realization.

Writing $\rho=|z|^2$ and $\tau=\operatorname{Re}z^{2N}$, every
invariant potential term below degree $2N$ is radial. Invariance of
$z^p\bar z^q$ requires $p-q$ to be divisible by $2N$, hence $p=q$
below that degree. Reflection selects the real combination at degree
$2N$. Therefore
\be\label{eq:LG-dihedral-general-potential}
V=\frac t2\rho+\sum_{k=2}^{N}u_k\rho^k-h\tau
+O(|z|^{2N+2})\,.
\ee
An explicit stable reduced realization is
\be\label{eq:LG-dihedral-matched-potential}
V=\frac t2\rho+\frac u4\rho^2
+\kappa(\operatorname{Im}z^N)^2,\qquad u,\kappa>0\,.
\ee
For $t\geq0$ the unique minimum is the origin. For $t<0$, the last
term vanishes at precisely the required $2N$ directions and the
radial minimum has $v^2=-t/u$. The Higgs representatives are
$z=v e^{\pi i j/N}$ with stabilizers conjugate to $H$.
The transition is precisely from $D(D_{2N})\to D(\Z_2)$.

\paragraf{$D_4$ LG* Parametrization.}
\label{sec:LG-D4-generators}
Denote by $I_k=\sum_{a\in X}\phi_a^k$ the $k$'th power of the coset fields, so that $I_3=c_{\cA}$. 
For $N=2$ the reduced theory has
\be\label{eq:LG-D4-potential}
\ba
V&=\frac t2\rho+\frac u4\rho^2+g x^2y^2,\\
I_4&=\frac12(x^4+y^4)\,.
\ea\ee
For $u>0$, $u+g>0$ and $t<0$, $g>0$ favors the axes and
$g<0$ the diagonals. The axis condensate contains the composite
charge $x^2-y^2$. The symmetry alone is consistent with the quartic invariants. 

\paragraf{$D_8$ LG* Parametrization.}
\label{sec:LG-D8-generators}
For $N=4$, $D_8$ has order sixteen. The two light fields generate
the other channels through $z^2,z^3,\operatorname{Re}z^4$, but
\be\label{eq:LG-D8-tensors}
I_4=\frac3{16}\rho^2,\qquad I_6=\frac5{128}\rho^3,\qquad
I_8=\frac{35\rho^4+\operatorname{Re}z^8}{4096}\,.
\ee
The quartic and sextic truncations are accidentally $O(2)$ invariant.
They cannot select the octagon of Higgs directions. Pinning first
appears at degree eight. Thus a cubic multiplication vertex in the
full theory need not remain a cubic, or even an anisotropic quartic,
after reduction to the light fields. Approximate continuous symmetry
and relevance of higher operators at criticality require a separate
RG analysis \cite{LouSandvikBalents:2007}.

\subsection{Discussion of Type  (ii) and (iii) Algebras}
\label{sec:LG-general-algebras}

The multiplicative structure is also relevant in the case of type (ii) and (iii) algebras. It is useful to formulate this a bit more concretely before discussing these two types.
{Let the algebra have anyon decomposition} $\cA=\bigoplus_a n_a a$. We choose embeddings
$\iota_{a\mu}:a\to\cA$, and choose fusion bases
$f_{ab}^{c,\lambda}:a\otimes b\to c$. Then
\be\label{eq:LG-fusion-vertex}
m\circ(\iota_{a\mu}\otimes\iota_{b\nu})
=\sum_{c,\rho,\lambda}
M_{a\mu,b\nu}^{c\rho,\lambda}
\,\iota_{c\rho}\circ f_{ab}^{c,\lambda}\,.
\ee
The labels $\mu,\nu,\rho$ distinguish copies of simple anyons and
$\lambda$ distinguishes fusion channels. The coefficients satisfy
associativity and braided commutativity, as well as the unit and
Frobenius conditions. They specify how condensate lines fuse, beyond
the information in the multiplicities $n_a$
\cite{Kong:2014condensation,Gai:2026hjk}.

With the map $\ell:\cA\to1$, which is the Frobenius counit, the corresponding trilinear morphism is 
\be\label{eq:LG-categorical-cubic}
c_{\cA}=\ell\circ m\circ(m\otimes\operatorname{id}_{\cA})
 :\cA^{\otimes3}\longrightarrow1\,.
\ee
In general
this is a fusion vertex for three anyons. 
Braided commutativity relates exchanges using the category's braiding. It does not identify  them with ordinary exchanges of commuting scalar components. Thus
\eqref{eq:LG-categorical-cubic} alone is not immediately a LG-type interaction. 
We will now discuss the remaining two algebra types in turn in light of these general considerations. 

\paragraf{Type (ii) Algebras.}
For type (ii) with bosonic M\"uger center
$\cE_{\cA}=\Rep(G_{\cA})$, the transparent subalgebra is
$\cB=\cA\cap\cE_{\cA}=\Fun(G_{\cA}/H_{\cA})$.
Condensing $\cB$ can be realized as an LG*-type transition  but does not in general complete the condensation of $\cA$
\cite{Schafer-Nameki:2026hgx}. 

For a super-Tannakian center, set $G_b=G_{\cA}/\langle z_{\cA}\rangle$
and $\cE_{\cA}^b=\Rep(G_b)$ as in
section~\ref{sec:MANA_super_Tannaka}. Since every simple summand of
$\cA$ has trivial twist, its transparent subalgebra is entirely bosonic:
\be\label{eq:LG-super-transparent-subalgebra}
\cB=\cA\cap\cE_{\cA}
=\cA\cap\cE_{\cA}^b
\cong\Fun(G_b/H_b)\,.
\ee
Bosonic de-equivariantization sends $\cA$ to $[G_b:H_b]$ connected
components in $\cM_{\cA}^f\subseteq\cT^f$, permuted transitively by
$G_b$. We choose the identity-coset component $\widehat\cA$ which has an 
$H_b$-equivariant algebra structure and has dimension 
$d_{\cA}/[G_b:H_b]$. we can then identify the TOs after condensation as follows
\be\label{eq:LG-super-endpoints}
\ba
\cT/\cB&\cong\cT^f/H_b^{(0)},\\
\cT/\cA&\cong(\cT^f/\widehat\cA)/H_b^{(0)}\,.
\ea\ee
Here $\cT^f$ is the full modular category of
section~\ref{sec:MANA_super_Tannaka}, and $\cT^f/\widehat\cA$ denotes its category of local modules.
The first step can be realized by an ordinary $G_b/H_b$ scalar order parameter. 
Again as in the bosonic case there is a remaining step for type (ii) algebras, that requires additional dynamical information.

For type (iii), $\langle\cA\rangle$ is already modular and M\"uger's theorem \cite{Muger:2003structure} implies the factorization 
\be\label{eq:LG-type-iii-factorization}
\ba
\cT&\cong\langle\cA\rangle\boxtimes\langle\cA\rangle',\\
\cT/\cA&\cong(\langle\cA\rangle/\cA)
 \boxtimes\langle\cA\rangle'\,.
\ea\ee
The second factor remains fixed. There is no nontrivial ordinary
symmetry supplied by the M\"uger center, so the first factor
requires its own dynamical description, such as a suitable
Chern-Simons-matter theory. The multiplication supplies the fusion vertices that such
a description must implement, but does not remove the need to
construct that description \cite{Schafer-Nameki:2026hgx}.

For twins of types (ii) and (iii), the identical anyon object still
has different multiplication tensors. 
In the data
$\cA(H,N,\gamma,\epsilon)$, the flux multiplication phases
$\gamma$ and the compatible charge data $\epsilon$ enter these
vertices \cite{Gai:2026hjk}. They can distinguish continuum
interactions once an appropriate realization has been supplied. 
Examples of twins with type (ii) and (iii) algebras exist. 
For example, the Lagrangian twins $\cA(G_{64}, 1)$, $\cA(G_{64}, \gamma_{\operatorname{Bog}})$ in $D(G_{64})$, 
where $G_{64}$ is a group of order 64 and $\gamma_{\operatorname{Bog}}$ is a Bogomolov multiplier (see \cite{Davydov:2013xov, Kobayashi:2025ykb, Gai:2026hjk} for details) are of type (ii). 
Finally, Lagrangian twins $\cA(S_3^{(1)}, 1)$ and $\cA(S_3^{(2)},1)$ in $D(GL(2,3))$, where $S_3^{(1)}$ and $S_3^{(2)}$ are two non-conjugate $S_3$ subgroups of $GL(2,3)$ (see \cite{Gai:2026hjk} for details) are of type (iii). It would be interesting so see how the distinct algebra structure enters in the proliferation transitions for these algebras. 

\subsection*{Acknowledgments}

We are grateful to Andrea Antinucci, Thomas Bartsch, Alison Warman, Rui Wen and Yunqin Zheng for discussions, and Andrea Antinucci  and Rui Wen for comments on the draft. 
We have used GAP \cite{GAP4} to perform various computations. Claude Opus 5.5 and ChatGPT Astra were used in some of the exploration, code writing for computing the fusion subcategory data, and sharpening of theorems, as well as text editing. We take full responsibility. 
This work is supported by the UKRI Frontier Research Grant, underwriting the ERC Advanced Grant ``Generalized Symmetries in Quantum Field Theory and Quantum Gravity''.

\appendix
 \renewcommand{\theHsection}{legacy.\Alph{section}}

\section{Preliminary: Symmetric Categories and (De-)Equivariantization}

\subsection{Symmetric Categories and Deligne's Theorem}
\label{append:DelignesTheorem}
This appendix collects the preliminaries used in the proof of Theorem~\ref{thm:Tannakian_group}. We assume familiarity with the basics of braided fusion categories. See \cite{Deligne:2002cat, DGNO, EGNO} for more. 

A braided fusion category $\cC$ is called \textbf{symmetric} if its centralizer~\eqref{eqn:defn_centralizer} is itself, $\cC'=\cC$. 

Let $G$ be a finite group and $z\in G$ a central element such that $z^2=1$. One defines a braiding $\sigma'$ on $\Rep(G)$ by
\begin{equation}
    \sigma_{X,Y}(x\otimes y) = (-1)^{mn} y \otimes x
\end{equation}
for $x\in X$, $y\in Y$ such that $zx=(-1)^m x$, $zy=(-1)^n y$. The fusion category $\Rep(G)$ equipped with this braiding is denoted by $\Rep(G, z)$.

In terms of the above notation, the category of super-vector spaces is $\sVec=\Rep(\Z_2, z)$, where $z$ is the non-trivial element in $\Z_2$.

For a symmetric fusion category $\cC$, a \textbf{super fiber functor} is a braided tensor functor
\begin{equation}
    F: \cC \rightarrow \sVec.
\end{equation}
For a super fiber functor, let $G_{F}:=\text{Aut}^{\otimes}(F)$. Let $z_{F}\in G_{F}$ denote the parity automorphism of $F$ (i.e., for each $X\in \cC$, $(z_{F})_{X}$ is the usual parity automorphism of $F(X)\in \sVec$). $z_{F}$ is central and $z_{F}^2=1$. $F$ has a lifting $\widetilde F: \cC \rightarrow \Rep(G_F, z_F)$.
\begin{theorem}[\cite{Deligne:2002cat}]
\label{thm:Deligne}
    Let $\cC$ be a symmetric fusion category. Then
    \begin{enumerate}
        \item there exists super fiber functors $\cC \rightarrow \sVec$;
        \item all super fiber functors $\cC \rightarrow \sVec$ are isomorphic to each other;
        \item for any super fiber functor $F:\cC \rightarrow \sVec$, the functor $\widetilde F: \cC \rightarrow \Rep(G_F, z_{F})$ constructed above is an equivalence.
    \end{enumerate}
\end{theorem}

\subsection{Equivariantization and De-equivariantization}

We follow \cite{DGNO} in order to review equivariantization and de-equivariantization, i.e., the mathematical structure of gauging in order to construct $\cM_{\cA}$ and $\cN_{\cA}$ stacked on the symmetry and physical boundaries, respectively. 
We start with the entirely bosonic case, then review the fermionic case in Appendix~\ref{append:fermion_de-equiv}.
Let $\cC$ be a fusion category. We denote by $\underline{\Aut^{\otimes}(\cC)}$ the monoidal category whose objects are tensor autoequivalences of $\cC$, arrows are tensor natural transformations and monoidal structure is composition of functors.

Let $G$ be a finite group, $\underline{G}$ is the discrete monoidal category whose objects are elements in $G$.
An \textbf{action} of $G$ on $\cC$ is a monoidal functor $F: \underline{G} \rightarrow \underline{\Aut^{\otimes}(\cC)}$.

Given a categorical $G$ action on $\cC$ $F: \underline{G} \rightarrow \underline{\Aut^{\otimes}(\cC)}$, for any $g,h\in G$, let $F_g$ be the image under $F$ and $\gamma_{g,h}$ be the isomorphism $F_{g} \circ F_{h} \cong F_{gh}$. The \textbf{$G$-equivariantization} $\cC^G$ (also denoted by $\cC/G$ in the main text) is the fusion category whose objects are pairs $(X, u)$, where $X$ is an object of $\cC$ and for each $g\in G$, $u_g:F_g(X) \cong X$ is an isomorphism such that
\begin{equation}
    \begin{tikzcd}
        F_g(F_h(X)) \arrow{r}{F_g(u_h)} \arrow[swap]{d}{\gamma_{g,h}(X)} & F_g(X) \arrow{d}{u_g}\\
        F_{gh}(X) \arrow{r}{u_{gh}} & X
    \end{tikzcd}
\end{equation}
commutes for all $g,h\in G$. Morphisms in $\cC^G$ are morphisms in $\cC$ commuting with $u_g$. Explicitly, a morphism $(X, u) \rightarrow (Y, v)$ is a morphism $f\in \cC(X,Y)$ such that $v_g \circ F_{g}(f) = f \circ u_g$.

There is an inverse to equivariantization.
Let $\cD$ be a fusion category over $\Rep(G)$. Denote by $A$ the regular algebra in $\Rep(G)$, the tensor functor $\Rep(G) \rightarrow D(\cD)$ makes sense of $A$-modules in $\cD$.
The \textbf{de-equivariantization} $\cD_{G}$ (also denoted by $\cD/\Rep(G)$ in the main text) is the tensor category of $A$-modules in $\cD$. 
$\cD_{G}$ naturally carries a $G$-action (the 0-form action in the main text) from the right translation action of $G$ on the regular algebra $A$.

If $\cD$ is a braided fusion category over $\Rep(G)$, its de-equivariantization naturally carries a $G$-crossed braided structure.

By Theorem 4.18 and Proposition 4.19 of \cite{DGNO}, equivariantization and de-equivariantization are inverse in the sense that they define a 2-equivalence between the 2-category of fusion categories with $G$-actions and fusion categories over $\Rep(G)$.

Denoting $G$-equivariantization by $\operatorname{eq}_G$ and $\Rep(G)$-de-equivariantization by $\operatorname{de-eq}_G$, for $N\trianglelefteq G$, the following commute
\begin{equation}
    \begin{tikzcd}[column sep=4em]
        \cC \arrow[yshift=0.5ex]{r}{\text{eq}_N} \arrow[yshift=1.8ex, bend left=20]{rr}{\text{eq}_{G}} & \cC^N \arrow[yshift=0.5ex]{r}{\text{eq}_{G/N}} \arrow[yshift=-0.5ex]{l}{\text{de-eq}_{N}} & \cC^G \arrow[yshift=-0.5ex]{l}{\text{de-eq}_{G/N}} \arrow[yshift=-1.8ex, bend left=20]{ll}{\text{de-eq}_{G}}
    \end{tikzcd}.
\end{equation}

\subsection{Fermionic (De-)Equivariantization}
\label{append:fermion_de-equiv}
We review the fermionic case following \cite{galindo2025categorical}.
A fermionic fusion category is a fusion category $\cC$ with a specified central fermion $(f,\sigma)\in \cZ(\cC)$, where $f\otimes f\cong1$ and $\sigma_{f,f}=-\id$. Equivalently, there is an embedding $\sVec \rightarrow \cZ(\cC)$ such that $\sVec \rightarrow \cZ(\cC) \rightarrow \cC$ is fully faithful.

Let $(\Gamma,z)$, where $\Gamma$ is a finite group and $z\neq1$, and let $G=\Gamma/\langle z\rangle$. $\Gamma$ is a central extension of $G$ and determines $\alpha\in H^2(G,\Z_2)$. Fix a normalized representative for $\alpha$. A fermionic $(G,\alpha)$-action consists of a categorical $G$-action $F_g$, with $\mu_{g,h}:F_gF_h\to F_{gh}$, preserving the central fermion, i.e., there is $\eta_g:F_g(f)\to f$ such that
\be
\label{eqn:append:phi_sVec}
\eta_g\circ F_g(\eta_h)
=(-1)^{\alpha(g,h)}\eta_{gh}\circ\mu_{g,h,f}\,.
\ee

For an equivariant tensor functor $T:\cC\to\cC'$ with $\kappa_g:TF_g\to F'_gT$ and fermion identification $\tau:T(f)\to f'$ satisfying
\be
\label{eq:fermionic-equivariant-functor-corrected}
\tau\circ T(\eta_g)
=\eta'_g\circ F'_g(\tau)\circ\kappa_{g,f}\,.
\ee
The following is proven in \cite{galindo2025categorical}.
\begin{theorem}[\cite{galindo2025categorical}, Theorem~3.16]
\label{append:thm:fermioniceq}
For $z\neq1$, equivariantization by $G$ and de-equivariantization by $\Rep(G)$ give a biequivalence between fermionic fusion categories with fermionic $(\Gamma,z)$-action and fusion categories over $\Rep(\Gamma,z)$.
\end{theorem}

\section{Categorical Data of $D(G, \omega)$}
\label{sec:math}

In this appendix we collect the relevant mathematical structures that are used in the main text.  
We start with a condensable algebra $\cA=\cA(H, N, \gamma, \epsilon)$ in Drinfeld center $D(G, \omega)$, in section~\ref{sec:GeneratedSubcat}, we compute its generated fusion subcategory $\langle \cA \rangle = S(K_{\cA}, L_{\cA}, B_{\cA})$ of $D(G, \omega)$. In section~\ref{sec:centralizer-muger}, we compute the M\"uger center $\cE_{\cA}=\cZ_2(\langle\cA \rangle) = S(R_{\cA}, P_{\cA}, \psi_{\cA})$ and determine its (super-)Tannakian structure. We explain the notations in the main text.
A reader interested in the physical statements may go directly to the summary in section~\ref{sec:math_summary}.

\subsection{Conventions: Cocycles and Transgressions}
Throughout, $G$ is a finite group and $\omega\in Z^3(G, U(1))$ is a normalized 3-cocycle.
In what follows, for $c\in C^2(G, U(1))$,
\be\label{eq:audit-cochain-convention}
(dc)(x,y,z)=\frac{c(y,z)c(x,yz)}{c(xy,z)c(x,y)}\,.
\ee
In $\Vect_G^\omega$, the associator
$U\otimes(V\otimes W)\to(U\otimes V)\otimes W$ acts on $U_x, V_y, W_z$ of degrees
$x,y,z$ by $\omega(x,y,z)$, as in~\cite[Eq.~(4)]{davydov2017lagrangian}. A fiber-functor $F$ has monoidal structure $\nu: F(U)\otimes F(V)\to F(U\otimes V)$ satisfying $d\nu=\omega^{-1}$. For $g,h\in G$, write ${}^gh :=ghg^{-1}$ and $h^g:=g^{-1}hg$. Following \cite{NNW:2009}, we set
\begin{equation}
    \beta_{a}(x,y) :=\frac{\omega(a,x,y) \omega(x,y,a^{xy})}{\omega(x, a^x, y)},
\end{equation}
and the transgression
\begin{equation}
\label{eqn:transgression}
    \tau_a(\omega)(x,y):=\frac{\omega(x,y,a)\omega({}^{xy}a, x, y)}{\omega(x, {}^y a, y)}.
\end{equation}
$\beta_a$ and $\tau_a(\omega)$ agree on the centralizer \cite{NNW:2009}:
\begin{equation}
    \beta_a|_{C_G(a)} = \tau_a(\omega)|_{C_G(a)},
\end{equation}
where $C_G(a)$ is the centralizer of $a$ in $G$.
From \cite[equation (20)]{NNW:2009}, the following identity is true:
\begin{equation}
\label{eqn:NNW_gamma_beta}
    \frac{\beta_{ab}(x,y)}{\beta_{b}(x^a, y^a) \beta_a(x,y)} = \frac{\beta_x(a,b) \beta_{y}(a,b)}{\beta_{xy}(a,b)}
\end{equation}
for $a,b\in C_G(x)\cap C_{G}(y)$. $\beta_a$ satisfies (twisted) cocycle condition \cite{NNW:2009}.
Explicitly, for all $a,x,y,z\in G$,
\be
\label{beta-twisted-cocycle}
\beta_a(x,y)\,\beta_a(xy,z)=\beta_{a^x}(y,z)\,\beta_a(x,yz) \,,
\ee
and the transgression satisfies
\be
\label{tau-twisted-cocycle}
\tau_a(\omega)(y,z)\,\tau_a(\omega)(x,yz)=\tau_{{}^za}(\omega)(x,y)\,\tau_a(\omega)(xy,z) \,,
\ee
which is the condition for linear maps $\rho(g):V_x\to V_{{}^gx}$ with $\rho(g)\rho(h)|_{V_x}=\tau_x(\omega)(g,h)\,\rho(gh)|_{V_x}$ to have associative compositions.

\subsection{Condensable Algebras in \texorpdfstring{$D(G, \omega)$}{D(G, \omega)}}
\label{sec:cond_alg}

We assume some familiarity with (group-theoretical) fusion categories, see e.g., \cite{Ostrikmodule, etingof2005fusion, EGNO} for background.

For any finite group $G$ and 3-cocycle $\omega\in H^3(G, U(1))$, denote by $D(G,\omega)$ the Drinfeld center of the pointed fusion category $\Vect_G^\omega$. 
We fix a set $R$ of representatives of conjugacy classes of $G$. 
Simple objects in $D(G,\omega)$ are the anyons in $D(G, \omega)$. This has a lattice realization in terms of  Kitaev quantum double \cite{Kitaev:1997wr}. In other words, $D(G, \omega)$ is the modular tensor category (MTC) capturing the topological order (TO) described by Kitaev's quantum double model $D(G, \omega)$.
More precisely, $D(G,\omega)\cong\Rep(D^{\omega}(G))$ is the MTC of the twisted quantum double. For $\omega=1$ it describes the TO of Kitaev's quantum double model \cite{Kitaev:1997wr}, and for $\omega\neq1$ the TO of the twisted quantum double model \cite{Hu:2012wx}. In both cases it is the deconfined phase of a 2+1d gauge theory with gauge group $G$: the anyon $(1,\irrep)$ is the Wilson line in the representation $\irrep$ (a charge), $(x,1)$ is a vortex with holonomy in the conjugacy class of $x$ (a flux), and a general anyon $(x,\irrep)$ is a dyon, a flux $x$ that carries a (projective) representation of the gauge transformations $C_G(x)$ that leave it invariant.

Each anyon in $D(G, \omega)$ is labeled by $(x, \irrep)$, where $x\in R$ and $\irrep$ is the character of an irreducible $\tau_x(\omega)$-projective representation of the centralizer $C_G(x)$ of $x$.
% , with projective 2-cocycle $\tau_x(\omega)$ the transgression of $\omega$
% \begin{equation}
%     \tau_x(\omega)(g,h):=\frac{\omega(g,h,x)\omega(^{gh}x, g, h)}{\omega(g, {}^hx, h)},
% \end{equation}
% where for any two group elements $g,h\in G$, we introduce notation ${}^gh:=ghg^{-1}$, $h^g:=g^{-1}hg$.

A collection of anyons which can proliferate in a TO carries an algebra structure \cite{Kong:2014condensation}. Mathematically, they form \textbf{condensable algebras} in the corresponding modular tensor category \cite{Frohlich:2003hm, Kong:2014condensation}. See the cited references for a precise definition.
Physically, a condensable algebra is a collection of bosons $\cA=\bigoplus_a n_a\, a$ that condense simultaneously \cite{Bais:2008ni, Eliens:2013epa, Burnell:2017otf}. The multiplication $m:\cA\otimes\cA\to\cA$ specifies how the condensed anyons fuse into each other inside the condensate. Connectedness, $\dim\Hom(1,\cA)=1$, gives a unique vacuum in the condensate. 

Condensable algebras in $D(G, \omega)$ have the following classification \cite{davydov2017lagrangian, Gai:2026hjk}.

\begin{theorem}
    A condensable algebra in $D(G, \omega)$ is labeled by $\cA(H,N,\gamma, \epsilon)$, where $N \trianglelefteq H \subseteq G$, $\gamma \in C^2(N, U(1))$ is such that $d\gamma = \omega|_{N}$ and $\epsilon:H \times N \rightarrow U(1)$ satisfies
    \begin{equation}
    \label{eqn:epsilon_compatibility}
        \begin{split}
            \epsilon(h_1h_2,\,n) &= \tau_n(\omega)(h_1,h_2)^{-1} \epsilon\big(h_1,\,{}^{h_2}n\big)\,\epsilon(h_2,n)\,,\\
\epsilon(h,\,n_1n_2) &= \Omega_h(n_1,n_2)\epsilon(h,n_1)\,\epsilon(h,n_2)\,
\frac{\gamma\big({}^hn_1,{}^hn_2\big)}{\gamma(n_1,n_2)}\,,\\
\epsilon(n_1,n_2) &= \frac{\gamma(n_1,n_2)}{\gamma\big({}^{n_1}n_2,\,n_1\big)}\,.
        \end{split}
    \end{equation}
        Two condensable algebras are isomorphic if and only if their classification data differ by conjugation or/and a coboundary.
    That is, $\cA(H, N, \gamma, \epsilon)$ and $\cA(H', N', \gamma', \epsilon')$ are isomorphic if and only if there exists $c:N\rightarrow U(1)$ and $g\in G$ such that
    \begin{equation}
        \ba
        H'& ={}^gH, & 
        \gamma'({}^gn_1, {}^g n_2) & = \frac{c(n_1 n_2)}{c(n_1) c(n_2)}\frac{\gamma(n_1, n_2)}{\Omega_g(n_1, n_2)} \\
        N'& = {}^g N, & \epsilon'({}^gh, {}^gn) & = \frac{c({}^hn)}{c(n)}\frac{\Omega_g({}^hn, h)}{\Omega_g(h,n)}\epsilon(h,n),
        \ea
    \end{equation}
    where
    \begin{equation}
    \label{eqn:Omega}
        \Omega_g(f,h):=\frac{\omega(g,f,h)\omega({}^gf, {}^gh, g)}{\omega({}^gf, g, h)}.
    \end{equation}
\end{theorem}
Physically, $\cA(H,N,\gamma,\epsilon)$ is obtained in two steps: a Higgs field in $\Fun(G/H)$ breaks the gauge group $G$ to $H$; and in the $H$ gauge theory the fluxes in $N$ condense, with the multiplication $\gamma$ and charges $\epsilon$.
The condensable algebra $\cA(H, N, \gamma, \epsilon)$ in $D(G, \omega)$ has quantum dimension
\begin{equation}
    \dim(\cA(H, N, \gamma, \epsilon)) = \frac{|G||N|}{|H|}
\end{equation}
and its charge-component (trivially graded component) is the coset algebra $\Fun(G/H)\cong \text{Ind}^G_H1$.

Each datum has a direct physical meaning:
\begin{itemize}
    \item $H$ is the \textbf{unbroken subgroup}: the charge condensate Higgses the 2+1d gauge group $G$ to $H$.
    \item $N$ consists of the \textbf{proliferated fluxes}: vortices with holonomy in $G$-conjugate of $N$ condense.
    \item $\gamma$ is a trivialization of $\omega|_{N}$, all possible $\gamma$ differ by an \textbf{$N$-SPT}: it modifies the multiplication of the condensed vortex operators, $V_n V_{n'}=\gamma(n,n')V_{nn'}$.
    \item $\epsilon$ is the \textbf{$H$-action phase}: the surviving gauge transformations act on the condensed vortex operators projectively:
    \begin{equation}
        U_h V_n U_{h}^{-1}=\epsilon(h,n) V_{{}^hn}.
    \end{equation}
\end{itemize}
Here $V_n$ is the operator that creates a flux $n$ at the base point of the condensate and $U_h$ is the gauge transformation by $h\in H$. 

For $\omega=1$, $\gamma$ is a 2-cocycle, the condensed flux operators form the twisted group algebra $\bC_\gamma[N]$, and inequivalent choices of $\gamma$ are classified by $H^2(N,U(1))$, the group that classifies 1+1d $N$-SPT phases. For $\omega=1$ and $\gamma=1$, $\epsilon(h,\cdot)$ restricted to $C_H(n)$ is the charge carried by the condensed flux $n$: the condensate contains the dyon $(n,\epsilon(\cdot,n))$ rather than the pure flux $(n,1)$.

Anyon proliferation results in a new TO, called the \textbf{reduced TO}. Mathematically, the reduced TO is the category of local modules of the condensable algebra \cite{Kong:2014condensation}. The reduced TO associated to $\cA(H,N,\gamma,\epsilon)$ in $D(G,\omega)$ is $D(H/N,\alpha)$ \cite{davydov2017lagrangian,Gai:2026hjk}. 
Physically, the charge condensate Higgses $G$ to $H$, and condensing the fluxes in $N$ leaves the gauge theory of $H/N$. Even for $\omega=1$ the twist $\alpha$, which is built from $\gamma$ and $\epsilon$, can be non-trivial. For instance $\cA(\Z_4,\Z_2,1,\epsilon_-)$ in $D(\Z_4)$, which condenses the dyon $m^2e^2$, has the double semion $D(\Z_2,\alpha)$ as its reduced TO \cite{Gai:2026hjk}.

Any $\cA$-proliferation, $\cA=\cA(H,N,\gamma, \epsilon)$, in $D(G, \omega)$, can be split into a two-step process \cite{Frohlich:2003hm, davydov2009modular, davydov2017lagrangian, Hannah:2023tae, Schafer-Nameki:2026hgx}: 
\begin{equation}
\label{eqn:two-step-condensation}
    D(G, \omega) \xrightarrow{\cA(H,1,1,1)} D(H, \omega|_{H}) \xrightarrow{\cA(H,N,\gamma, \epsilon)} D(H/N, \alpha),
\end{equation}
where the arrows labeled by $\cA$ represent $\cA$-proliferation processes.
In the second step $\cA(H,N,\gamma,\epsilon)$ denotes the image of $\cA$ in $D(H,\omega|_H)$, a condensable algebra of $D(H,\omega|_H)$. Physically, the first step Higgses $G$ to $H$, and the second step condenses the fluxes in $N$ in the $H$ gauge theory.

% \begin{proposition}
%     Let $\cA=\cA(H, N, \gamma, \epsilon)$ be a condensable algebra in $D(G, \omega)$. Its charge component $\cA_1=\Fun(G/H)=\cA(H,1,1,1)$ is a subalgebra and proliferating $\cA$ into a pure-charge step followed by a pure-flux 
% \end{proposition}

\subsection{Fusion Subcategories of $D(G, \omega)$}
\label{sec:NNW}
Fusion subcategories of $D(G, \omega)$ are classified in \cite{NNW:2009}. We first review the definition of ($G$-invariant) twisted bicharacter:
\begin{definition}
    Let $L$, $K$ be normal subgroups of a group $G$ that centralize each other, and $\omega\in H^3(G, U(1))$. A function $B:K\times L \rightarrow \bC^\times$ is a \textbf{$\omega$-bicharacter} if
    \begin{equation}
    \label{eqn:defn_bicharacter}
        \begin{split}
            B(x,yz) & = \beta_x^{-1}(y,z) B(x,y) B(x,z),\\
            B(wx,y) & = \beta_y(w,x) B(w,y) B(x,y)
        \end{split}
    \end{equation}
    for all $w,x\in K$ and $y,z\in L$. Equivalently, $B(x,\cdot)$ is a $\beta_x$-character of $L$ and $B(\cdot,y)$ is a $\beta_y^{-1}$-character of $K$.
\end{definition}

\begin{theorem}[\cite{NNW:2009}]
\label{thm:NNW}
    Fusion subcategories of $D(G, \omega)$ are in bijection with triples $(K,L,B)$ where $K$, $L$ are normal subgroups of $G$ centralizing each other and $B:K\times L \rightarrow \bC^\times$ is a $G$-invariant $\omega$-bicharacter.
\end{theorem}

For a triple $(K,L,B)$ as in the above theorem, the fusion subcategory of $D(G, \omega)$ is denoted by $S(K,L,B)$. It is defined to be the full Abelian subcategory of $D(G, \omega)$ generated by $(x,\irrep)$ where $x\in K\cap R$ and $\irrep$ is an irreducible $\beta_x$-character of $C_G(x)$ such that $\irrep(l)=B(x,l)\text{deg}(\irrep)$ for all $l\in L$. Being $\beta_x$-character of $C_G(x)$, since $\beta_x|_{C_G(x)}=\tau_{x}(\omega)|_{C_G(x)}$, $\irrep$ is indeed a $\tau_x(\omega)$-projective character.

For any fusion subcategory $\cD$ of $D(G, \omega)$, one can recover the triple $(K_\cD,L_\cD,B_\cD)$ in Theorem~\ref{thm:NNW} as follows \cite{NNW:2009}: normal subgroups $K_\cD$ and $L_\cD$ are defined as
\begin{equation}
\label{eqn:NNW_KL_from_subcat}
    \begin{split}
        K_\cD & =\{ {}^g a \,|\, g\in G, \, (a,\irrep) \in \cD \text{ for some } \irrep \},\\
        L_\cD & = \bigcap_{\irrep: (1,\irrep)\in \cD} \text{ker}(\irrep)
    \end{split}
\end{equation}
where $\text{ker}(\irrep):=\{g\in G\,|\, \irrep(g) = \irrep(1)\}$. Here, $L_\cD$ is the normal subgroup such that
\begin{equation}
\label{eqn:NNW_defn_of_L}
    \cD\cap \text{Rep}(G)=\text{Rep}(G/L_\cD)
\end{equation}
where $\text{Rep}(G)$ is viewed as a subcategory of $D(G, \omega)$. The $G$-invariant $\omega$-bicharacter $B_{\cD}:K_\cD\times L_\cD 
\rightarrow U(1)$ is defined as follows. For $(x, \irrep) \in \cD$, $g\in G$, $l\in L_\cD$
\begin{equation}
\label{eqn:NNW_B}
    B_{\cD}(x^g, l)=\frac{\beta_x(g,l)\beta_x(gl,g^{-1})}{\beta_x(g,g^{-1})}\frac{\irrep({}^gl)}{\text{deg}(\irrep)}.
\end{equation}
It is shown to be well-defined and does not depend on $\irrep$ in \cite{NNW:2009}. Conceptually, identifying $D(G, \omega)$ as the $G$-equivariantization of $\Vect_G^\omega$ ($G$ acts by conjugation), the category $S(K,L,B)$ can be identified to be the subcategory of $D(G, \omega)=(\Vect_G^\omega)^G$ consisting of vector spaces supported on $K$, where for $x\in K$, the action of $l\in L$ on the $x$-graded component is the scalar multiplication by $B(x,l)$.
Physically, $S(K,L,B)$ consists of the anyons whose fluxes lie in $K$ and on which the gauge transformations $l\in L$ act by the scalar $B(\cdot,l)$. Its pure charges are the representations of $G/L$.

We review the following \cite[Lemma 3.10, Lemma 5.9]{naidu2007categorical}
\begin{lemma}
\label{lem:NNW_dim_subcat}
    The dimension of the subcategory $S(K,L,B)$ in $D(G, \omega)$ is $|K|[G:L]$.
\end{lemma}

The fusion subcategory $S(K,L,B)$ of $D(G, \omega)$ is an equivariantization of a pointed category \cite[Remark 7.1, Theorem 7.2]{NNW:2009}:
$S(K,L,B)$ of $D(G, \omega)$ contains subcategory $\Rep(G/L)$. %and is identified with category of equivariant vector bundles on $K$. 
Let $F:S(K,L,B) \rightarrow \Vect_{K}^{\omega|_{K}}$ denote the restriction of the forgetful functor $D(\Vect_G^\omega) \rightarrow \Vect_G^\omega$, and let $I$ denote the left adjoint to $F$. $A:=I(1)=\Fun(G/L)$ is the regular algebra in $\Rep(G/L)$ and $I(X)$ has a structure of an $A$-module for any object $X$ in $\Vect_{K}^{\omega|_{K}}$. The functor $X \mapsto I(X)$ is a tensor equivalence between $\Vect_{K}^{\omega|_{K}}$ and the category of $A$-modules in $S(K,L,B)$. Equivalently, 
\begin{equation}
\label{eqn:S_equivr_pointed}
    S(K,L,B) \cong \lb \Vect_{K}^{\omega|_{K}}\rb/{(G/L)}.
\end{equation}
In this paper, $/$ has multiple meanings which are either made explicit or inferred from the context: it could denote the $G$-equivariantization $\cC/G$, or de-equivariantization (e.g., $\cC/\Rep(G)$), or condensation, i.e., taking local modules ($\cC/\cA$). In the above equation, it denotes equivariantization.
The above says that $S(K,L,B)$ is equivalent to a $G/L$-equivariantization of $\Vect_{K}^{\omega|_{K}}$. 
In particular, $S(K,L,B)/{\Rep(G/L)}$, the corresponding de-equivariantization is the pointed category $\Vect_{K}^{\omega|_{K}}$. The same logic implies that we obtain a pointed category from any fusion subcategory of $D(G, \omega)$ via (the proof also follows Theorem 4.8 of \cite{Galindo2021EquivariantFS}): 
\begin{lemma}
\label{lem:pointedness_de-equiv}
    Let $\cD$ be a fusion subcategory of $D(G, \omega)$ such that 
    \begin{equation}
        \cD \cap \Rep(G) = \Rep(G/N)
    \end{equation}
    for some normal subgroup $N$. The de-equivariantization $\cD/\Rep(G/N)$ is pointed.
\end{lemma}
\begin{proof}
    Here we sketch the proof. Let $\cC:=\Vect_G^\omega$. Given a fusion subcategory $\cD$ of $\cC^G=D(G, \omega)$ such that $\cD\cap \Rep(G) = \Rep(G/N)$, let $\cE$ be the full fusion subcategory of $\cC$ generated by the image of $\cD$ under the forgetful functor
    \begin{equation}
        F_{G}: \cC^G \rightarrow \cC,
    \end{equation}
    and similarly, let $\widetilde\cE$ be the full fusion subcategory of $\cC^N$ generated by the image of $\cD$ under
    \begin{equation}
        F_{G/N}:\cC^G \rightarrow \cC^{N}.
    \end{equation}
    From the equivalence $(\cC^G)_{G/N}\cong \cC^N$, $\widetilde\cE$ is the de-equivariantization $\widetilde\cE=\cD/\Rep(G/N)$. It was shown in \cite{Galindo2021EquivariantFS} that $F_{N}$ induces an equivalence $\widetilde\cE\cong \cE$. 

    Now, as a subcategory of $\cC$, $\cE$ is pointed. We see that the de-equivariantization $\cD/\Rep(G/N)$ is pointed.
\end{proof}

Recall the following definition of centralizer.
\begin{definition}
    Let $\cB$ be a braided fusion category with braiding $c$. For any full subcategory $\cC$ of $\cB$, its \textbf{centralizer} in $\cB$ is the fusion subcategory
    \begin{equation}
    \label{eqn:defn_centralizer}
        \cC':=\{ x\in \cB \,|\, c_{y,x}\circ c_{x,y}=\text{id}_{x\otimes y}, \, \forall y\in \text{ob}(\cC) \}.
    \end{equation}
\end{definition}
Physically, $\cC'$ is the category of anyons in $\cB$ that braid trivially with anyons in $\cC$.

The following lemma is shown in \cite[Lemma 5.10]{naidu2007categorical}
\begin{lemma}
\label{lem:NNW_centralizer}
    The centralizer of the fusion subcategory $S(K,L,B)$ in $D(G, \omega)$ is 
\begin{equation}
\label{eqn:NNW_centralizer}
    S(K,L,B)'=S(L,K,(B^{\mathrm{op}})^{-1}),
\end{equation}
where $B^{\mathrm{op}}(l,k):=B(k,l)$ for $k\in K$ and $l\in L$.
\end{lemma}

\subsection{Generated Subcategories}
\label{sec:GeneratedSubcat}
Recall from \cite{davydov2009modular, Hannah:2023tae, Gai:2026hjk} the explicit description of $\cA(H,N,\gamma, \epsilon)$ in $D(G, \omega)$ in terms of a $G$-graded vector space with $G$-action (we omit the multiplication because the fusion subcategory generated by such an algebra does not depend on the multiplication structure). $\cA:=\cA(H,N, \gamma, \epsilon)$ is spanned by vectors $v_{g,n}$, $g\in G$, $n\in N$, with $G$-grading $|v_{g,n}|={}^gn$ and $G$-action
\begin{equation}
\label{eqn:Alg_G-action}
    x\cdot v_{g,n}=\tau_{n}(\omega)(x,g)v_{xg, n},
\end{equation}
where $\tau_{n}(\omega)$ is the transgression defined in \eqref{eqn:transgression}
subject to relation
\begin{equation}
\label{eqn:Alg_span_relation}
    v_{gh,n} = \epsilon(h,n)\tau_{n}(\omega)(g,h)^{-1}v_{g,{}^hn}
\end{equation}
where $g\in G$, $h\in H$ and $n\in N$. Thus a set $R_H$ of representatives for $G/H$ gives a basis $\{v_{r,n}\}_{r\in R_H, n\in N}$.

The multiplication is
\begin{equation}
    v_{g,n} v_{g', n'} = \delta_{gH, g'H}\frac{\gamma(n,n')}{\Omega_g(n,n')}v_{g,nn'},
\end{equation}
$g, g'\in R_{H}$.

In particular, the trivially graded component of $\cA$ is
\begin{equation}
    \cA_1\cong \text{Fun}(G/H)
\end{equation}
as algebras in $\Rep(G)$, with $v_{r,1}$, $r\in R_H$ its primitive idempotents. As an object in $\Rep(G)$, it is $\text{Ind}^G_{H}1$.
Physically, $\cA_1=\Fun(G/H)$ is the Higgs field, whose condensate breaks $G$ to $H$. The set of vacua is $G/H$. The components $v_{g,n}$ with $n\neq 1$ create the flux ${}^gn$ in the vacuum labeled by $gH$.

We now answer the question which fusion subcategory of $D(G, \omega)$ is generated by a condensable algebra $\cA:=\cA(H, N, \gamma, \epsilon)$, we set 
\begin{equation}
    K_{\cA}:=\langle {}^gN \mid g\in G\rangle,
\end{equation}
to be the group closure of the support of $\cA$. This will be the support of the fusion subcategory.
Physically, $K_{\cA}$ is the group generated by all condensed fluxes.

For the second classification data, the group elements acting as a scalar, we define
\begin{equation}
    \Sigma:=G\times N.
\end{equation}
For $a=(g,n)\in \Sigma$ and $l\in \Core_G(C_H(N))$ (for a group $G$, the \textbf{core} of a subgroup $N$ with respect to a subset $H\subset G$ is $\Core_H(N):=\cap_{h\in H} N^h$), define
\begin{equation}
    \sigma_{a}:={}^gn,
\end{equation}
and the scalar
\be\label{eq:audit-lambda}
\lambda_{(g,n)}(l):=
\frac{\tau_n(\omega)(l,g)}{\tau_n(\omega)(g,l^g)}
\epsilon(l^g,n)\,.
\ee

For $\omega=1$,
\be
\label{lambda-omega1}
\lambda_{(g,n)}(l)=\epsilon(l^g,n) \,,
\ee
the charge $\epsilon(\cdot,n)$ of the condensed flux $n$, evaluated on the gauge transformation $l$ transported to the base point.
More generally, if $w=(a_1, \cdots, a_r)$ is a word in $\Sigma$, define
\begin{equation}
\label{eqn:defn_Lambda}
    \begin{split}
        p_j & := \sigma_{a_1} \cdots \sigma_{a_j},\\
        \Lambda_{l}(w) & := \lb \prod_{j=1}^{r-1} \beta_{l}(p_j, \sigma_{a_{j+1}}) \rb \lb \prod_{j=1}^{r} \lambda_{a_j}(l) \rb,
    \end{split}
\end{equation}
and $\Lambda_{l}(\emptyset):=1$. With this convention, define
\begin{equation}
\label{eqn:defn_LA}
    \begin{split}
        L_{\cA}:=\{ l\in \Core_G(C_H(N)) \mid {} & \Lambda_{l}(w)=1 \text{ for any word } \\
        & w \text{ such that } p_r=1 \}.
    \end{split}
\end{equation}
Here $p_r$ denotes the product $\sigma_{a_1} \cdots \sigma_{a_r}$ of the word $w$. We call word $w$ a \textbf{relator} if $p_r=1$.
Group $L_{\cA}$ contains elements that acts as a scalar determined by $\Lambda$ (hence the $\omega$-bicharacter is determined by $\Lambda$) on each graded component. We prove this as:
\begin{lemma}
\label{lem:B_well_defined}
On objects in $\langle \cA \rangle$, the group elements acting via a scalar and the scalar are determined by the following:
\begin{enumerate}
    \item Let $l\in G$, $l$ preserve $\bC v_{g,n}\subset \cA$ for every $g\in G$ and $n\in N$ if and only if $l\in \Core_G(C_H(N))$, in which case, 
    \begin{equation}
    \label{eqn:Core_action_scalar}
        l\cdot v_{g,n} = \lambda_{(g,n)}(l)v_{g,n}.
    \end{equation}
    
    \item Element $l$ in $L_{\cA}$ has the property that for $k\in K_{\cA}$, choosing any word $w$ with $p_r=k$, 
\begin{equation}
\label{eqn:defn_B_cA}
    B_\cA(k,l):=\Lambda_{l}(w)
\end{equation}
is well-defined, i.e., independent on the choice of word $w$. 
\end{enumerate}
\end{lemma}
\begin{proof}
For the first claim, equation~\eqref{eqn:Alg_G-action} gives
\be\label{eq:audit-scalar-proof}
\ba
l\cdot v_{g,n}&=\tau_n(\omega)(l,g)v_{lg,n}\,,\\
v_{lg,n}&=\epsilon(l^g,n)\tau_n(\omega)(g,l^g)^{-1}
v_{g,{}^{l^g}n}\,,
\ea\ee
$l^g\in H$.
$v_{lg,n}$ is proportional to $v_{g,n}$ only when $l^g\in H$.
Under that condition, $v_{g,{}^{l^g}n}$ is proportional to $v_{g,n}$ if and only if $l^g$ centralizes $n$, in which case $l$ acts by the scalar~\eqref{eq:audit-lambda}.

    Finally, requiring $l\cdot v_{g,n}$ to be proportional to $v_{g,n}$ for every $g\in G$ and $n\in N$, we conclude that this is equivalent to $l\in \cap_{g\in G}(C_H(N))^g =  \Core_G(C_H(N))$.

    For the second claim, let $l\in L_{\cA}$, and let $u = (a_1, \cdots, a_r)$ and $v = (b_1, \cdots, b_s)$ be words in $\Sigma$, with partial products denoted by $p_i=\sigma_{a_1}\cdots \sigma_{a_j}$ and $q_j=\sigma_{b_1}\cdots \sigma_{b_j}$, respectively. From the definition of $\Lambda$, we have
    \begin{equation}
    \label{eqn:proof_B_well_defined_telescope}
        \frac{\Lambda_{l}(uv)}{\Lambda_l(u)\Lambda_{l}(v)} = \beta_l(p_r,\sigma_{b_1})\prod_{j=1}^{s-1}\frac{\beta_l(p_r q_j, \sigma_{b_{j+1}})}{\beta_l(q_j, \sigma_{b_{j+1}})}.
    \end{equation}
    The 2-cocycle condition on $\beta$ implies
    \begin{equation}
        \frac{\beta_l(p_r q_j, \sigma_{b_{j+1}})}{\beta_l(q_j, \sigma_{b_{j+1}})} = \frac{\beta_l(p_r, q_j \sigma_{b_{j+1}})}{\beta_l(p_r, q_j)}.
    \end{equation}
    Substituting into \eqref{eqn:proof_B_well_defined_telescope}, the RHS telescopes, and we have
    \begin{equation}
    \label{eqn:proof_B_well_defined_productrule}
        \frac{\Lambda_{l}(uv)}{\Lambda_l(u)\Lambda_{l}(v)} = \beta_{l}(p_r, q_s).
    \end{equation}

    Now if two words $u$, $v$ have the same product $p_r=q_s=k$, then for $b_j=(g_j,n_j)\in \Sigma$, define $b_j'=(g_j, n_j^{-1})$ and $v':=(b_{s}', \cdots, b_{1}')$. Now, by definition \eqref{eqn:defn_LA} of $L_{\cA}$,
    \begin{equation}
        \Lambda_l(uv') = \Lambda_l(vv')=1.
    \end{equation}
    Using \eqref{eqn:proof_B_well_defined_productrule}, we have
    \begin{equation}
        \begin{split}
            \beta_l(k, k^{-1}) \Lambda_l(u) \Lambda_l(v') & =1, \\
            \beta_l(k, k^{-1}) \Lambda_l(v) \Lambda_l(v') & =1.
        \end{split}
    \end{equation}
    Dividing gives
    \begin{equation}
        \Lambda_l(u)=\Lambda_l(v).
    \end{equation}
    Hence, $B_{\cA}(k,l)$ is well-defined.
\end{proof}

% More generally, an element $k$ in $K_\cA:=\langle \Sigma_N \rangle$ is a word $k=a_1\cdots a_r$, $r\in \Z_{\geq 0}$, $a_i\in \Sigma_N$. 

% Element $l\in L_0$ acts on $k$-graded components 

Using the above notation, we have:
\begin{theorem}
\label{thm:fusion_subcat_from_alg}
    The fusion subcategory generated by the condensable algebra $\cA:=\cA(H,N, \gamma, \epsilon)$ in $D(G, \omega)$ is
    \begin{equation}
        \langle \cA \rangle=S(K_{\cA}, L_{\cA}, B_{\cA}),
    \end{equation}
    where
    \begin{equation}
        \begin{split}
            K_{\cA} & :=\langle {}^gN \mid g\in G\rangle, \\
            & \hspace{-1.343em}\begin{split}
        L_{\cA} :=\{ l\in \Core_G(C_H(N)) \mid {} & \Lambda_{l}(w)=1 \text{ for word } \\
        & w \text{ such that } p_r=1 \},
    \end{split}
        \end{split}
    \end{equation}
    and the bicharacter $B_{\cA}$ is~\eqref{eqn:defn_B_cA}.
\end{theorem}
\begin{proof}
Let $\cD=\langle\cA\rangle=S(K_{\cD},L_{\cD},B_{\cD})$. Take grading of fusion product of any objects in $\cA$ gives
\be\label{eqn:proof-KA-equals-KD}
K_{\cD}=\langle{}^gN\mid g\in G\rangle=K_{\cA}\,.
\ee
If $l\in L_{\cD}$, its action on every degree-$k$ component is the scalar $B_{\cD}(k,l)$. Lemma~\ref{lem:B_well_defined}(1) therefore forces $l \in \Core_G(C_H(N))$. Iterating the first-variable bicharacter identity gives $B_{\cD}(p_r,l)=\Lambda_l(w)$. Every relator has $p_r=1$, so $L_{\cD}\subseteq L_{\cA}$.

Conversely, let $l\in L_{\cA}$. It commutes with $K_{\cA}$, and its action on a homogeneous tensor $v_{g_1, n_1} \otimes \cdots \otimes v_{g_r, n_r}$ with word $w =(g_1, n_1) \cdots (g_r, n_r)$ is $\Lambda_l(w)$.
It acts as the identity on the degree-one part of every tensor power of $\cA$. Every simple object of $\cD$ occurs in a tensor power of $\cA$, including every $(1, \chi) \in \cD$. Thus $l$ lies in the common kernel of these $\chi$, which is $L_{\cD}$ by~\eqref{eqn:NNW_KL_from_subcat}.
We obtain $L_{\cA}=L_{\cD}$, and $B_{\cA}=B_{\cD}$ follows.
\end{proof}

We now prove the claim that $B_{\cA}$ is indeed well-defined and is a $G$-invariant $\omega$-bicharacter.
\begin{lemma}
\label{lem:B_G-inv_omega_bichr}
    The set $L_{\cA}$ defined in \eqref{eqn:defn_LA} is a normal subgroup of $G$ centralizing $K_{\cA}$, and
    $B_\cA$ as defined in \eqref{eqn:defn_B_cA} is a $G$-invariant $\omega$-bicharacter.
\end{lemma}
\begin{proof}
The proof of Theorem~\ref{thm:fusion_subcat_from_alg} identifies
$(K_{\cA},L_{\cA},B_{\cA})$ with the triple $(K_{\cD}, L_{\cD}, B_{\cD})$ of
$\cD=\langle\cA\rangle$. Normality, mutual centralization and the $G$-invariant $\omega$-bicharacter properties therefore follow from Theorem~\ref{thm:NNW}. (They can also be directly verified.)
\end{proof}

\begin{corollary}
    In the special case of trivial $\omega$ and $\epsilon$, i.e., for condensable algebra $\cA:=\cA(H,N,\gamma, 1)$ in $D(G)$. 
    \begin{equation}
        \langle \cA \rangle = S(K_{\cA}, \Core_G(C_H(N)), 1).
    \end{equation}
\end{corollary}
\begin{proof}
    When $\omega=1$ and $\epsilon=1$, 
    $\lambda_{(g,n)}(l)=1$ for every $l\in \Core_G(C_H(N))$, and any $g\in G$, $n\in N$. Hence
    \begin{equation}
        L_{\cA}=\Core_G(C_H(N)).
    \end{equation}
    Also, $\beta_a=1$, hence $B_{\cA}=1$ is constant.
\end{proof}

Finally, we note that none of $K_{\cA}$, $L_{\cA}$ or $B_{\cA}$ involves $\gamma$, this is consistent with the fact that $\cA(H, N, \gamma, \epsilon) \cong \cA(H, N, \gamma', \epsilon)$ are twin algebras. Defined on the same simple objects, twin algebras generate the same fusion subcategory.
The object underlying $\cA(H,N,\gamma,\epsilon)$ depends only on $(H,N,\epsilon)$. If $(\gamma,\epsilon)$ and $(\gamma',\epsilon)$ both satisfy \eqref{eqn:epsilon_compatibility}, the two algebras are isomorphic as objects, and they are twins if they are not isomorphic as algebras.

\subsection{Centralizer and M\"uger Center}
\label{sec:centralizer-muger}

\noindent\textbf{Centralizer.}
Results in \cite{NNW:2009} can be applied to obtain centralizer and the M\"uger center of a fusion subcategory generated by a condensable algebra in $D(G, \omega)$ from the classification data.

In what follows, we fix a condensable algebra $\cA:=\cA(H, N, \gamma, \epsilon)$ in $D(G, \omega)$. Denote its generated fusion subcategory by
\begin{equation}
    \cC_{\cA}:=\langle\cA\rangle.
    \label{eqn:muger-abbreviations}
\end{equation}

By Lemma~\ref{lem:NNW_centralizer}, we obtain its centralizer
\begin{equation}
\label{eqn:centralizer-generated-category}
    \cC_{\cA}'=S\bigl(L_{\cA},K_{\cA},(B_{\cA}^{\mathrm{op}})^{-1}\bigr),
\end{equation}
where recall
\begin{equation}
    (B_{\cA}^{\mathrm{op}})^{-1}(l,k)=B_{\cA}(k,l)^{-1}.
\end{equation}
For any word $w_k$ whose product is $k$, 
\begin{equation}
    (B_{\cA}^{\mathrm{op}})^{-1}(l,k) = \Lambda_{l}(w_k)^{-1}.
\end{equation}

On the other hand, we give a criterion in terms of the algebra data $(H, N, \gamma, \epsilon)$ to determine which simple objects are in the centralizer $\cC_{\cA}'$.

\begin{proposition}
    A simple object $(x,\irrep)$ in $D(G,\omega)$ is an object of $\cC_{\cA}'$ if and only if $x\in \Core_G(C_H(N))$ and
    \begin{equation}
    \label{eqn:chi_centralizing_object}
        \irrep({}^gn) = \lambda_{(g,n)}(x)^{-1}\deg(\irrep)
    \end{equation}
    for every $g\in G$, $n\in N$.
\end{proposition}
\begin{proof}
    By definition, an object $(x, \irrep)$ is in $\cC_{\cA}'$ if and only if it centralizes objects in $\cA$. From the description of braiding in \cite{davydov2017lagrangian}, for $u$ an element in $x$-graded $V_{x}$ and $v_{g,n}$ of degree $b={}^gn$, the double braiding 
    \begin{equation}
        u\otimes v_{g,n} \xmapsto{c^2} {}^{xg}n\cdot u \otimes x\cdot v_{g,n}.
    \end{equation}
    The above double braiding is identity for all $g\in G$, $n\in N$ only if $x\in \cap_{g\in G}C_{G}({}^gN)$, and the action of $x$ on each homogeneous component of $\cA$ and the action of ${}^{xg}n={}^gn$ on $V_x$ are inverse scalar operators. As before, $x$ acting as a scalar on every $\bC v_{g,n}$ is equivalent to $x\in \Core_G(C_H(N))$, and the scalar is $\lambda_{(g,n)}(x)$, see \eqref{eqn:Core_action_scalar} in Lemma~\ref{lem:B_well_defined} 1. On the other hand, ${}^gn$ acts on $u$ by $\pi_\irrep({}^gn)$, where $\pi_{\irrep}$ denotes the irrep whose character is $\irrep$, hence we conclude that $\irrep({}^gn) = \lambda_{(g,n)}(x)^{-1}\deg(\irrep)$. 

    The reverse direction is straightforward: given $x\in \Core_G(C_H(N))$ and $\irrep$ with $\irrep({}^gn)=\lambda_{(g,n)}(x)^{-1}\deg(\irrep)$, then $(x,\irrep)$ centralizes $\cA$.
\end{proof}

As a consequence, the centralizer $\cC_{\cA}'$ is generated by objects $(x,\irrep)$ where $x\in \Core_G(C_H(N))\cap R$ and $\irrep$ such that \eqref{eqn:chi_centralizing_object} holds for every $g\in G$, $n\in N$.

\vspace{2mm}
\noindent\textbf{M\"uger center.}
From definition, the M\"uger center of a fusion subcategory $\cD$ of $\cC$ can be computed from the intersection with its centralizer:
\begin{equation}
    \cZ_2(\cD) = \cD\cap \cD'.
\end{equation}
We use this to write down the M\"uger center of $\langle \cA \rangle$ for a condensable algebra $\cA$ in $D(G, \omega)$, following the same logic as in \cite{NNW:2009}.

Given a condensable algebra $\cA:=\cA(H,N,\gamma, \epsilon)$ in $D(G, \omega)$, define
\begin{equation}
    \varphi_{\cA}(a,b):=B_{\cA}(a,b) B_{\cA}(b,a)
\end{equation}
for $a,b\in K_\cA\cap L_{\cA}$. Note that $\varphi_{\cA}$ is a symmetric bicharacter on $K_\cA\cap L_{\cA}$: for $a,b,c\in K_\cA\cap L_{\cA}$,
\begin{equation}
    \varphi_{\cA}(ab,c) = B_{\cA}(ab,c) B_{\cA}(c,ab) = \varphi_{\cA}(a,c) \varphi_{\cA}(b,c),
\end{equation}
where in the final step, the twisted factors in \eqref{eqn:defn_bicharacter} cancel. 
Let 
\begin{equation}
\label{eqn:defn_RA}
    R_{\cA}:=\{ a\in K_\cA\cap L_{\cA} \mid \varphi_{\cA}(a,\cdot) = 1 \},
\end{equation}
where $1$ denotes the constant map from $K_\cA\cap L_{\cA}$ to $\bC^\times$. 
Because $K_{\cA}$ and $L_{\cA}$ are normal in $G$ and centralize each other, $R_{\cA}$ is an abelian subgroup of $G$. $R_{\cA}$ is further normal because $\varphi_{\cA}$ is $G$-invariant.
Also denote by
\begin{equation}
    P_{\cA}:=K_{\cA}L_{\cA}=\{kl \mid k\in K_{\cA}, l\in L_{\cA}\}.
\end{equation}
For $a\in R_{\cA}$, $p=lk$ for $k\in K_\cA$ and $l\in L_\cA$, define $\psi_{\cA}:R_{\cA} \times P_{\cA} \rightarrow \bC^\times$ by
\begin{equation}
\label{eqn:defn_psiA}
    \psi_{\cA}(a,lk) := \beta_{a}(l,k)^{-1} B_{\cA}(a,l) B_{\cA}(k,a)^{-1}.
\end{equation}
It was shown in \cite{NNW:2009} that $\psi_{\cA}$ is well-defined and is a $G$-invariant $\omega$-bicharacter on $R_{\cA} \times P_{\cA}$. With these, applying Proposition~6.6 in \cite{NNW:2009}, we have the following:
\begin{proposition}
\label{prop:Muger-center}
    Given a condensable algebra $\cA$ in $D(G, \omega)$, the M\"uger center of its generated fusion subcategory is
    \begin{equation}
        \cZ_2(\langle \cA \rangle ) = S(R_{\cA}, P_{\cA}, \psi_{\cA}).
    \end{equation}
\end{proposition}
\begin{proof}
    By definition, $\cZ_2(\langle \cA \rangle) = \langle \cA \rangle\cap \langle \cA \rangle'$. Substituting $\langle \cA \rangle = S(K_{\cA}, L_{\cA}, B_{\cA})$ and $\langle \cA \rangle' = S(L_{\cA}, K_{\cA}, (B_{\cA}^{\text{op}})^{-1})$ into \cite[Proposition 6.3]{NNW:2009} computes the intersection subcategory as $S(\ker \varphi_{B,B'}, LL', \psi_{B,B'})$. One obtains the first entry:
    \begin{equation}
        \begin{split}
            R_{\cA} & =\{ a\in K_{\cA}\cap L_{\cA} \mid B_{\cA}(a,h) = (B_{\cA}^{\text{op}})^{-1}(a,h)\, \forall h \} \\
            & = \{ a\in K_{\cA}\cap L_{\cA} \mid B_{\cA}(a,h)B_{\cA}(h,a) = 1\, \forall h \},
        \end{split}
    \end{equation}
    the second entry is $L_{\cA}K_{\cA}=P_{\cA}$, and the pairing is $\psi_{\cA}$ as defined in \eqref{eqn:defn_psiA}.
\end{proof}

\begin{corollary}
    Given a condensable algebra $\cA$ in $D(G, \omega)$, $\langle \cA \rangle$ is nondegenerate, i.e., modular, if and only if
    \begin{equation}
        K_{\cA}L_{\cA} = G \quad \text{and} \quad R_{\cA}=1.
    \end{equation}
\end{corollary}
\begin{proof}
    This follows from the dimension of a fusion subcategory $S(R_{\cA}, P_{\cA}, \psi_{\cA})$, see Lemma~\ref{lem:NNW_dim_subcat}:
    \begin{equation}
        \dim \cZ_2(\langle \cA \rangle ) = |R_{\cA}| [G: P_{\cA}],
    \end{equation}
    which is 1 if and only if $P_{\cA} = G$ and $R_{\cA}=1$.
\end{proof}

In the special case where $\omega$ and $\epsilon$ are trivial, we obtain the following simplification:
\begin{corollary}
    Given $\cA=\cA(H,N,\gamma,1)$ in $D(G)$, 
    \begin{equation}
        \begin{split}
            \langle \cA \rangle' & = S(C_{\cA}, K_{\cA}, 1),\\
            \cZ_2(\langle \cA \rangle ) & = S(K_{\cA}\cap C_{\cA}, K_{\cA}C_{\cA}, 1),
        \end{split}
    \end{equation}
    where $C_{\cA}:=\Core_G(C_H(N))$.
\end{corollary}

\vspace{2mm}
\noindent\textbf{Tannakian Criterion.}
% A symmetric fusion category is in general of the form $\Rep(G,z)$ for a central involution $z\in G$. 
Throughout this discussion, fix a condensable algebra $\cA=\cA(H, N, \gamma, \epsilon)$ in $D(G, \omega)$ and define
\begin{equation}
    \cE_{\cA}:=\cZ_2(\langle\cA\rangle)
       =S(R_{\cA},P_{\cA},\psi_{\cA}).
    \label{eqn:defn-EA}
\end{equation}

% Being a symmetric fusion category, $\cZ_2(\langle \cA \rangle)$ is super-Tannakian, i.e., it is equivalent to $\Rep(\Gamma,z)$ for some finite group $\Gamma$ and a central involution $z\in \Gamma$. Result of \cite{DGNO} says a fusion subcategory consisting of objects of trivial twists is Tannakian. Although a condensable algebra $\cA$ consists of trivially-twisted objects, $\cZ_2(\langle \cA \rangle)$ is not guaranteed to be Tannakian. See \cite{Schafer-Nameki:2026hgx} for a condensable algebra $A_{E_6}$ in $SU(2)_{10}$ whose generated subcategory has a transparent fermion.
We give the following criterion for determining when $\cZ_2(\langle \cA \rangle)$ is Tannakian or super-Tannakian.
\begin{proposition}
\label{prop:Tannakian_criterion}
    Given a condensable algebra $\cA$ in $D(G, \omega)$.
    The symmetric fusion category $\cE_{\cA}$ is Tannakian if and only if 
    \begin{equation}
    \label{eqn:B_alternating}
        B_{\cA}(r,r)=1
    \end{equation}
    for every $r\in R_{\cA}$. Otherwise it is super-Tannakian.
\end{proposition}
\begin{proof}
    Let $\Theta$ denote the ribbon twist of $D(G, \omega)$. For a simple object $X:=(r, \irrep)$ of $\cE_{\cA}$, write
    \begin{equation}
        \Theta_{X} = \theta_{X} \text{id}_{X}.
    \end{equation}
    Here $r\in R_{\cA}$ is a representative of the conjugacy class that is the support of $X$. The ribbon twist in $\Rep(D(G, \omega))$ is
    \begin{equation}
        \theta_{X} = \frac{\irrep(r)}{\deg(\irrep)}.
    \end{equation}
    By definition of $S(R_{\cA}, P_{\cA}, \psi_{\cA})$, we have $\irrep(p) = \psi_{\cA}(r,p) \deg(\irrep)$ for $p\in P_{\cA}$ (see \cite{NNW:2009}). Since $r\in R_{\cA} \subseteq P_{\cA}$,
    \begin{equation}
        \theta_{X}=\psi_{\cA}(r,r) = B_{\cA}(r,r).
    \end{equation}
    Note that it is well defined because $r\in R_{\cA}$, and it forces $\theta_{X}\in \{\pm 1\}$.
    Now, %to see that $B_{\cA}(r,r) = 1$ for all $r\in R_{\cA}$ is equivalent to $\cE_{\cA}$ being Tannakian, 
    since $\cE_{\cA}$ is symmetric, the ribbon twist $\Theta$ is a monoidal automorphism of the identity functor, i.e., 
    \begin{equation}
        \Theta_{X\otimes Y} = \Theta_{X} \otimes \Theta_{Y}
    \end{equation}
    for $X,Y$ objects in $\cE_{\cA}$. 
    Results of \cite{DGNO} apply and we have a $\Z_2$ grading 
    \begin{equation}
        \cE_{\cA} = (\cE_{\cA})_{+} \oplus (\cE_{\cA})_{-}
    \end{equation}
    where $(\cE_{\cA})_{\pm}$ is the full abelian subcategory of $\cE_{\cA}$ generated by irreducible object $X$ in $\cE_{\cA}$ with ribbon twist $\theta_{X}=\pm \text{id}_{X}$. Moreover, $(\cE_{\cA})_{+}$ is the maximal Tannakian subcategory of $\cE_{\cA}$. 

    Hence $\cE_{\cA}$ is Tannakian when $\theta_{X}=\text{id}_{X}$ for all of its simple object $X$. %By Proposition~6.5 (ii) of \cite{NNW:2009} this is equivalent to $\psi_{\cA}|_{R_{\cA}\times R_{\cA}}$ is alternating, meaning that $\psi_{\cA}(r,r)=1$ for all $r\in R_{\cA}$.
    
    Otherwise if $\cE_{\cA}$ is not Tannakian, i.e., $\psi_{\cA}(r,r)\neq 1$ for some $r\in R_{\cA}$, Deligne's theorem implies $\cE_{\cA}\cong \Rep(G_{\cA}, z_{\cA})$ for some element $z_{\cA}\neq 1$. 
\end{proof}

Since the ribbon twist is monoidal on a symmetric category, all simple summands of a fusion product of bosons in $\cE_{\cA}$ are bosons. If $\cA$ is of type (i), $\langle\cA\rangle=\cE_{\cA}$ is generated by the anyons of $\cA$, which are bosons, and $\cE_{\cA}$ is therefore Tannakian. A super-Tannakian M\"uger center can only occur for algebras of type (ii). Moreover, we find that type (i) algebras are equivalent to the following algebra:
\begin{theorem}
    \label{thm:type-i_LG}
    Let $\cA$ be a condensable algebra in $D(G, \omega)$. $\cA$ is of type (i) if and only if there is a finite group $G'$, $\omega'\in H^3(G', U(1))$, subgroup $H'\subseteq G'$ and a braided equivalence $F: D(G, \omega) \rightarrow D(G', \omega')$ under which
    \begin{equation}
        F(\cA) \cong \Fun(G'/H')
    \end{equation}
    as algebras in the subcategory $\Rep(G')\subset D(G', \omega')$. 
    If this is the case, then
    \begin{equation}
        G_{\cA} \cong G'/\Core_{G'}(H').
    \end{equation}
\end{theorem}
The content of this theorem says that the proliferation transition induced by a type (i) algebra is always a Higgsing of $G'$ to $H'$.
\begin{proof}
    For the forward direction, $\cA$ is of type (i), hence $(R_{\cA}, P_{\cA}) = (K_{\cA}, L_{\cA})$ and $\cE_{\cA}$ is Tannakian. Moreover, 
    \begin{equation}
        K_{\cA}\subseteq L_{\cA},
    \end{equation}
    and since $K_{\cA}$ and $L_{\cA}$ centralize each other, $K_{\cA}$ is abelian. Consider fusion subcategory $S(K_{\cA}, K_{\cA}, B_{\cA})$ (here $B_{\cA}$ is restricted to $K_{\cA}\times K_{\cA}$), which contains $\cA$. Because $B_{\cA}(a,b) B_{\cA}(b,a) = 1$, it is symmetric
    \begin{equation}
        S(K_{\cA}, K_{\cA}, B_{\cA})' = S(K_{\cA}, K_{\cA}, B_{\cA}).
    \end{equation}
    Furthermore, the twist of a simple object of $S(K_{\cA}, K_{\cA}, B_{\cA})$ is $B_{\cA}(a,a)=1$ for some $a\in K_{\cA}$, $S(K_{\cA}, K_{\cA}, B_{\cA})$ is Tannakian and
    \begin{equation}
        S(K_{\cA}, K_{\cA}, B_{\cA}) \cong \Rep(G')
    \end{equation}
    for some finite group $G'$. In the language of \cite{DGNO}, $S(K_{\cA}, K_{\cA}, B_{\cA})$ is a Lagrangian subcategory of $ D(G, \omega)$, and by \cite[Thm.~4.64]{DGNO}, there exists a finite group $G'$ and $\omega'\in H^3(G', U(1))$ and braided equivalence
    \begin{equation}
        F: D(G, \omega) \rightarrow  D(G', \omega')
    \end{equation}
    such that $F(S(K_{\cA}, K_{\cA}, B_{\cA})) = \Rep(G') \subset  D(G', \omega')$. Hence, the connected algebra
    $F(\cA)\cong \Fun(G'/H')$ for some $H'\subseteq G'$. Moreover, $\Core_{G'}(H')$ is the kernel of the action of $G'$ on $G'/H'$, hence
    \begin{equation}
        F(\langle \cA \rangle) = \Rep(G'/\Core_{G'}(H')).
    \end{equation}
For the converse, $F(\langle\cA\rangle)\subseteq\Rep(G')$ is symmetric.
Since $F$ preserves braiding, $\langle\cA\rangle\subseteq\langle\cA\rangle'$, so $\cE_{\cA}=\langle\cA\rangle$ and $\cA$ is of type (i).
\end{proof}

\vspace{2mm}
\noindent\textbf{Tannakian Group.}
Next, we determine the Tannakian group $G_{\cA}$ and its central involution $z_{\cA}$ such that $\cE_{\cA}\cong \Rep(G_{\cA}, z_{\cA})$ as symmetric fusion categories. We define
\begin{equation}
    Q_{\cA}:=G/P_{\cA}.
\end{equation}
It is the group such that the conjugation action of $G$ on $R_{\cA}$ factors through $Q_{\cA}$. $K_{\cA}$ and $L_{\cA}$ commute, hence for $r\in R_{\cA}$, $P_{\cA}$ acts trivially on $R_{\cA}$ by conjugation, $P_{\cA}\subseteq C_G(r)$, and $\overline{g}$ fixes $r$ if and only if $g\in C_G(r)$,
\begin{equation}
\label{eqn:StabQA(r)}
    \text{Stab}_{Q_{\cA}}(r) = C_{G}(r)/P_{\cA}.
\end{equation}
Denote the Pontryagin dual of any abelian group $A$ by $\widehat A:=\Hom(A, U(1))$.
Note that $Q_{\cA}$ acts on $\widehat R_{\cA}$ by conjugation:
\begin{equation}
    \lb {}^{\overline{g}}\chi \rb (r) :=\chi(r^g),
\end{equation}
where $g$ is a representative for $\overline{g}\in G/P_{\cA}$. This is well-defined again because $K_{\cA}$ and $L_{\cA}$ commute.

There is a character $q_{\cA}\in \widehat R_{\cA}$ (proved below), defined as 
\begin{equation}
\label{eqn:defn_qcA}
    q_{\cA}(r) := B_{\cA}(r,r).
\end{equation}
\begin{lemma}
\label{lem:qA_character}
    $q_{\cA}$ as defined above is indeed a character and has order at most 2. That is
    \begin{equation}
        q_{\cA}(r_1 r_2) = q_{\cA}(r_1) q_{\cA}(r_2)
    \end{equation}
    for all $r_1, r_2\in R_{\cA}$, and $q_{\cA}^2=1$.
\end{lemma}
\begin{proof}
$q_{\cA}(r) = \theta_{\delta_r}$ is the ribbon twist of simple object $\delta_r$ graded by $r\in G$.
The twist on a symmetric category $\cE_{\cA}$ is monoidal, so it acts on $\delta_{rs}\cong \delta_r \otimes \delta_s$ by scalar $q_{\cA}(rs) = q_{\cA}(r)q_{\cA}(s)$. Thus $q_{\cA}$ is a character. Furthermore $q_{\cA}(r) = B_{\cA}(r,r) \in \{\pm 1\}$, hence $q_{\cA}^2=1$.
\end{proof}

Note that by definition of $R_{\cA}$, $B_{\cA}(r,r)\in \{\pm 1 \}$. It can be directly checked from its definition that $q_{\cA}$ is $Q_{\cA}$-invariant. %3-cocycle on $(a,x,r,x^{-1})$ and $(a,x,x^{-1},a)$, $a={}^xr$ 

For any section $s:Q_{\cA} \rightarrow G$ of the projection $G\twoheadrightarrow Q_{\cA}$ such that $s(\overline{1}) = 1$, denote the factor set by
\begin{equation}
    f_s(\overline{g}, \overline{h}) := s(\overline{g}) s(\overline{h}) s(\overline{gh})^{-1}.
\end{equation}
Factor set $f_s$ is the extension class of $1 \rightarrow P_{\cA} \rightarrow G \rightarrow Q_{\cA} \rightarrow 1$. 

\begin{theorem}
\label{thm:Tannakian_group}
    Let $\cA=\cA(H, N, \gamma, \epsilon)$ be a condensable algebra in $D(G, \omega)$. There is a short exact sequence of finite groups
    \begin{equation}
        1 \rightarrow \widehat R_{\cA} \rightarrow G_{\cA} \rightarrow Q_{\cA} \rightarrow 1
    \end{equation}
    whose extension class is a normalized twisted 2-cocycle $\alpha_{\cA}\in Z^2(Q_{\CA}, \widehat R_{\cA})$ such that the finite group
    \begin{equation}
    \label{eqn:GA}
        G_{\cA}:=\widehat R_{\cA} \rtimes_{\alpha_{\cA}} Q_{\cA}
    \end{equation}
    with the central involution
    \begin{equation}
        z_{\cA}:=(q_{\cA}, \overline{1}) \in G_{\cA}
    \end{equation}
    fits in the equivalence of symmetric fusion categories
    \begin{equation}
        \cE_{\cA}\cong \Rep(G_{\cA}, z_{\cA}).
    \end{equation}
\end{theorem}
Here $\widehat R_{\cA}\rtimes_{\alpha_{\cA}}Q_{\cA}$ denotes the set $\widehat R_{\cA}\times Q_{\cA}$ with the product $(\chi,\overline g)(\zeta,\overline h)=(\chi\,({}^{\overline g}\zeta)\,\alpha_{\cA}(\overline g,\overline h),\overline{gh})$. 

Before stating the proof, we note an immediate corollary (follows from the above theorem and Prop.~\ref{prop:Tannakian_criterion}):
\begin{corollary}
    If $B_{\cA}(r,r)=1$ for all $r\in R_{\cA}$, then
    \begin{equation}
        \cE_{\cA}\cong \Rep(G_{\cA}).
    \end{equation}
\end{corollary}

\begin{proof}[Proof of Theorem~\ref{thm:Tannakian_group}]
    From Deligne's theorem, there exists a super fiber functor
    \begin{equation}
        F_{\cA}: \cE_{\cA} \rightarrow \sVec
    \end{equation}
    and that $\cE_{\cA}\cong \Rep(G_{\cA}, z_{\cA})$, where $G_{\cA}=\text{Aut}^{\otimes}(F_{\cA})$ and $z_{\cA}\in Z(G_{\cA})$, $z_{\cA}^2=1$. (See Appendix~\ref{append:DelignesTheorem} for a review.) 
    
    To determine $G_{\cA}$, recall that by definition \eqref{eqn:NNW_defn_of_L}, $P_{\cA}$ is such that $\cE_{\cA} \cap \Rep(G) \cong \Rep(G/P_{\cA})=\Rep(Q_{\cA})$. The inclusion 
    \begin{equation}
        \Rep(Q_{\cA}) \hookrightarrow \Rep(G_{\cA}, z)
    \end{equation}
    corresponds to a surjection
    \begin{equation}
        \pi: G_{\cA} \twoheadrightarrow Q_{A}.
    \end{equation}
    De-equivariantization gives
    \begin{equation}
        \cV_{\cA}:=\cE_{\cA}/\Rep(Q_{\cA})\cong \Rep(\ker(\pi)).
    \end{equation}
    $\cV_{\cA}$ is pointed by Lemma~\ref{lem:pointedness_de-equiv}, moreover, $\cV_{\cA}$ is equivalent to its image under the forgetful functor $F_{G}:D(G, \omega) \rightarrow \Vect^\omega_G$ (see proof of Lemma~\ref{lem:pointedness_de-equiv} for notation). Hence $\cV_{\cA}$ has simple objects $\delta_r$ with $r\in R_{\cA}$. Fusion rules follow from group multiplication. Pointedness says that $\ker(\pi)$ is abelian, whose Pontryagin dual is $R_{\cA}$. Hence $\ker(\pi)$ is naturally isomorphic to its double Pontryagin dual, being
    \begin{equation}
        \ker(\pi) \cong \widehat R_{\cA}.
    \end{equation}
    This gives rise to the short exact sequence
    \begin{equation}
        1 \rightarrow \widehat R_{\cA} \rightarrow G_{\cA} \rightarrow Q_{\cA} \rightarrow 1.
    \end{equation}
    We have $G_{\cA} \cong \widehat R_{\cA}\rtimes_{\alpha_{\cA}} Q_{\cA}$, where $\alpha_{\cA}$ is the extension class, as claimed. As a consistency check, $\dim\Rep(G_{\cA}) = |G_{\cA}| = |R_{\cA}||Q_{\cA}| = \dim\cE_{\cA}$ where the last equality is from Lemma~\ref{lem:NNW_dim_subcat}.
    
    To identify the parity element $z_{\cA}$, on the object $\delta_r$ in $\cV_{\cA}\cong \Rep(\ker(\pi))$, the ribbon twist is
    \begin{equation}
        q_{\cA}(r) = B_{\cA}(r,r).
    \end{equation}
    By Lemma~\ref{lem:qA_character}, $q_{\cA}\in \widehat R_{\cA}$, $q_{\cA}^2=1$ and $q_{\cA}$ is $Q_{\cA}$-invariant. Hence $(q_{\cA}, \overline{1})$ is a central element of $G_{\cA}$ of order at most two.
    There exists a super fiber functor of whom $z_{\cA}$ is the parity automorphism: indeed, the symmetric braiding is the one for which $\theta_{X} = \rho_{X}(z)$ for every object $X$. We have computed $\theta_{\delta_{r}} = q_{\cA}(r)$ and under the isomorphism $\ker \pi \cong \widehat R_{\cA}$, the object $\delta_{r}$ corresponds to the character $\text{ev}_r:\widehat R_{\cA} \rightarrow U(1)$, so $\text{ev}_r(q_{\cA}) = q_{\cA}(r) = \theta_{\delta_r}$. Hence the parity element restricted to $\ker \pi$ is $q_{\cA}$. Being central (and $Q_{\cA}$-invariant) pins it down to $(q_{\cA},\overline{1})$ in $G_{\cA}$.
    We conclude that $\cE_{\cA} \cong \Rep(G_{\cA}, z_{\cA})$. 
\end{proof}

The above theorem shows that $G_{\cA}$ is an extension of $Q_{\cA}$ by $\widehat R_{\cA}$, but does not specify the extension class $\alpha_{\cA}$. We now turn to derive a formula for $\alpha_{\cA}$. 

Consider the composite tensor functor
\be\label{eq:audit-fiber-functor}
F_{\cA}^{\Vect}:\cE_{\cA}\longrightarrow\cV_{\cA}
\cong\Vect_{R_{\cA}}^{\omega|_{R_{\cA}}}
\xrightarrow{U_\nu}\Vect\,.
\ee
Here the first arrow is de-equivariantization by $\Rep(Q_{\cA})$, and $U_\nu$ is the forgetful functor. Its tensor structure $U_\nu(V)\otimes U_\nu(W)\to U_\nu(V\otimes W)$ acts by
$\nu(r,s)$ on degrees $r,s$. It satisfies $d\nu=\omega|_{R_{\cA}}^{-1}$ and
\be\label{eqn:nu_Bgrading}
\frac{\nu(r,s)}{\nu(s,r)}=B_{\cA}(r,s)(-1)^{|r||s|}\,,
\ee
where
\be\label{eqn:parity_q}
(-1)^{|r|}:=q_{\cA}(r)\,.
\ee
$U_{\nu}$ can be factorized as a symmetric functor $\Vect_{R_{\cA}}^{\omega|_{R_{\cA}}} \rightarrow \sVec$ preserving the parity \eqref{eqn:parity_q} composed with the forgetful functor $\sVec \rightarrow \Vect$.
From this symmetric functor $\Vect_{R_{\cA}}^{\omega|_{R_{\cA}}} \rightarrow \sVec$, $\cV_{\cA}\cong\Rep(\widehat R_{\cA},q_{\cA})$ as a symmetric category.

\begin{lemma}
\label{lem:nu_exist_unique}
A normalized $\nu$ satisfying $d\nu=\omega|_{R_{\cA}}^{-1}$
and~\eqref{eqn:nu_Bgrading} exists and is unique up to a coboundary.
Consequently $F_{\cA}^{\Vect}$ is determined up to tensor isomorphism.
\end{lemma}
\begin{proof}
The symmetric category $\cV_{\cA}\cong\Rep(\widehat R_{\cA},q_{\cA})$ admits a super fiber functor. Its monoidal structure is given by $\nu$. Coherence gives $d\nu=\omega|_{R_{\cA}}^{-1}$ and preservation of the symmetry gives~\eqref{eqn:nu_Bgrading}, hence existence.
If $\nu'$ is another 2-cochain such that $d\nu' = \omega|_{R_{\cA}}^{-1}$ and satisfies \eqref{eqn:nu_Bgrading}, $d(\nu'/\nu)=1$. Thus $\nu'/\nu$ is a coboundary.
\end{proof}

% We now fix a 2-cochain $\nu\in C^2(R_{\cA}, U(1))$ trivializing $\omega$ on $R_{\cA}$. This exists because $\cV_{\cA} \cong \Rep(\widehat R_{\cA})$ is symmetric and pointed hence has trivial associator, and on the other hand, $\cV_{\cA} \cong \Vect_{R_{\cA}}^{\omega|_{R_{\cA}}}$. Define
Define a 2-cocycle (checked by direct computation)
\begin{equation}
    \kappa_{g}(r_1, r_2) := \Omega_{g}(r_1, r_2) \frac{\nu(r_1, r_2)}{\nu({}^gr_1, {}^g r_2)}.
\end{equation}

With these, define 
\begin{equation}
    \cG_{\cA}:=\{(\varphi, g) \mid g\in G,\, \varphi:R_{\cA}\rightarrow U(1),\, d\varphi = \kappa_g\}
\end{equation}
where $d\varphi_{\overline{g}}(r_1, r_2) = \varphi(r_1)\varphi(r_2)/\varphi(r_1 r_2)$. $\cG_{\cA}$ is a group with multiplication
\begin{equation}
\label{eqn:cG_multiplication}
    (\varphi, g) (\varphi', g') = (\varphi' \cdot ({}^{g'}\varphi)\cdot \tau_{\bullet}(\omega)(g, g'), gg').
\end{equation}
Here, $({}^{g'}\varphi)(r) = \varphi({}^{g'}r)$ and $\tau_{\bullet}(\omega)(g,g')(r) = \tau_r(\omega)(g,g')$. 
Note that $\kappa_1 = 1$, and $\widehat R_{\cA}$ is contained in $\cG_{\cA}$ as $\{(\varphi, 1)\}$. 

Also define
\begin{equation}
    \Gamma_{\cA}:=\{ (\psi_{\cA}(-, p)^{-1}, p) \mid p\in P_{\cA} \}.
\end{equation}

\begin{proposition}
\label{prop:GcA=cG/Gamma}
    $\cG_{\cA}$ is a group of order $|R_{\cA}||G|$, contains $\Gamma_{\cA} \cong P_{\cA}$ as a normal subgroup and
    \begin{equation}
        G_{\cA} \cong \cG_{\cA}/\Gamma_{\cA}.
    \end{equation}
\end{proposition}
\begin{proof}
    Recall that an object of $D(G, \omega)$ is a $G$-graded vector space $V=\bigoplus_{x\in G}V_x$ with $G$-action, i.e., linear maps
    \begin{equation}
        \rho(g):V_{x} \rightarrow V_{{}^gx},
    \end{equation}
    such that $\rho(1)=1$ and
    \begin{equation}
        \rho(g) \rho(h)|_{V_x} = \tau_x(\omega)(g,h) \rho(gh)|_{V_{k}}.
    \end{equation}
    On a tensor product,
    \begin{equation}
    \label{eqn:rho_tensor}
        \rho_{V\otimes W}(g)|_{V_r \otimes W_{s}} = \Omega_g(r,s)(\rho_V(g) \otimes \rho_W(g)),
    \end{equation}
    where $\Omega_g$ \eqref{eqn:Omega} makes the above two expressions compatible:
    \begin{equation}
        \begin{split}
            & \Omega_g({}^hr, {}^hs) \Omega_h(r,s) \tau_{r}(\omega)(g,h) \tau_{s}(\omega)(g,h) \\
            & \quad = \tau_{rs}(\omega)(g,h) \Omega_{gh}(r,s).
        \end{split}
    \end{equation}
    In terms of these, $\cE_{\cA}=S(R_{\cA}, P_{\cA}, \psi_{\cA})$ is the full fusion subcategory of $D(G, \omega)$ of objects supported on $R_{\cA}$ with
    \begin{equation}
        \rho(p)|_{V_{r}} = \psi_{\cA}(r,p) \id
    \end{equation}
    for $p\in P_{\cA}$ and $r\in R_{\cA}$.
    
    We recover $G_{\cA}$, the group of automorphisms on the fiber functor, by constructing a surjective homomorphism $\Phi: \cG_{\cA} \twoheadrightarrow G_{\cA}$ and computing $\ker\Phi$ explicitly. 
    
Let $F$ forget the grading, with tensor constraint $\nu$ satisfying
$d\nu=\omega|_{R_{\cA}}^{-1}$ and~\eqref{eqn:nu_Bgrading}. 
    For $g\in G$ and a map $\varphi:R_{\cA} \rightarrow U(1)$ with $\varphi(1) =1$, define a linear endomorphism $\Phi(\varphi, g)_{V}$ on $F(V)$ for each object $V$ in $\cE_{\cA}$ as
    \begin{equation}
        \Phi(\varphi, g)_{V}(v):=\varphi(r) \rho(g)v \in V_{{}^gr}
    \end{equation}
    for $v\in V_r$. This is natural in $V$ since any morphism of $\cE_{\cA}$ is grading-preserving and commutes with $\rho$. It is monoidal if and only if $d\varphi = \kappa_g$: for $v\otimes w\in V_r \otimes W_s$, from the monoidal structure $\nu$ of $F$ and \eqref{eqn:rho_tensor}, we have
    \begin{equation}
        \varphi(rs) \nu(r,s) \Omega_g(r,s) = \nu({}^gr, {}^gs) \varphi(r) \varphi(s),
    \end{equation}
    i.e., $d\varphi(r,s) = \Omega_g(r,s) \nu(r,s)/\nu({}^gr, {}^gs) = \kappa_g(r,s)$. Hence $\Phi(\varphi, g) \in \Aut^{\otimes}(F)$ if and only if $(\varphi,g) \in \cG_{\cA}$.

    We now show that $\cG_{\cA}$ is a group of order $|R_{\cA}||G|$ and $\Phi$ defines a group homomorphism.
    For each $g$, there exists $\varphi$ such that $d\varphi = \kappa_g$: 
    for $g\in G$, let $T_g$ denote the braided conjugation $G$-action on $\cV_{\cA}$, then both $U_{\nu}$ and $U_{\nu} \circ T_g$ are super fiber functors on $\cV_{\cA}$, and are isomorphic by Deligne's Theorem~\ref{thm:Deligne}. Hence there exists scalar $\varphi(r)$ for each $r\in R_{\cA}$ such that the natural transformation whose $\delta_r$-component is $\varphi(r)$ defines the isomorphism from $U_{\nu}$ to $U_{\nu} \circ T_g$, which satisfies $d\varphi = \kappa_g$.
    Moreover, the possible $\varphi$'s form a torsor under $\widehat R_{\cA} = Z^1(R_{\cA}, U(1))$, hence $|\cG_{\cA}|=|R_{\cA}||G|$.

    The composition can be computed directly, for $v\in V_r$,
    \begin{equation}
        \begin{split}
            \Phi(\varphi, g) \circ \Phi(\varphi', g') v & = \varphi({}^{g'}r)\varphi'(r) \rho(g) \rho(g')v\\
            & = \varphi({}^{g'}r)\varphi'(r) \tau_{r}(\omega)(g,g') \rho(gg')v\\
            & =\Phi(\varphi'\cdot ({}^{g'} \varphi) \cdot \tau_{\bullet}(\omega)(g,g'), gg') v.
        \end{split}
    \end{equation}
    Hence $\Phi$ is a group homomorphism.

    We are left to show that $\Phi$ is surjective with kernel $\Gamma_{\cA}$. 
    An object $V$ of $\Rep(Q_{\cA}) = \cE_{\cA} \cap \Rep(G)$ is trivially graded $V_1$.
    %, on which $\rho(p) = \psi_{\cA}(1,p)\id = \id$, so $\rho$ factors through $Q_{\cA}$ via a $Q_{\cA}$-representation $\overline{\rho}$. 
    Since $\varphi(1)=1$, for $v\in V_{1}$
    % Concretely, restricting the automorphisms along $\Rep(Q_{\cA}) = \cE_{\cA}$
    % $\Phi(\varphi, g)$ to $\Rep(Q_{\cA})\hookrightarrow \cE_{\cA}$ acts on objects with trivial grading, 
    % \begin{equation}
    %     \Phi(\varphi, g)(v) = \varphi(1) u_g(v) = \rho_X(\overline{g})(v).
    % \end{equation}
    \begin{equation}
        \Phi(\varphi, g)(v) = \varphi(1) \rho(g)v = {\rho}({g})(v).
    \end{equation}
    It follows that $\pi(\Phi(\varphi, g)) = \overline{g}$ hence $\pi \circ \Phi$ surjects to $Q_{\cA}$, where $\pi:G_{\cA} \twoheadrightarrow Q_{\cA}$ is from the proof of Theorem~\ref{thm:Tannakian_group}. 
    $\Phi$ restricted to $\{(\chi, 1) \mid \chi \in \widehat R_{\cA}\}$ is an isomorphism onto $\ker \pi = \widehat R_{\cA}$.
    Moreover, $\im \Phi$ surjects onto $G_{\cA}/\ker \pi$: for $x\in G_{\cA}$, choose $y\in \im \Phi$ such that $\pi(y) = \pi(x)$, then $xy^{-1}\in \ker \pi \subseteq \im \Phi$.
    So $\text{Im}\Phi$ is a subgroup of $G_{\cA}$ containing $\ker\pi$ and surjects onto $G_{\cA}/\ker \pi$. Such a subgroup is the whole group and $\Phi$ is surjective onto $G_{\cA}$.

    Finally, to show $\ker\Phi = \Gamma_{\cA}$. %Any element $(\psi_{\cA}(-,p)^{-1},p)$ in $\Gamma_{\cA}$ is indexed by $p\in P_{\cA}$. $\psi_{\cA}(-,p)$ is an $\omega$-bicharacter hence $d(\psi_{\cA}(-,p)^{-1}) = \beta_{p} = \kappa_{p}$ on $R_{\cA}$
    For $(\varphi, g)\in \ker \Phi$, applying $\pi$ gives $\overline{g}=\overline{1}$, so $g\in P_{\cA}$, $\rho(g)|_{V_r} = \psi_{\cA}(r,g)$ and for $v\in V_r$, 
    \begin{equation}
        \Phi(\varphi, g)(v) = \varphi(r) \psi_{\cA}(r,g) v.
    \end{equation}
    This is the identity on every object of $\cE_{\cA}$ if and only if $\varphi(r) \psi_{\cA}(r,g)=1$ for every $r\in R_{\cA}$, $\varphi = \psi_{\cA}(-,g)^{-1}$ and $\ker \Phi \subseteq \Gamma_{\cA}$. Conversely, the same computation shows $\Phi(\varphi, g)$ is identity for $(\varphi, g) \in \Gamma_{\cA}$. Hence $\ker \Phi = \Gamma_{\cA}$ and we obtain the claim $G_{\cA} \cong \cG_{\cA}/\Gamma_{\cA}$.
\end{proof}

As a corollary, we obtain a formula for $\alpha_{\cA}$.
\begin{corollary}
    Let $\tilde{s}(\overline{g}) := (\varphi_{\overline{g}}, s(\overline{g}))$ be a section along $\pi \circ \Phi:\cG_{\cA} \twoheadrightarrow Q_{\cA}$ from Proposition~\ref{prop:GcA=cG/Gamma}, 
    \begin{equation}
    \label{eqn:alpha_formula}
        \begin{split}
            \alpha_{\cA}(\overline{g}, \overline{h})(r) & = \psi_{\cA}(r, f_s(\overline{g}, \overline{h})) \frac{\tau_u(\omega)(s(\overline{g}), s(\overline{h}))}{\tau_u(\omega)(f_s(\overline{g}, \overline{h}), s(\overline{gh}))} \\
            & \quad \times\frac{\varphi_{\overline{h}}(u)\varphi_{\overline{g}}({}^{s(\overline{h})}u)}{\varphi_{\overline{gh}}(u)}
        \end{split}
    \end{equation}
    where $u:=r^{s(\overline{gh})}$.
\end{corollary}
\begin{proof}
    Write $f=f_s(\overline{g}, \overline{h})$, from the group multiplication of $\cG_{\cA}$ \eqref{eqn:cG_multiplication}, section $\tilde{s}$ satisfies
    \begin{equation}
    \label{eqn:alpha_formula_sgsh}
        \tilde{s}(\overline{g}) \tilde{s}(\overline{h}) = \lb \varphi_{\overline{h}} \cdot ({}^{s(\overline{h})}\varphi_{\overline{g}}) \, \tau_{\bullet}(\omega)(s(\overline{g}), s(\overline{h})), f\, s(\overline{gh}) \rb, 
    \end{equation}
    while $\tilde{s}(\overline{gh}) = (\varphi_{\overline{gh}}, s(\overline{gh}))$. To compute $\alpha_{\cA}$, we multiply by $\gamma_{f}=(\psi_{\cA}(-,f)^{-1}, f) \in \Gamma_{\cA}$ and obtain $\gamma_{f}\tilde{s}(\overline{gh}) = \lb \varphi_{\overline{gh}}\cdot ({}^{s(\overline{gh})}\psi_{\cA}(-,f)^{-1}) \, \tau_{\bullet}(\omega)(f, s(\overline{gh})), f \, s(\overline{gh}) \rb$, whose second entry is the same as that in \eqref{eqn:alpha_formula_sgsh}. The first entries of \eqref{eqn:alpha_formula_sgsh} and $\gamma_{f}\tilde{s}(\overline{gh})$ differ by an element of $\widehat R_{\cA}$, which is $\alpha_{\cA}(\overline{g}, \overline{h})$ (this can be explicitly checked by applying $\Phi$).
\end{proof}

For $\omega=1$, equation~\eqref{eqn:alpha_formula} becomes
\be\label{eq:audit-untwisted-alpha}
\ba
\alpha_{\cA}(\overline g,\overline h)(r)
&=\psi_{\cA}(r,f_s(\overline g,\overline h))\frac{\varphi_{\overline h}(u)
\varphi_{\overline g}({}^{s(\overline h)}u)}
{\varphi_{\overline{gh}}(u)}\,,
\ea
\ee
where $u=r^{s(\overline{gh})}$. If $\nu$ can be chosen to be $G$-invariant, then $\kappa_g=1$ and we fix $\varphi_{\overline g}=1$, the above reduces to
\be\label{eq:audit-invariant-nu-alpha}
\alpha_{\cA}(\overline g,\overline h)
=\psi_{\cA}(-,f_s(\overline g,\overline h))\,.
\ee
This is indeed the case when $B_{\cA}(r,s)(-1)^{|r||s|}=1$ on $R_{\cA}$, by choosing $\nu=1$, but is not automatic in general.

\subsection{Proliferation Transition and $\cM_{\cA}$ $\cN_{\cA}$ Construction}
\label{sec:MANA}
The proliferation transition construction of \cite{Schafer-Nameki:2026hgx} starts from a condensable algebra $\cA$ in an MTC $\cT$, and builds a SymTFT from the transparent lines $\cE_{\cA}:=\cZ_2(\langle \cA \rangle)$. When the M\"uger center $\cE_{\cA}$ is Tannakian, hence equivalent to $\Rep(G_{\cA})$ for some finite group $G_{\cA}$, one defines
\begin{equation}
    \cM_\cA:=\langle \cA \rangle /\cE_{\cA}, \quad \cN_{\cA}:=\langle \cA \rangle' / \cE_{\cA},
\end{equation}
where ``$/$'' denotes de-equivariantization by the Tannakian category $\cE_{\cA}$. That is, if $\cE_{\cA}\cong \Rep(G_{\cA})$, denote by $\bC^{G_{\cA}}$ its regular algebra, $\cM_{\cA}$ is the category of $\bC^{G_{\cA}}$-modules in $\langle \cA \rangle$. Similarly, $\cN_{\cA}$ is the category of $\bC^{G_{\cA}}$-modules in $\langle \cA \rangle'$.
Physically, $\cM_{\cA}$ consists of the lines of $\langle\cA\rangle$ that are not transparent: they braid non-trivially with some anyon generated by the condensate, hence cannot be lifted into the bulk of the SymTFT, and lives on the symmetry boundary. $\cN_{\cA}$ consists of the anyons of $\cT$ that braid trivially with all of $\langle\cA\rangle$ and lives on the physical boundary. In terms of these data \cite{Schafer-Nameki:2026hgx},
\be
\label{T-and-TA-from-MN}
\cT\cong\frac{\cM_{\cA}\boxtimes\cN_{\cA}}{G_{\cA}^{(0)}}\,,\qquad \frac{\cT}{\cA}\cong\frac{\cM_{\cA}/\widehat\cA\boxtimes\cN_{\cA}}{H_{\cA}^{(0)}} \,,
\ee
where $\widehat\cA$ is the imega of $\cA$ under the condensation map $\langle \cA \rangle \rightarrow \cM_{\cA}$.

Since $\cE_{\cA}$ is the M\"uger center of both $\langle \cA \rangle$ and $\langle \cA \rangle'$, both $\cM_{\cA}$ and $\cN_{\cA}$ are %non-degenerate 
modular \cite{Bruguieres2000, MUGER2004256}. 
Moreover, $\cM_{\cA}$ and $\cN_{\cA}$ have dual $G_{\cA}^{(0)}$ $0$-form symmetry in \cite{Schafer-Nameki:2026hgx}. Mathematically, this follows from that the de-equivariantization carries a $G_{\cA}$-action \cite{DGNO}.

% Conceptually, $\cM_{\cA}$ is the non-transparent content generated by the proliferating algebra, and $\cN_{\cA}$ is the complementary modular content in the starting TO $\cT$.

Now, consider the MTC $\cT=D(G, \omega)$ for some finite group $G$ and a (normalized) 3-cocycle $\omega$. Recall $\langle \cA \rangle = S(K_{\cA}, L_{\cA}, B_{\cA})$ for a condensable algebra $\cA$ in $D(G, \omega)$. We further denote the intersection 
\begin{equation}
    I_{\cA}:=K_{\cA}\cap L_{\cA}
\end{equation}
and the product group $P_{\cA}:=K_{\cA}L_{\cA}$. In what follows, we assume $\cZ_2(\langle \cA \rangle )$ is Tannakian, i.e., (by Prop~\ref{prop:Tannakian_criterion}) $B_{\cA}(r,r)=1$ for every $r\in R_{\cA}$, $R_{\cA} \subseteq I_{\cA}$ defined in \eqref{eqn:defn_RA}. $\cM_{\cA}$ and $\cN_{\cA}$ have the following description.

\begin{proposition}
\label{prop:MANA}
    Given a condensable algebra $\cA$ in $D(G, \omega)$ such that $\cZ_2(\langle \cA \rangle)$ is Tannakian. Then
    \begin{equation}
        \begin{split}
            \cM_{\cA} & \cong \lb \Vect_{K_{\cA}/R_{\cA}}^{\overline{\omega}_{\cM}} \rb^{K_{\cA}/I_{\cA}}, \\
            \cN_{\cA} & \cong \lb \Vect_{L_{\cA}/R_{\cA}}^{\overline{\omega}_{\cN}} \rb^{L_{\cA}/I_{\cA}},
        \end{split}
    \end{equation}
    for some normalized 3-cocycles $\overline{\omega}_{\cM}$ and $\overline{\omega}_{\cN}$. Here the categorical $K_{\cA}/I_{\cA}$- and $L_{\cA}/I_{\cA}$-actions are induced by conjugation.
\end{proposition}
\begin{proof}
    We prove the case for $\cM_{\cA}$, the proof for the case of $\cN_{\cA}$ follows after swapping the role of $K_{\cA}$ and $L_{\cA}$.
    
    Recall from \eqref{eqn:S_equivr_pointed}, 
    \begin{equation}
        \langle \cA \rangle = S(K_{\cA}, L_{\cA}, B_{\cA}) \cong \lb \Vect_{K_{\cA}}^{\omega|_{K_{\cA}}} \rb^{G/L_{\cA}}. 
    \end{equation}
    We compute its de-equivariantization in two steps:
    \begin{equation}
        \cM_{\cA} = \langle \cA \rangle /\cE_{\cA} = (\langle \cA \rangle /\Rep(G/P_{\cA}))/\Rep(\widehat R_{\cA})
    \end{equation}
    i.e., de-equivariantize with respect to $\Rep(G/P_{\cA}) = \Rep(G_{\cA}/\widehat R_{\cA})$ first, then de-equivariantize with respect to $\Rep(\widehat R_{\cA})$.

    Because $\cE_{\cA}$ is transparent in $\langle \cA \rangle$ (also in $\langle \cA \rangle'$), the de-equivariantization $\cM_{\cA}$ (also $\cN_{\cA}$) is braided with $G_{\cA}$ acting by braided autoequivalences \cite[Thm.~4.18, Prop.~4.22]{DGNO}. $\cM_{\cA}$ and $\cN_{\cA}$ are non-degenerate by \cite[Cor.~4.31]{DGNO}. Hence both $\cM_{\cA}$ and $\cN_{\cA}$ are modular. Next, we examine their structure as fusion categories.

    The intersection of $\cE_{\cA} = S(R_{\cA}, P_{\cA}, \psi_{\cA})$ with $\Rep(G)$ is $\cE_{\cA} \cap \Rep(G) = \Rep(G/P_{\cA})$. By the isomorphism theorems of groups, $P_{\cA}/L_{\cA}$ is embedded in $G/L_{\cA}$ as the kernel of the projection $G/L_{\cA} \twoheadrightarrow G/P_{\cA}$ and
    \begin{equation}
        P_{\cA}/L_{\cA} \cong K_{\cA}/I_{\cA}.
    \end{equation}
    Hence, the de-equivariantization 
    \begin{equation}
        \langle\cA \rangle/\Rep(G/P_{\cA}) \cong \lb \Vect_{K_{\cA}}^{\omega|_{K_{\cA}}} \rb^{K_{\cA}/I_{\cA}}. 
    \end{equation}
    We now de-equivariantize with respect to $\Rep(\widehat R_{\cA})\cong \cE_{\cA}/\Rep(G/P_{\cA})$. Under the identification $\Rep(\widehat R_{\cA}) \cong \Vect_{R_{\cA}} \hookrightarrow \Vect_{K_{\cA}}^{\omega|_{K_{\cA}}}$, the regular algebra decomposes as
    \begin{equation}
        A_R:= \bigoplus_{r\in R_{\cA}} \delta_r.
    \end{equation}
    $R_{\cA}$ centralizes both $K_{\cA}$ and $L_{\cA}$, hence %the de-equivariantization, i.e., the category of $A_{R}$ modules in 
    the category of $A_R$-modules in $\Vect_{K_{\cA}}^{\omega|_{K_{\cA}}}$ is pointed: its simple objects are 
    \begin{equation}
        X_{\overline{k}} := \bigoplus_{r\in R_{\cA}} \delta_{kr}
    \end{equation}
    where $k$ is a representative of $\overline{k}\in K_{\cA}/R_{\cA}$. Note that the monoidal structure agrees with the group multiplication $X_{\overline{k}} \otimes_{A_R} X_{\overline{h}} \cong X_{\overline{kh}}$. 
This category is equivalent to
$\Vect_{K_{\cA}/R_{\cA}}^{\overline\omega_{\cM}}$ for a normalized
3-cocycle $\overline\omega_{\cM}$, whose construction uses $\omega$ together
with the multiplication and central structure of $A_R$ inherited from
$\cE_{\cA}$. Similarly for $\overline{\omega}_{\cN}$.

    Concretely, for multiplication $m$ with component
    \begin{equation}
        m_{\overline{k}, \overline{h}} : X_{\overline{k}} \otimes_{A_R} X_{\overline{h}} \rightarrow X_{\overline{kh}},
    \end{equation}
    $\overline{\omega}(\overline{g}, \overline{h}, \overline{k})$ is the difference between $m_{\overline{gh}, \overline{k}}\circ (m_{\overline{g}, \overline{h}} \otimes \id)$ and $m_{\overline{g}, \overline{hk}}\circ (\id \otimes m_{\overline{h}, \overline{k}}) \circ \omega'$, where $\omega'$ is induced from $\omega$. 

    The $K_{\cA}/I_{\cA}$-action fixes $A_R$, hence descends to the category of $A_R$-modules $\Vect_{K_{\cA}/R_{\cA}}^{\overline{\omega}}$. This shows
    \begin{equation}
        \cM_{\cA} \cong \lb \Vect_{K_{\cA}/R_{\cA}}^{\overline{\omega}_{\cM}} \rb^{K_{\cA}/I_{\cA}}.
    \end{equation}
    The same argument with $K_{\cA}$ and $L_{\cA}$ interchanged proves the result for $\cN_{\cA}$.
\end{proof}

The dimensions are
\be
\label{dim-MA-NA}
\dim\cM_{\cA}=\frac{|K_{\cA}|^2}{|R_{\cA}|\,|I_{\cA}|}\,,\qquad \dim\cN_{\cA}=\frac{|L_{\cA}|^2}{|R_{\cA}|\,|I_{\cA}|}\,,
\ee
in agreement with $\dim\cM_{\cA}=\dim\langle\cA\rangle/|G_{\cA}|$ and $\dim\cN_{\cA}=\dim\langle\cA\rangle'/|G_{\cA}|$.

The following special cases are immediate consequence of the above proof.
\begin{corollary}
Given condensable algebra $\cA$ in $D(G, \omega)$ such that $\cZ_2(\langle \cA \rangle)$ is Tannakian.
    \begin{enumerate}[leftmargin=3.7ex]
        \item if $R_{\cA}=1$, then
    \begin{equation}
        \cM_{\cA} \cong \lb \Vect_{K_{\cA}}^{\omega|_{K_{\cA}}} \rb^{K_{\cA}/I_{\cA}}, \quad \cN_{\cA} \cong \lb \Vect_{L_{\cA}}^{\omega|_{L_{\cA}}} \rb^{L_{\cA}/I_{\cA}}.
    \end{equation}
    
    \item If $\omega=1$ and $\epsilon=1$, then $B_{\cA}=1$ and $R_{\cA}=I_{\cA}$, 
    \begin{equation}
        \cM_{\cA} \cong D(K_{\cA}/I_{\cA}), \quad \cN_{\cA} \cong D(L_{\cA}/I_{\cA}).
    \end{equation}
    \end{enumerate}
\end{corollary}
We emphasize that when $\omega=1$ and $\epsilon=1$, the above expresses $\cM_{\cA}$ and $\cN_{\cA}$ as Drinfeld centers $D(G)$ with trivial $\omega$. We give one more criterion which guarantees $\cM_{\cA}$ and $\cN_{\cA}$ to be of the form $D(G, \omega)$.
\begin{proposition}
    The canonical Tannakian subcategory $\Rep(K_{\cA}/I_{\cA})$ in $\cM_{\cA}$ has centralizer $\Rep(K_{\cA}/I_{\cA})' = \Rep(K_{\cA}/I_{\cA})$ (such $\Rep(K_{\cA}/I_{\cA})$ is called \textbf{Lagrangian}) if and only if
    \begin{equation}
        R_{\cA} = I_{\cA},
    \end{equation}
    equivalently, this is true if and only if
    \begin{equation}
        \varphi_{\cA}|_{I_{\cA}\times I_{\cA}} = 1.
    \end{equation}
    When this is the case, 
    \begin{equation}
        \begin{split}
            \cM_{\cA} & \cong D(K_{\cA}/I_{\cA}, \overline{\omega}_{\cM}), \\
            \cN_{\cA} & \cong D(L_{\cA}/I_{\cA}, \overline{\omega}_{\cN}).
        \end{split}
    \end{equation}
\end{proposition}
\begin{proof}
    By Proposition 4.58 from \cite{DGNO}, $\Rep(K_{\cA}/I_{\cA})$ in $\cM_{\cA}$ is Lagrangian, i.e., $\Rep(K_{\cA}/I_{\cA})' = \Rep(K_{\cA}/I_{\cA})$, if and only if the neutral component (as a $K_{\cA}/I_{\cA}$-graded category) $\cC_1$ of the de-equivariantization $\cC=\cM_{\cA}/\Rep(K_{\cA}/I_{\cA})$ is trivial.
    Here $\cC=\Vect_{K_{\cA}/R_{\cA}}^{\overline{\omega}_{\cM}}$ with the $K_{\cA}/I_{\cA}$-grading induced from the projection $K_{\cA}/R_{\cA} \twoheadrightarrow K_{\cA}/I_{\cA}$, so the neutral component
    \begin{equation}
        \cC_1 = \Vect_{I_{\cA}/R_{\cA}},
    \end{equation}
    which is trivial if and only if $R_{\cA}=I_{\cA}$.

    By construction of $R_{\cA} \subseteq I_{\cA}$ \eqref{eqn:defn_RA}, $R_{\cA}=I_{\cA}$ if and only if $\varphi_{\cA}|_{I_{\cA}\times I_{\cA}} = 1$.

    When $R_{\cA}=I_{\cA}$, \cite[Prop.~4.61]{DGNO} shows a braided equivalence 
    \begin{equation}
        \cM_{\cA} = \lb \Vect_{K_{\cA}/I_{\cA}}^{\overline{\omega}_{\cM}} \rb^{K_{\cA}/I_{\cA}} \cong D(K_{\cA}/I_{\cA}, \overline{\omega}_{\cM}),
    \end{equation}
    and similarly $\cN_{\cA}\cong D(L_{\cA}/I_{\cA}, \overline{\omega}_{\cN})$.
\end{proof}

\subsection{Summary}
\label{sec:math_summary}
We now summarize each structure studied in this section and their corresponding physics meaning.

Classification data of a condensable algebra $\cA = \cA(H, N, \gamma, \epsilon)$ in $D(G, \omega)$:
\begin{itemize}
    \item $H\subseteq G$. The (pure) charge part of $\cA$ (its trivially-graded component) is $\cA_1 = \Fun(G/H)$, the algebra of functions on the $G$-set $G/H$.

    \item $N\trianglelefteq H$. The flux part of $\cA$.

    \item $\gamma\in C^2(N, U(1))$ with $d\gamma = \omega|_N$. An $N$-SPT stacked before gauging.

    \item $\epsilon:H\times N \rightarrow U(1)$ satisfying extra compatibility conditions \eqref{eqn:epsilon_compatibility}. $H$-equivariant factor \eqref{eqn:Alg_span_relation}.
\end{itemize}

Data for fusion subcategory $\langle \cA \rangle = S(K_{\cA}, L_{\cA}, B_{\cA})$ in $D(G, \omega)$ generated by $\cA$:
\begin{itemize}
    \item $K_{\cA} = \langle {}^gN \mid g\in G \rangle$. Flux sectors present in $\langle \cA\rangle$.
    
    \item $L_{\cA} \subset \Core_G(C_H(N))$ defined as \eqref{eqn:defn_LA}. It consists of elements that acts on each flux in $\cA$ by a scalar, equivalently, $\langle \cA \rangle \cap \Rep(G) = \Rep(G/L_{\cA})$.

    \item $B_{\cA}:K_{\cA} \times L_{\cA}\rightarrow U(1)$ defined in \eqref{eqn:defn_B_cA}. The scalar $B_{\cA}(k,l)$ by which $l\in L_{\cA}$ acts on the $k$-component of the algebra.
\end{itemize}

M\"uger center $\cE_{\cA}:=\cZ_2(\langle \cA \rangle ) = S(R_{\cA}, P_{\cA}, \psi_{\cA})$ (the transparent lines) as a fusion subcategory of $D(G, \omega)$:
\begin{itemize}
    \item $R_{\cA} \subseteq K_{\cA} \cap L_{\cA}$ defined in \eqref{eqn:defn_RA}. Flux for the transparent lines.
    \item $P_{\cA}= K_{\cA}L_{\cA}$. Elements acting on each flux by a scalar.

    \item $\psi_{\cA}$ defined in \eqref{eqn:defn_psiA}. Scalar by which $P_{\cA}$ acts on $R_{\cA}$.
\end{itemize}

Data for the M\"uger center as a symmetric (Tannakian or super-Tannakian) category $\cE_{\cA} \cong \Rep(G_{\cA}, z_{\cA})$, hence the symmetry:
\begin{itemize}
    \item $G_{\cA} = \widehat R_{\cA} \rtimes_{\alpha_{\cA}} (G/P_{\cA})$, $\alpha_{\cA}\in Z^2(G/P_{\cA}, \widehat R_{\cA})$. The symmetry group for the bulk SymTFT.
    \item $z_{\cA} = (q_{\cA}, 1) \in G_{\cA}$, $q_{\cA}(r) = B_{\cA}(r,r)$: a distinguished central element of order at most two, which is the fermion parity element. Note that $\cZ_2(\langle \cA \rangle)$ is bosonic/Tannakian (i.e., $z_{\cA}=1$) if and only if $B_{\cA}(r,r)=1$ for all $r\in R_{\cA}$.
\end{itemize}

When $\cE_{\cA}$ is Tannakian/bosonic, denote by $\langle \cA \rangle'$ the centralizer \eqref{eqn:defn_centralizer} of $\langle \cA\rangle$, then (see Prop.~\ref{prop:MANA} for details)
\begin{itemize}
    \item $\cM_{\cA}=\langle \cA \rangle /\cE_{\cA} \cong \lb \Vect_{K_{\cA}/R_{\cA}}^{\overline{\omega}_{\cM}} \rb^{K_{\cA}/I_{\cA}}$,
    \item $\cN_{\cA} = \langle \cA \rangle'/\cE_{\cA} \cong \lb \Vect_{L_{\cA}/R_{\cA}}^{\overline{\omega}_{\cN}} \rb^{L_{\cA}/I_{\cA}}$,
\end{itemize}
are two $G_{\cA}$-symmetric MTCs, which can be stacked on a Dirichlet boundary before gauging the $G_{\cA}$ 0-form symmetry to obtain a symmetry boundary.

%%%%%%%%%%%%%%%%%%%%%%%%%%%%%%%%%%%%%

\twocolumngrid
\bibliographystyle{ytphys}
\small 
\baselineskip=.7\baselineskip
\let\bbb\bibitem\def\bibitem{\itemsep3.3pt\bbb}
\bibliography{ref}

\providecommand{\href}[2]{#2}\begingroup\raggedright\begin{thebibliography}{10}

\bibitem{Moore:1989yh}
G.~W. Moore and N.~Seiberg, ``{Taming the Conformal Zoo},''
  \href{http://dx.doi.org/10.1016/0370-2693(89)90897-6}{{\em Phys. Lett. B}
  {\bfseries 220} (1989) 422--430}.

\bibitem{Bais:2008ni}
F.~A. Bais and J.~K. Slingerland, ``{Condensate induced transitions between
  topologically ordered phases},''
  \href{http://dx.doi.org/10.1103/PhysRevB.79.045316}{{\em Phys. Rev. B}
  {\bfseries 79} (2009) 045316},
  \href{http://arxiv.org/abs/0808.0627}{{\ttfamily arXiv:0808.0627
  [cond-mat.mes-hall]}}.

\bibitem{Kong:2014condensation}
L.~Kong, ``{Anyon condensation and tensor categories},''
  \href{http://dx.doi.org/10.1016/j.nuclphysb.2014.07.003}{{\em Nucl. Phys. B}
  {\bfseries 886} (2014) 436--482},
  \href{http://arxiv.org/abs/1307.8244}{{\ttfamily arXiv:1307.8244
  [cond-mat.str-el]}}.

\bibitem{Lan:2017grs}
T.~Lan and X.-G. Wen, ``{Hierarchy Construction and Non-Abelian Families of
  Generic Topological Orders},''
  \href{http://dx.doi.org/10.1103/PhysRevLett.119.040403}{{\em Phys. Rev.
  Lett.} {\bfseries 119} no.~4, (2017) 040403},
  \href{http://arxiv.org/abs/1701.07820}{{\ttfamily arXiv:1701.07820
  [cond-mat.str-el]}}.

\bibitem{Hsin:2018vcg}
P.-S. Hsin, H.~T. Lam, and N.~Seiberg, ``{Comments on One-Form Global
  Symmetries and Their Gauging in 3d and 4d},''
  \href{http://dx.doi.org/10.21468/SciPostPhys.6.3.039}{{\em SciPost Phys.}
  {\bfseries 6} no.~3, (2019) 039},
  \href{http://arxiv.org/abs/1812.04716}{{\ttfamily arXiv:1812.04716
  [hep-th]}}.

\bibitem{Burnell:2017otf}
F.~J. Burnell, ``{Anyon condensation and its applications},''
  \href{http://dx.doi.org/10.1146/annurev-conmatphys-033117-054154}{{\em Ann.
  Rev. Condensed Matter Phys.} {\bfseries 9} (2018) 307--327},
  \href{http://arxiv.org/abs/1706.04940}{{\ttfamily arXiv:1706.04940
  [cond-mat.str-el]}}.

\bibitem{Schafer-Nameki:2023jdn}
S.~Schafer-Nameki, ``{ICTP lectures on (non-)invertible generalized
  symmetries},'' \href{http://dx.doi.org/10.1016/j.physrep.2024.01.007}{{\em
  Phys. Rept.} {\bfseries 1063} (2024) 1--55},
  \href{http://arxiv.org/abs/2305.18296}{{\ttfamily arXiv:2305.18296
  [hep-th]}}.

\bibitem{Shao:2023gho}
S.-H. Shao, ``{What's Done Cannot Be Undone: TASI Lectures on Non-Invertible
  Symmetries},'' in {\em {Theoretical Advanced Study Institute in Elementary
  Particle Physics 2023}: {Aspects of Symmetry}}.
\newblock 8, 2023.
\newblock \href{http://arxiv.org/abs/2308.00747}{{\ttfamily arXiv:2308.00747
  [hep-th]}}.

\bibitem{Cheng:2026qax}
M.~Cheng and N.~Seiberg, ``{Proliferation transitions from a topological phase
  in $2+1$ dimensions},'' \href{http://arxiv.org/abs/2603.00245}{{\ttfamily
  arXiv:2603.00245 [cond-mat.str-el]}}.

\bibitem{Schafer-Nameki:2026hgx}
S.~Schafer-Nameki, Y.~Zheng, and A.~Antinucci, ``{Proliferation Transitions for
  Non-Abelian Anyons},'' \href{http://arxiv.org/abs/2608.12303}{{\ttfamily
  arXiv:2608.12303 [cond-mat.str-el]}}.

\bibitem{Cheng:2026dxn}
M.~Cheng and N.~Seiberg, ``{Peeling off Abelian Anyons and Proliferation
  Transitions},'' \href{http://arxiv.org/abs/2610.04704}{{\ttfamily
  arXiv:2610.04704 [cond-mat.str-el]}}.

\bibitem{Zhang:2024bye}
C.~Zhang, A.~Vishwanath, and X.-G. Wen, ``{Hierarchy construction for
  non-Abelian fractional quantum Hall states via anyon condensation},''
  \href{http://dx.doi.org/10.1103/jndb-435f}{{\em Phys. Rev. B} {\bfseries 112}
  no.~12, (2025) 125116}, \href{http://arxiv.org/abs/2406.12068}{{\ttfamily
  arXiv:2406.12068 [cond-mat.str-el]}}.

\bibitem{Ji:2026yfj}
W.~Ji, R.~A. Lanzetta, Z.~Zhou, and C.~Wang, ``{Self-dual Higgs transitions:
  Toric code and beyond},'' \href{http://arxiv.org/abs/2601.20945}{{\ttfamily
  arXiv:2601.20945 [cond-mat.str-el]}}.

\bibitem{Cordova:2025eim}
C.~Cordova, D.~Garc{\'\i}a-Sep{\'u}lveda, and K.~Ohmori, ``{Higgsing
  Transitions from Topological Field Theory {\&} Non-Invertible Symmetry in
  Chern-Simons Matter Theories},''
  \href{http://arxiv.org/abs/2504.03614}{{\ttfamily arXiv:2504.03614
  [hep-th]}}.

\bibitem{Cordova:2025zkz}
C.~Cordova, D.~Garc{\'\i}a-Sep{\'u}lveda, and J.~A. Harvey, ``{Generalized
  Level-Rank Duality, Holomorphic Conformal Field Theory, and Non-Invertible
  Anyon Condensation},'' \href{http://arxiv.org/abs/2512.24419}{{\ttfamily
  arXiv:2512.24419 [hep-th]}}.

\bibitem{Lu:2026fid}
D.-C. Lu, C.~Wang, and A.~Vishwanath, ``{Self-dual $S_3$ gauge theory in 2+1d:
  lattice model and topological phase transitions},''
  \href{http://arxiv.org/abs/2608.05294}{{\ttfamily arXiv:2608.05294
  [cond-mat.str-el]}}.

\bibitem{Garcia-Sepulveda:2026fud}
D.~Garc{\'\i}a-Sep{\'u}lveda and D.-C. Lu, ``{Anyon Proliferation and Anyon
  Superconductivity in Higgsing Transitions via Conformal Embeddings},''
  \href{http://arxiv.org/abs/2610.00452}{{\ttfamily arXiv:2610.00452
  [cond-mat.str-el]}}.

\bibitem{Hung:2015hfa}
L.-Y. Hung and Y.~Wan, ``{Generalized ADE classification of topological
  boundaries and anyon condensation},''
  \href{http://dx.doi.org/10.1007/JHEP07(2015)120}{{\em JHEP} {\bfseries 07}
  (2015) 120}, \href{http://arxiv.org/abs/1502.02026}{{\ttfamily
  arXiv:1502.02026 [cond-mat.str-el]}}.

\bibitem{Zhao:2024ilc}
Y.~Zhao and Y.~Wan, ``{Nonabelian anyon condensation in 2+1d topological
  orders: A string-net model realization},''
  \href{http://dx.doi.org/10.1007/JHEP05(2025)156}{{\em JHEP} {\bfseries 05}
  (2025) 156}, \href{http://arxiv.org/abs/2409.05852}{{\ttfamily
  arXiv:2409.05852 [cond-mat.str-el]}}.

\bibitem{Zhao:2025zsb}
Y.~Zhao and Y.~Wan, ``{Landau-Ginzburg Paradigm of Topological Phases},''
  \href{http://arxiv.org/abs/2506.05319}{{\ttfamily arXiv:2506.05319
  [cond-mat.str-el]}}.

\bibitem{Cheng:2026hgu}
M.~Cheng, ``{Note on proliferation transitions of non-Abelian anyons},''
  \href{http://arxiv.org/abs/2608.20604}{{\ttfamily arXiv:2608.20604
  [cond-mat.str-el]}}.

\bibitem{Hung:2025gcp}
L.-Y. Hung, K.~Ji, C.~Shen, Y.~Wan, and Y.~Zhao, ``{A 2D-CFT Factory: Critical
  Lattice Models from Competing Anyon Condensation Processes in
  SymTO/SymTFT},'' \href{http://arxiv.org/abs/2506.05324}{{\ttfamily
  arXiv:2506.05324 [cond-mat.str-el]}}.

\bibitem{Bombin:2007qv}
H.~Bombin and M.~A. Martin-Delgado, ``{A Family of Non-Abelian Kitaev Models on
  a Lattice: Topological Confinement and Condensation},''
  \href{http://dx.doi.org/10.1103/PhysRevB.78.115421}{{\em Phys. Rev. B}
  {\bfseries 78} (2008) 115421},
  \href{http://arxiv.org/abs/0712.0190}{{\ttfamily arXiv:0712.0190
  [cond-mat.str-el]}}.

\bibitem{Kang:2020avc}
B.~Kang, S.~A. Parameswaran, A.~C. Potter, R.~Vasseur, and S.~Gazit,
  ``{Superuniversality from disorder at two-dimensional topological phase
  transitions},'' \href{http://dx.doi.org/10.1103/PhysRevB.102.224204}{{\em
  Phys. Rev. B} {\bfseries 102} no.~22, (2020) 224204},
  \href{http://arxiv.org/abs/2008.09617}{{\ttfamily arXiv:2008.09617
  [cond-mat.str-el]}}.

\bibitem{Flores-Calderon:2026bue}
R.~Flores-Calder{\'o}n, F.~Pollmann, and M.~Knap, ``{Non-Abelian Anyon
  Condensation: a Path-Integral Monte Carlo Approach},''
  \href{http://arxiv.org/abs/2609.19282}{{\ttfamily arXiv:2609.19282
  [cond-mat.str-el]}}.

\bibitem{Bhardwaj:2023ayw}
L.~Bhardwaj and S.~Schafer-Nameki, ``{Generalized charges, part II:
  Non-invertible symmetries and the symmetry TFT},''
  \href{http://dx.doi.org/10.21468/SciPostPhys.19.4.098}{{\em SciPost Phys.}
  {\bfseries 19} no.~4, (2025) 098},
  \href{http://arxiv.org/abs/2305.17159}{{\ttfamily arXiv:2305.17159
  [hep-th]}}.

\bibitem{Bhardwaj:2024qiv}
L.~Bhardwaj, D.~Pajer, S.~Schafer-Nameki, A.~Tiwari, A.~Warman, and J.~Wu,
  ``{Gapped phases in (2+1)d with non-invertible symmetries: Part I},''
  \href{http://dx.doi.org/10.21468/SciPostPhys.19.2.056}{{\em SciPost Phys.}
  {\bfseries 19} no.~2, (2025) 056},
  \href{http://arxiv.org/abs/2408.05266}{{\ttfamily arXiv:2408.05266
  [hep-th]}}.

\bibitem{Wen:2024qsg}
R.~Wen, ``{String condensation and topological holography for 2+1D gapless
  SPT},'' \href{http://arxiv.org/abs/2408.05801}{{\ttfamily arXiv:2408.05801
  [cond-mat.str-el]}}.

\bibitem{Bhardwaj:2025piv}
L.~Bhardwaj, S.~Schafer-Nameki, A.~Tiwari, and A.~Warman, ``{Gapped Phases in
  (2+1)d with Non-Invertible Symmetries: Part II},''
  \href{http://arxiv.org/abs/2502.20440}{{\ttfamily arXiv:2502.20440
  [hep-th]}}.

\bibitem{Bhardwaj:2025jtf}
L.~Bhardwaj, Y.~Gai, S.-J. Huang, K.~Inamura, S.~Schafer-Nameki, A.~Tiwari, and
  A.~Warman, ``{Gapless Phases in (2+1)d with Non-Invertible Symmetries},''
  \href{http://arxiv.org/abs/2503.12699}{{\ttfamily arXiv:2503.12699
  [cond-mat.str-el]}}.

\bibitem{Deligne:2002cat}
P.~Deligne, ``{Cat\'egories tensorielles},'' {\em Mosc. Math. J.} {\bfseries 2}
  (2002) 227--248.

\bibitem{Dijkgraaf:1989pz}
R.~Dijkgraaf and E.~Witten, ``{Topological Gauge Theories and Group
  Cohomology},'' \href{http://dx.doi.org/10.1007/BF02096988}{{\em Commun. Math.
  Phys.} {\bfseries 129} (1990) 393}.

\bibitem{davydov2009modular}
A.~Davydov, ``Modular invariants for group-theoretical modular data. {I},''
  {\em J. Algebra} {\bfseries 323} (2010) 1321--1348,
  \href{http://arxiv.org/abs/0908.1044}{{\ttfamily arXiv:0908.1044 [math.QA]}}.

\bibitem{davydov2017lagrangian}
A.~Davydov and D.~Simmons, ``On lagrangian algebras in group-theoretical
  braided fusion categories,'' {\em Journal of Algebra} {\bfseries 471} (2017)
  149--175.

\bibitem{Gai:2026hjk}
Y.~Gai, S.~Schafer-Nameki, and A.~Warman, ``{Twin Algebras: Condensable
  Algebras beyond Anyons},'' \href{http://arxiv.org/abs/2605.31602}{{\ttfamily
  arXiv:2605.31602 [cond-mat.str-el]}}.

\bibitem{Warman:2026gfz}
A.~Warman, Y.~Gai, and S.~Schafer-Nameki, ``{Twin Phases: Intrinsic Deconfined
  Quantum Criticality},'' \href{http://arxiv.org/abs/2605.31601}{{\ttfamily
  arXiv:2605.31601 [cond-mat.str-el]}}.

\bibitem{Pollmann:2012}
F.~Pollmann and A.~M. Turner, ``Detection of symmetry-protected topological
  phases in one dimension,''
  \href{http://dx.doi.org/10.1103/PhysRevB.86.125441}{{\em Physical Review B}
  {\bfseries 86} no.~12, (Sept., 2012) }.

\bibitem{Davydov:2013xov}
A.~Davydov, ``{Bogomolov multiplier, double class-preserving automorphisms, and
  modular invariants for orbifolds},''
  \href{http://dx.doi.org/10.1063/1.4895764}{{\em J. Math. Phys.} {\bfseries
  55} (2014) 092305}, \href{http://arxiv.org/abs/1312.7466}{{\ttfamily
  arXiv:1312.7466 [math.CT]}}.

\bibitem{Kobayashi:2025ykb}
R.~Kobayashi and M.~Barkeshli, ``{Soft symmetries of topological orders},''
  \href{http://arxiv.org/abs/2501.03314}{{\ttfamily arXiv:2501.03314
  [cond-mat.str-el]}}.

\bibitem{Kobayashi:2025pxs}
R.~Kobayashi and H.~Watanabe, ``{Projective Representations, Bogomolov
  Multiplier, and Their Applications in Physics},''
  \href{http://arxiv.org/abs/2507.12515}{{\ttfamily arXiv:2507.12515
  [cond-mat.str-el]}}.

\bibitem{Eliens:2013epa}
I.~S. Eli{\"e}ns, J.~C. Romers, and F.~A. Bais, ``{Diagrammatics for Bose
  condensation in anyon theories},''
  \href{http://dx.doi.org/10.1103/PhysRevB.90.195130}{{\em Phys. Rev. B}
  {\bfseries 90} no.~19, (2014) 195130},
  \href{http://arxiv.org/abs/1310.6001}{{\ttfamily arXiv:1310.6001
  [cond-mat.str-el]}}.

\bibitem{Etingof2009FusionCA}
P.~Etingof, D.~Nikshych, V.~Ostrik, and E.~Meir, ``{Fusion categories and
  homotopy theory},'' \href{http://arxiv.org/abs/0909.3140}{{\ttfamily
  arXiv:0909.3140 [math.QA]}}.

\bibitem{galindo2025categorical}
C.~Galindo and C.~F. Venegas-Ram{\'\i}rez, ``Categorical fermionic actions and
  minimal modular extensions,'' \href{https://sigma-journal.com/2025/085/}{{\em
  SIGMA} {\bfseries 21} (2025) 085},
  \href{http://arxiv.org/abs/1712.07097}{{\ttfamily arXiv:1712.07097
  [math.QA]}}.

\bibitem{NNW:2009}
D.~Naidu, D.~Nikshych, and S.~Witherspoon, ``{Fusion subcategories of
  representation categories of twisted quantum doubles of finite groups},''
  {\em Int. Math. Res. Not.} {\bfseries 2009} no.~22, (2009) 4183--4219,
  \href{http://arxiv.org/abs/0810.0032}{{\ttfamily arXiv:0810.0032 [math.QA]}}.

\bibitem{DGNO}
V.~Drinfeld, S.~Gelaki, D.~Nikshych, and V.~Ostrik, ``On braided fusion
  categories. {I},'' \href{https://doi.org/10.1007/s00029-010-0017-z}{{\em
  Selecta Math. (N.S.)} {\bfseries 16} no.~1, (2010) 1--119}.

\bibitem{Muger:2003structure}
M.~M{\"u}ger, ``{On the Structure of Modular Categories},''
  \href{http://dx.doi.org/10.1112/S0024611503014187}{{\em Proc. Lond. Math.
  Soc.} {\bfseries 87} no.~2, (2003) 291--308},
  \href{http://arxiv.org/abs/math/0201017}{{\ttfamily arXiv:math/0201017}}.

\bibitem{Lan:2016rcq}
T.~Lan, L.~Kong, and X.-G. Wen, ``{Modular Extensions of Unitary Braided Fusion
  Categories and 2+1D Topological/SPT Orders with Symmetries},''
  \href{http://dx.doi.org/10.1007/s00220-016-2748-y}{{\em Commun. Math. Phys.}
  {\bfseries 351} no.~2, (2017) 709--739},
  \href{http://arxiv.org/abs/1602.05936}{{\ttfamily arXiv:1602.05936
  [math.QA]}}.

\bibitem{Walker:2012mcd}
K.~Walker and Z.~Wang, ``{(3+1)-TQFTs and topological insulators},''
  \href{http://dx.doi.org/10.1007/s11467-011-0194-z}{{\em Front. Phys.
  (Beijing)} {\bfseries 7} no.~2, (2012) 150--159}.

\bibitem{JohnsonFreydReutter:2024}
T.~Johnson-Freyd and D.~Reutter, ``{Minimal nondegenerate extensions},''
  \href{http://dx.doi.org/10.1090/jams/1023}{{\em J. Am. Math. Soc.} {\bfseries
  37} no.~1, (2023) 81--150}, \href{http://arxiv.org/abs/2105.15167}{{\ttfamily
  arXiv:2105.15167 [math.QA]}}.

\bibitem{Fidkowski:2022fermioniccode}
L.~Fidkowski, J.~Haah, and M.~B. Hastings, ``Gravitational anomaly of
  $3+1$-dimensional {$\mathbb{Z}_2$} toric code with fermionic charges and
  fermionic loop self-statistics,''
  \href{http://dx.doi.org/10.1103/PhysRevB.106.165135}{{\em Phys. Rev. B}
  {\bfseries 106} (2022) 165135},
  \href{http://arxiv.org/abs/2110.14654}{{\ttfamily arXiv:2110.14654
  [cond-mat.str-el]}}.

\bibitem{GAP4}
``{GAP -- Groups, Algorithms, and Programming, Version 4.15.1},'' 2025.
\newblock \url{https://www.gap-system.org}.

\bibitem{TambaraYamagami}
D.~Tambara and S.~Yamagami, ``Tensor categories with fusion rules of
  self-duality for finite abelian groups,''
  \href{https://www.sciencedirect.com/science/article/pii/S0021869398975585}{{\em
  Journal of Algebra} {\bfseries 209} no.~2, (1998) 692--707}.

\bibitem{AlgebrasTQD}
A.~Warman, ``{AlgebrasTQD: Computing Algebras and Interfaces in Twisted Quantum
  Doubles},'' 2026.
\newblock \url{GitHub - AlisonWarman/AlgebrasTQD: GAP code for algebras in
  Twisted Quantum Doubles}. GAP code.

\bibitem{Teo:2015xla}
J.~C.~Y. Teo, T.~L. Hughes, and E.~Fradkin, ``{Theory of Twist Liquids: Gauging
  an Anyonic Symmetry},''
  \href{http://dx.doi.org/10.1016/j.aop.2015.05.012}{{\em Annals Phys.}
  {\bfseries 360} (2015) 349--445},
  \href{http://arxiv.org/abs/1503.06812}{{\ttfamily arXiv:1503.06812
  [cond-mat.str-el]}}.

\bibitem{WittGroup}
A.~{Davydov}, M.~{Mueger}, D.~{Nikshych}, and V.~{Ostrik}, ``{The Witt group of
  non-degenerate braided fusion categories},'' {\em arXiv e-prints} (Sept.,
  2010) arXiv:1009.2117, \href{http://arxiv.org/abs/1009.2117}{{\ttfamily
  arXiv:1009.2117 [math.QA]}}.

\bibitem{LouSandvikBalents:2007}
J.~Lou, A.~W. Sandvik, and L.~Balents, ``Emergence of {U(1)} symmetry in the
  {3D XY} model with {$Z_q$} anisotropy,''
  \href{http://dx.doi.org/10.1103/PhysRevLett.99.207203}{{\em Phys. Rev. Lett.}
  {\bfseries 99} (2007) 207203},
  \href{http://arxiv.org/abs/0704.1472}{{\ttfamily arXiv:0704.1472
  [cond-mat.str-el]}}.

\bibitem{EGNO}
P.~Etingof, S.~Gelaki, D.~Nikshych, and V.~Ostrik,
  \href{http://dx.doi.org/10.1090/surv/205}{{\em Tensor categories}}, vol.~205
  of {\em Mathematical Surveys and Monographs}.
\newblock American Mathematical Society, Providence, RI, 2015.
\newblock \url{https://doi.org/10.1090/surv/205}.

\bibitem{Ostrikmodule}
V.~{Ostrik}, ``{Module categories over the Drinfeld double of a finite
  group},'' {\em arXiv Mathematics e-prints} (Feb., 2002) math/0202130,
  \href{http://arxiv.org/abs/math/0202130}{{\ttfamily arXiv:math/0202130
  [math.QA]}}.

\bibitem{etingof2005fusion}
P.~Etingof, D.~Nikshych, and V.~Ostrik, ``On fusion categories,'' {\em Annals
  of Mathematics} (2005) 581--642.

\bibitem{Kitaev:1997wr}
A.~Y. Kitaev, ``{Fault tolerant quantum computation by anyons},''
  \href{http://dx.doi.org/10.1016/S0003-4916(02)00018-0}{{\em Annals Phys.}
  {\bfseries 303} (2003) 2--30},
  \href{http://arxiv.org/abs/quant-ph/9707021}{{\ttfamily
  arXiv:quant-ph/9707021}}.

\bibitem{Hu:2012wx}
Y.~Hu, Y.~Wan, and Y.-S. Wu, ``{Twisted quantum double model of topological
  phases in two dimensions},''
  \href{http://dx.doi.org/10.1103/PhysRevB.87.125114}{{\em Phys. Rev. B}
  {\bfseries 87} no.~12, (2013) 125114},
  \href{http://arxiv.org/abs/1211.3695}{{\ttfamily arXiv:1211.3695
  [cond-mat.str-el]}}.

\bibitem{Frohlich:2003hm}
J.~Frohlich, J.~Fuchs, I.~Runkel, and C.~Schweigert, ``{Correspondences of
  ribbon categories},'' \href{http://dx.doi.org/10.1016/j.aim.2005.04.007}{{\em
  Adv. Math.} {\bfseries 199} (2006) 192--329},
  \href{http://arxiv.org/abs/math/0309465}{{\ttfamily arXiv:math/0309465}}.

\bibitem{Hannah:2023tae}
S.~Hannah, R.~Laugwitz, and A.~R. Camacho, ``{Frobenius Monoidal Functors of
  Dijkgraaf-Witten Categories and Rigid Frobenius Algebras},''
  \href{http://dx.doi.org/10.3842/SIGMA.2023.075}{{\em SIGMA} {\bfseries 19}
  (2023) 075}, \href{http://arxiv.org/abs/2303.04493}{{\ttfamily
  arXiv:2303.04493}}.

\bibitem{naidu2007categorical}
D.~Naidu, ``Categorical morita equivalence for group-theoretical categories,''
  {\em Communications in Algebra{\textregistered}} {\bfseries 35} no.~11,
  (2007) 3544--3565.

\bibitem{Galindo2021EquivariantFS}
C.~Galindo and C.~Jones, ``Equivariant fusion subcategories,''
  \href{https://api.semanticscholar.org/CorpusID:244269972}{{\em Transformation
  Groups} {\bfseries 30} (2021) 267 -- 282}.

\bibitem{Bruguieres2000}
A.~Bruguières, ``Catégories prémodulaires, modularisations et invariants des
  variétés de dimension 3,''
  \href{http://dx.doi.org/10.1007/s002080050011}{{\em Mathematische Annalen}
  {\bfseries 316} (02, 2000) 215--236}.

\bibitem{MUGER2004256}
M.~M{\"u}ger, ``Galois extensions of braided tensor categories and braided
  crossed g-categories,'' {\em Journal of Algebra} {\bfseries 277} no.~1,
  (2004) 256--281.

\end{thebibliography}\endgroup

\end{document}